\documentclass[twocolumn]{aastex62}
 \pdfoutput=1
\usepackage{graphicx,float,subfigure} 
\usepackage[T1]{fontenc}
\usepackage{ae}
\usepackage{aecompl}
\usepackage{amsmath, amssymb, amsfonts, braket, gensymb} 
\usepackage{newtxtext,newtxmath} 
\usepackage{enumitem}
\def\url#1{\expandafter\string\csname #1\endcsname}

\received{\today}
\revised{???}
\accepted{???}
\submitjournal{ApJ}

\shorttitle{Light element abundance variations in GCs}
\shortauthors{Nataf et al.}

\begin{document}

\title{The Relationship Between Globular Cluster Mass, Metallicity, and Light Element Abundance Variations}
\correspondingauthor{David M. Nataf}
\email{dnataf1@jhu.edu, david.nataf@gmail.com}
\author{David M. Nataf}
\affiliation{Center for Astrophysical Sciences and Department of Physics and Astronomy,  The Johns Hopkins University, Baltimore, MD 21218}
\author{Rosemary Wyse}
\affiliation{Center for Astrophysical Sciences and Department of Physics and Astronomy,  The Johns Hopkins University,  Baltimore, MD 21218}
\author{Ricardo P. Schiavon}
\affiliation{Astrophysics Research Institute, Liverpool John Moores University,  146
Brownlow Hill, Liverpool L3 5RF, UK}
\author{Yuan-Sen Ting}
\affiliation{Institute for Advanced Study, Princeton, NJ 08540, USA}
\affiliation{Department of Astrophysical Sciences, Princeton University, Princeton, NJ 08544, USA}
\affiliation{University, Princeton, Observatories of the Carnegie Institution of Washington, 813 Santa Barbara Street, Pasadena, CA 91101, USA}
\author{Dante Minniti} 
\affiliation{Departamento de Ciencias Fisicas, Facultad de Ciencias Exactas, Universidad Andres Bello, Av. Fernandez Concha 700, Las Condes, Santiago, Chile}
\affiliation{Millennium Institute of Astrophysics, Av. Vicuna Mackenna 4860, 782-0436, Santiago, Chile}
\affiliation{Vatican Observatory, V00120 Vatican City State, Italy}
\author{Roger E. Cohen}
\affiliation{Space Telescope Science Institute, 3700 San Martin Drive, Baltimore, MD 21218, USA}
\author{Jos\'e G. Fern\'andez-Trincado}
\affiliation{Instituto de Astronom\'ia y Ciencias Planetarias, Universidad de Atacama, Copayapu 485, Copiap\'o, Chile}
\affiliation{Institut Utinam, CNRS UMR 6213, Universit\'e Bourgogne-Franche-Comt\'e, OSU THETA Franche-Comt\'e, Observatoire de Besan\c{c}on, \\ BP 1615, F-25010 Besan\c{c}on Cedex, France}
\affiliation{Departamento de Astronom\'\i a, Casilla 160-C, Universidad de Concepci\'on, Concepci\'on, Chile}
\author{Douglas Geisler}
\affiliation{Departamento de Astronom\'\i a, Casilla 160-C, Universidad de Concepci\'on, Concepci\'on, Chile}
\affiliation{Instituto de Investigaci\'on Multidisciplinario en Ciencia y Tecnolog\'ia, Universidad de La Serena. Avenida Ra\'ul Bitr\'an S/N, La Serena, Chile}
\affiliation{Departamento de F\'isica y Astronom\'ia, Facultad de Ciencias, Universidad de La Serena. Av. Juan Cisternas 1200, La Serena, Chile}
\author{Christian Nitschelm}
\affiliation{Centro de Astronom{\'i}a (CITEVA), Universidad de Antofagasta, Avenida Angamos 601, Antofagasta 1270300, Chile}
\author{Peter M. Frinchaboy}
\affiliation{Department of Physics \& Astronomy, Texas Christian University, TCU Box 298840, Fort Worth, TX 76129, USA}


\begin{abstract}
We investigate aluminum abundance variations in the stellar populations of globular clusters using both literature measurements of sodium and aluminum and APOGEE measurements of nitrogen and aluminum abundances. For the latter, we show that the Payne is the most suitable of the five available abundance pipelines for our purposes.  Our combined sample of 42 globular clusters spans approximately 2 dex in [Fe/H] and 1.5 dex in $\log{M_{GC}/M_{\odot}}$. We find no fewer than five globular clusters with significant internal variations in nitrogen and/or sodium with little-to-no corresponding variation in aluminum, and that the minimum present-day cluster mass for aluminum enrichment in metal-rich systems is  $\log{M_{GC}/M_{\odot}} \approx 4.50 + 2.17(\rm{[Fe/H]}+1.30)$. We demonstrate that the slopes of the [Al/Fe] vs [Na/Fe] and [Al/Fe] vs [N/Fe] relations for stars without field-like abundances are approximately log-linearly dependent on both the metallicity and the stellar mass of the globular clusters. In contrast, the relationship between [Na/Fe] and [N/Fe] shows no evidence of such dependencies. This suggests that there were (at least) two classes of non-supernovae chemical polluters that were common in the early universe, and that their relative contributions within globular clusters somehow scaled with the metallicity and mass of globular clusters. The first of these classes is predominantly responsible for the CNO and NeNa abundance variations, and likewise the second for the MgAl abundance variations. 47 Tuc and M4 are particularly striking examples of this dichotomy. 
As an auxiliary finding, we argue that abundance variations among Terzan 5 stars are consistent with it being a normal globular cluster. 
\end{abstract}

\section{Introduction}  \label{sec:Introduction}

It is now firmly established that stars showing light-element abundance variations are  ubiquitous or nearly ubiquitous within globular clusters \citep{1971Obs....91..223O,2009A&A...505..117C,2015AJ....149..153M}. The most commonly measured of these is the sodium-oxygen anti-correlation, whereby stars within a cluster will often have greater [Na/Fe] and lower [O/Fe] values relative to that characteristic of field stars with the same [Fe/H] as the cluster. Observations have separately demonstrated that the [C,N,O/Fe]  abundances of globular cluster stars vary both relative to one another \citep{1999AJ....118.1273I} and sometimes in total abundance \citep{2012ApJ...746...14M}, with many of the stars having field-like abundances. Similarly, stars with relatively lower abundances of magnesium (likely ${}^{24}$Mg) also tend to have relatively elevated abundances of aluminum, with the total abundance of aluminum and magnesium not varying within individual globular clusters \citep{1996AJ....112.2639S,2003A&A...402..985Y}.

There is thus a wealth of observational constraints as to the nature of these abundance variations, which have long been argued to date back to the gas from which the stars formed \citep{1981ApJ...245L..79C}. Nevertheless, the literature currently contains no consistent model of globular cluster formation which matches all of these constraints. For example, it is frequently posited that the globular clusters first formed a population of stars with [X/Fe] abundance ratios similar to those of most $\alpha$-enhanced halo, thick disk, and bulge field stars, and that the gaseous ejecta of these stars was subsequently recycled to form a population with chemically anomalous abundances. These stellar groups respectively form a ``first" and ``second" generation \citep{2011A&A...533A.120V,2012ApJ...758...21C,2018ApJ...869...35K}. 

These abundance anti-correlations can in principal be matched nuclear processing within a high-temperature gas \citep{2007A&A...470..179P}.  An ongoing challenge to this model is that the observed abundance variations are not matched by predicted yields of the main hypothesized chemical polluters -- which will necessarily contain gradients in temperature, density, and chemistry -- asymptotic giant branch stars (AGB, \citealt{2001ApJ...550L..65V,2009A&A...499..835V}) and the winds of fast-rotating massive stars (WFRMS, \citealt{2007A&A...464.1029D}).  Some of that offset may be due to uncertainties in the stellar models and their input physics. 

\citet{2018MNRAS.477..421K}, who computed the predicted yields of low-metallicity, $1 \leq M/M_{\odot} \leq 7$ AGB stars, showed that the predicted yields of AGB stars provide a qualitative match to the CNO abundance variations observed in globular cluster stars, but not to the Na-O anti-correlation. For the latter, the models predict a $\sim$0.60 \textit{depletion} in [Na/Fe], in contrast to the $\sim$0.70 dex \textit{enhancement} found in clusters such as 47 Tuc.  \citet{2017MNRAS.465.4817S} predicted the AGB yields of $3 \leq M/M_{\odot} \leq 6$ stars with ${}^{22}$Na(p,$\alpha$)${}^{23}$Ne nuclear reaction rates updated from experimental data from the Laboratory for Underground Nuclear Astrophysics \citep{2015PhRvL.115y2501C}, which are up to 2,000\% higher than those of \citet{2010NuPhA.841...31I}, that were employed by \citet{2018MNRAS.477..421K}. They found that their models could provide a satisfying fit to both the CNO and Na-O abundance trends, but only if they both included the new reaction rate and made other substantial changes to their treatment of stellar evolution, dredge-up, and imposed an arbitrary 80\% reduction in the ${}^{23}$Na(p,$\alpha$)${}^{20}$Ne reaction rate. \citet{2018PhRvL.121q2701F} investigated whether the inclusion of three newly-identified resonances to the computation of the  ${}^{23}$Na(p,$\alpha$)${}^{20}$Ne reaction rate could alleviate these discrepancies. They found that the predicted yields of sodium are increased by $\sim$0.20 dex.

Similarly, models of the WFRMS can also produce gas similar to that observed in chemically anomalous globular cluster stars, but only after fine-tuning the assumed treatments of convection, nuclear reaction rates, etc. For example, \citet{2016A&A...593A..36C} showed that various prescriptions for the injection of helium-burning products into the hydrogen-burning zone could increase the production of nitrogen and aluminum by two orders of magnitude. They did identify some results that were robust to such modifications: all of their models predicted greater enrichment of aluminum than of magnesium, and of nitrogen than of carbon or oxygen. 

 Independently of the challenge posed by chemical abundance measurements, the ``multiple generations" framework for globular cluster formation also has a mass normalization issue. The total stellar mass currently in globular cluster populations with anomalous abundances is $\sim 1-2 \times$ higher than in globular cluster populations with field-like abundances \citep{2009A&A...505..117C}. Thus, the gaseous ejecta of the first generation can only have been sufficiently massive to give birth to the second generation if: the first generation was 10$+$ times more massive at birth; either the first or second generations were both with a non-standard initial mass function; or both \citep{2008MNRAS.391..825D,2011A&A...533A.120V,2012ApJ...758...21C,2018arXiv180702309B}. This issue is referred to as the ``mass budget problem" \citep{2008MNRAS.391..354R,2015MNRAS.454.4197R}. 

These and other outstanding issues related to the question of the origin of globular clusters are discussed in greater detail by \citet{2015MNRAS.454.4197R} and \citet{2018ARA&A..56...83B}.

Though the currently available globular cluster formation models leave a lot to be desired, the same cannot be said of the available data. The latter are spectacular. Their color-magnitude diagrams show the presence of distinct populations, with the morphology of the separation being a function of which filters are used. For example, the fact that the relatively metal-rich ([Fe/H]$\approx -1.20$) main sequence of $\omega$ Cen is bluer than its metal-poor ([Fe/H]$\approx -1.60$) counterpart can be interpreted as a helium enhancement of order ${\Delta} Y \approx 0.15$ \citep{2004ApJ...612L..25N,2005ApJ...621..777P}, as enhanced helium is predicted to result in higher temperatures at fixed luminosity for main sequence and red giant branch stars, and thus bluer broadband colours \citep{2008ApJS..178...89D}
. Similarly, a triple main sequence is seen in the massive globular cluster NGC 2808 \citep{2007ApJ...661L..53P}. The helium values inferred from studies of the main sequence colours are consistent with those inferred from modelling of the horizontal branch morphology \citep{2004ApJ...611..871D,2008MNRAS.390..693D}. Separately,   CNO abundance variations are predicted to affect the absorption of several molecular features, leading to particularly strong photometric offsets in the widely used \textit{Hubble Space Telescope (HST)} $F275W$ and $F336W$ filters. (See Section 3 \citet{2017MNRAS.464.3636M}, and Figure 6 of \citealt{2018MNRAS.475.4088L}). 

Meanwhile, the spectroscopic data are vast. The \textit{Very Large Telescope (VLT)} program of  \citet{2009A&A...505..117C} measured sodium, oxygen, and iron abundances for 1,409 stars in 15 globular clusters with the  FLAMES-GIRAFFE spectrograph, and \citet{2009A&A...505..139C} measured magnesium, silicon, and aluminum abundances for 202 stars in 17 globular clusters with the FLAMES-UVES spectrograph. These studies helped establish the diagnostic potential of multi-object spectrographs for the study of globular cluster populations, the ubiquity of chemically anomalous stars within, and the correlation of these anomalies with globular cluster parameters such as metallicity and integrated luminosity.

\citet{2017A&A...601A.112P}, who analyzed a sample of  globular clusters clusters observed by the Gaia-ESO survey, showed that the Na-O anti-correlation operates independently of the Mg-Al anti-correlation. Whereas observations are consistent with the Na-O anti-correlation being present in all globular clusters (with the possible exception of Ruprecht 106, see \citealt{2013ApJ...778..186V} and \citealt{2018ApJ...865L..10D}), the Mg-Al anti-correlation is smaller or non-existent in higher-metallicity and in lower-mass globular clusters. The different anti-correlations do not always appear together, and thus need not be explained by a single origin. 

The APOGEE survey \citep{2017AJ....154...94M} has also yielded numerous insights on the chemistry of globular clusters. \citet{2017MNRAS.466.1010S} and \citet{2018arXiv180107136F} have identified and measured the trends of multiple populations for several Bulge globular clusters. \citet{2017MNRAS.465...19T} identified two groups in the chemical abundance space of NGC 6553, and \citet{2018ApJ...855...38T} studied the aluminum, magnesium, and silicon variations in the metal-poor globular cluster NGC 5053. \citet{2015AJ....149..153M} used earlier APOGEE data to explore the abundance variations in 10 northern globular clusters, a study recently followed up and expanded upon by \citet{2018arXiv181208817M}. Two major strengths of the APOGEE data result from its spectral window being in the $H$-band: the presence of numerous CNO molecular feature enables the consistent measure of all three of these abundances, and the lower sensitivity to extinction facilitates the study of Disk and Bulge globular clusters. 

The three  nuclear reactions most relevant to the abundance trends previously discussed here are plausibly ${}^{23}$Na(p,$\alpha$)${}^{20}$Ne, ${}^{14}$N(p,$\gamma$)${}^{15}$O, and ${}^{24}$Mg(p,$\gamma$)${}^{25}$Al \citep{2007A&A...464.1029D,2010ARNPS..60..381W}. Their rates have different dependencies on temperature \citep{2001ApJS..134..151I,2004PhRvC..70d5802H}, and thus, given the plausibility that different kinds of stars contribute to chemical enrichment in different kinds of globular clusters, there should be no expectation that the different abundance correlations occur in lockstep, which is indeed what \citet{2017A&A...601A.112P} has demonstrated. 

In this investigation, we studied the largest sample of [Al/Fe] variations in globular clusters that we could assemble, so as to better calibrate if and how the Mg-Al abundance anti-correlation separates from the CNONa abundance variations. The result is a meta-analysis of available literature measurements as well as those from the APOGEE survey \citep{2017AJ....154...94M}. The latter could not be trivially incorporated. We wanted to include the full list of globular cluster stars observed by APOGEE, not just those explicitly targeted but also those serendipitously observed, for which there was no prior census. We also had to determine which, if any, of the five available pipelines was best suited to study abundance variations among globular cluster stars, and in which atmospheric parameter regime. That analysis had not yet been performed, nor can one pipeline be \textit{a priori} assumed to be superior. Our investigation thus necessarily contains an investigation-within-an-investigation. That is the census of globular cluster stars observed by APOGEE, and the assessment of this sample and of the five spectroscopic abundance pipelines for suitability to study chemical abundance variations among globular cluster stars. 

The structure of this paper is as follows. We describe the data assembled in Section \ref{sec:Data}. We describe our census of globular cluster stars measured by APOGEE, the range of stellar atmospheric parameter space in which they are suitable for chemical abundance studies, and their observed chemical abundance trends, in Section \ref{sec:Census}. We describe the meta-analysis of aluminum abundance variations in the combined literature and APOGEE samples in Section \ref{sec:Analysis}. We present what are arguably the principal findings of this investigation in Section \ref{subsec:fullsample}, and discuss the possible physical implications in Section \ref{subsec:Physics}.  We apply our results to the globular cluster Terzan 5 in Section \ref{sec:Terzan5}. We discuss our findings and present our conclusions in Section \ref{sec:Conclusion}. 

\subsection{The Need to Understand Nitrogen-Enriched Stars in the Field}
\label{subsec:Martell}
A part of our motivation for this study is the recent discovery that stars with abundances typical of chemically anomalous globular cluster stars are common in the field. These were first discovered by \citet{2010A&A...519A..14M} toward the Milky Way's Halo, with subsequent findings having followed toward both the Halo \citep{2012ApJ...757..164R,2016ApJ...825..146M,2016ApJ...833..132F} and toward the Bulge / inner Halo \citep{2017MNRAS.465..501S,2017ApJ...846L...2F}. At this time, it is not known if these stars are former members of surviving globular clusters, former members of now fully dissolved globular clusters, or if they formed via a different channel. 

We aim to study those stars in greater detail, using the APOGEE spectroscopic catalog of field stars. However, as a prerequisite, we must first evaluate how well the spectroscopic pipelines that have been applied to APOGEE data perform at measuring globular cluster abundances, and to try and ascertain a physical meaning for the abundances that are being reliably measured. 

\section{Data} \label{sec:Data}

\subsection{Globular Cluster Parameters} \label{subsec:GCparameters}

We used the Harris catalog \citep{1996AJ....112.1487H} for initial estimates of the tidal radii, values of the positions, and all but two of the metallicities [Fe/H] (hereafter referred to as [Fe/H]$_{\rm{Harris}}$) of the globular clusters. 

We used the recent analysis and compilation of  \citet{2018MNRAS.478.1520B} for updated values of the  radial velocity dispersions, tidal radii. This latter compilation includes parameters for all clusters for which we found candidate members other than Terzan 12. The cluster masses and structural parameters were determined by comparing the observed velocity dispersion, surface density, and stellar mass function profiles against a grid of N-body simulations which assumed that the globular clusters contain no dark matter. The method  is more fully described by \citet{2017MNRAS.464.2174B}. 

The estimated values for cluster radial velocities, mean proper motions in right ascension and declination, and central proper motion dispersion were taken from \citet{2019MNRAS.482.5138B}, where available. That work makes use of data and results by \citet{2015ApJ...803...29W}, \citet{2017MNRAS.464.2174B}, \citet{2018MNRAS.473.5591K}, and \citet{2018MNRAS.478.1520B}. 

Most of these estimates are available for download on a website\footnote{https://people.smp.uq.edu.au/HolgerBaumgardt/globular/} maintained by Holger Baumgardt, which lists up-to-date fundamental parameters for over 150 globular clusters, including the stellar mass estimates used in this work. 

We make a few targeted changes to the adopted parameters. For the globular cluster NGC 6522, we use the same parameters as \citet{2018arXiv180107136F}, as their diligent investigation of that cluster's overlap with APOGEE has already been vetted. The adopted metallicity for that cluster, [Fe/H]$=-1.0$, is from \citet{2009A&A...507..405B} and \citet{2014A&A...570A..76B}. For the globular cluster NGC 3201, we adopted the metallicity [Fe/H] = $-1.46$ \citep{2015ApJ...801...69M}. For the globular cluster Terzan 5, we use the tidal radii from the website linked above, which assumes Equation 8 of \citet{2013ApJ...764..124W}. 



\subsection{Sodium and Aluminum Literature Compilation} \label{subsec:aluminum}
We conducted a literature search for abundance measurements of sodium and aluminum in well-sampled globular clusters. For each globular cluster, we kept the largest available sample. 

The largest component of our literature compilation is that of \citet{2009A&A...505..139C}, who reported measurements of  [O/Fe], [Na/Fe], [Mg/Fe], [Al/Fe], and [Si/Fe] for 202 red giants in 17 globular clusters. That work is part of a series, from which additional measurements, obtained with the same methodology, are available for the clusters NGC 6441 \citep{2006A&A...455..271G} and  NGC 6388  \citep{2007A&A...464..967C}. The same instrument and a similar methodology was used for published measurements of stars in NGC 1851 \citep{2012A&A...543A.117C}, NGC 362 \citep{2013A&A...557A.138C}, NGC 6093 (M80)  \citep{2015A&A...578A.116C}, NGC 2808 \citep{2015ApJ...810..148C,2018A&A...615A..17C}, NGC 6139 \citep{2015A&A...583A..69B}, and additional measurements for NGC 6388 \citep{2018A&A...614A.109C}.


The second largest component of our compilation are the abundance measurements of NGC 5986, NGC 6229, and NGC 6569 by \citet{2017ApJ...842...24J}, \citet{2017AJ....154..155J}, and \citet{2018AJ....155...71J} respectively. We also include measurements for NGC 362 and NGC 1904 (M79) \citep{2015MNRAS.449.4038D}; Gaia-ESO measurements \citep{2012Msngr.147...25G} of NGC 5927 and NGC 4833 which were previously investigated by \citet{2017A&A...601A.112P};  measurements for NGC 5897 \citep{2014A&A...565A..23K}; for NGC 6681 \citep{2017ApJ...846...23O} and NGC 6584 \citep{2018ApJ...856..130O};  NGC 6362 \citep{2016ApJ...824...73M,2017MNRAS.468.1249M}; NGC 6397 \citep{2018MNRAS.475..257M}; NGC 6440 \citep{2017A&A...605A..12M};  NGC 6528 \citep{2018A&A...620A..96M}; NGC 4147 \citep{2016MNRAS.460.2351V}  NGC 6626 \citep{2017MNRAS.464.2730V}; and  NGC 6266 (M62) \citep{2015ApJ...813...97L}. 

We exclude from this compilation data for the globular clusters $\omega$ Cen (NGC 5139), NGC 6273 (M19), and Terzan 5 due to the enormous internal variations in [Fe/H] (see respective measurements by\citealt{2010ApJ...722.1373J}, \citealt{2017ApJ...836..168J}, and \citealt{2014ApJ...795...22M}); data where a large fraction of the measurements are upper or lower bounds; and data published earlier than January 1st, 2000. 

Table 1 of \citet{2016MNRAS.460.1869C}, which also compiles a literature sample of aluminum abundance variations, helped inform our sample.

\subsection{APOGEE Data} \label{subsec:apogeedata}

We use APOGEE and APOGEE-2 \citep{2017AJ....154...94M} data products (stellar atmosphere parameters and abundances) from Data Release 14 (DR14, \citealt{2018ApJS..235...42A}) of the Sloan Digital Sky Survey \citep{2006AJ....131.2332G}. APOGEE was a component of SDSS-III \citep{2011AJ....142...72E}, and APOGEE-2 is part of SDSS-IV \citep{2017AJ....154...28B}. Observations were taken from  the 2.5m Sloan Telescope at Apache Point Observatory \citep{2006AJ....131.2332G}, which is coupled to a 300-fiber, high-resolution (R $\sim$ 22,000) $H$-band spectrograph \citep{2012SPIE.8446E..0HW}.

We make partial use of previously unpublished APOGEE DR16 data, for which the analysis is restricted to materials discussed in Section \ref{sec:Analysis}. As these data have not been as strictly vetted, we make use of strict inclusion criteria: $v_{\rm{Macro}} \leq$ 10 km/s, $\chi^2 \leq$ 20, [N/Fe] $\geq -0.20$, $|\rm{[Ca/Fe]}+0.15| \leq 0.35$, $|\rm{[Ti/Fe]}+0.30| \leq 0.30$, and  $\rm{|Ni/Fe]+0.10|} \leq 0.30$, where all values are estimated by the Payne pipeline (described in Section \ref{subsec:stellarparam}). 

APOGEE targets were selected predominantly on the basis of 2MASS photometry \citep{2006AJ....131.1163S} to lie in the brightness range $7 \lesssim H \lesssim 13.8$. A full description of the target selection can be found in \citet{2013AJ....146...81Z} and \citet{2017AJ....154..198Z}. The data reduction and radial velocity pipelines are described by \citet{2015AJ....150..173N}. 



\subsection{Stellar Abundances and Parameters} \label{subsec:stellarparam}

We investigated the potential of five pipelines for the measurement of atmospheric parameters and chemical abundances of globular cluster stars. These are the five pipelines with results published for the full DR14 sample.  For all cases in this Section, where the pipelines are being compared, we use DR14-calibrated data, and restrict our analysis to the publicly available DR14 data. Given the challenging nature of comparing different pipelines \citep{2018arXiv181108041J}, this restriction is desirable, as it facilitates a fairer comparison.

The APOGEE Stellar Parameter and Abundances Pipeline (ASPCAP, \citealt{2016AJ....151..144G}), derives atmospheric parameters  ($T_{\rm{eff}}$, $\log{g}$, $v_{mic}$, [M/H], [$\alpha$/M]), with these values used for the subsequent derivation of individual chemical abundances [X$_{i}$/H]. Each spectra is analyzed independently in an automated manner. 

The Cannon \citep{2015ApJ...808...16N,2016arXiv160303040C} derives atmospheric parameters and abundances by means of a data-driven method. A subsample of spectra with ASPCAP-derived parameters deemed to be particularly reliable form a ``training set," from which the parameters of other stars are derived via quadratic interpolation. 

The Payne \citep{2018arXiv180401530T} simultaneously derives best-fit values for all atmospheric parameters and abundances using neural networks as an emulator. 
The training set is composed of synthetic spectra, which are constructed from a different suite of model atmospheres and assume a different line list than ASPCAP, see \citet{2018arXiv180401530T} for details. A disadvantage of the first Payne-derived data release of APOGEE abundances is that the parameter space of the training set was restricted to [Fe/H] $\geq -$1.50. That was a choice of the analysis, not intrinsic to the Payne itself, and can be adjusted for future data releases.  Thus, we thus restrict our evaluation of the Payne to globular clusters with [Fe/H]$_{\rm{Harris}}\geq -1.55$. The additional 0.05 dex is a small extrapolation that we allow ourselves to include NGC 6205 (M13) in the sample. We show, later in this work, that the measured relative abundance trends for NGC 6205 are reasonable.

AstroNN uses artificial neural networks with dropouts to simultaneously fit for atmospheric parameters and abundances \citep{2019MNRAS.483.3255L}. The theory of their method is described in Section 2 of that work, and the selection of their training set is described in Section 3 of that work. Similarly to the Cannon, their training set is selected from stars with ASPCAP-derived atmospheric parameters deemed to be particularly reliable, with the values of those parameters assumed by the calibration. In contrast to the other methods, astroNN does not perform a fit, but rather, it maps spectra onto spectral labels. It thus derives labels for nearly all stars. In principle, it should report larger uncertainty estimates for stars with spectra that are somehow problematic. 

The Brussels Automatic Code for Characterizing HighaccUracy Spectra (BACCHUS) \citep{2016ascl.soft05004M} determines metallicity, microturbulence and macroturbulence/$v\sin i$ as well as several relative abundances. The BACCHUS-derived parameters used in this work assume photometric temperatures and gravities, and several of the atmospheric parameters and relative abundances are subsequently determined using a sequential and iterative process more fully described by \citet{2018arXiv181208817M}. This temperature scale gives more consistent CNO abundances with the  optical studies, most likely because it is closer to the temperatures derived from the  optical spectra, as otherwise APOGEE-derived temperatures and those derived from optical spectra have a typical offset of $\sim$100 K \citep{2018AJ....156..126J}. 

Within the work of \citet{2018arXiv181208817M}, 885 red giant stars observed by APOGEE were associated to ten globular clusters. We restrict our evaluation of the BACCHUS pipeline and its implementation to those stars. 


The DR14 data tables have 277,371 stellar targets, but many of these are duplicates and thus there are in fact 258,475 distinct stars. For the stars that are observed two or more times, we keep the best measurement (defined below) from each of ASPCAP, the Cannon, and astroNN, none of which necessarily correspond to one another or to the measurement with the highest signal-to-noise ratio (SNR). The first pass is the reporting  of a physical value, 
where the pipelines converge to a solution. When multiple observations of a duplicate yield physical estimates, we keep that measured with the highest signal-to-noise ratio (SNR). For astroNN, we implement a different duplicate-removal method due to the different philosophy of that pipeline. 
We instead first remove the duplicates where one star has the unlikely [Fe/H] $< -5.0$, then we remove the duplicates with the larger estimated error in [Fe/H]. This is not an issue for the Payne catalog, as for each source it only reports values for the measurement with the highest SNR.  The Payne reports measurements for 222,707 of the 258,475 stars in the APOGEE DR14 catalog. 




\section{Census and Evaluation of APOGEE Stars Within Globular Clusters} \label{sec:Census}

\subsection{Selection of an APOGEE Globular Cluster Sample} \label{subsec:GCsample}

\begin{table*}
\centering
\small\addtolength{\tabcolsep}{-1.0pt}
\caption{Name, metallicities, and physical parameters for the 28 globular clusters with stars measured in APOGEE DR14, and 11 clusters selected from DR16. The references for the globular cluster parameters are given in Section \ref{subsec:GCparameters}.  The mean velocity $V_{r}$ and velocity dispersion $\sigma_{Vr}$  are listed in units of km/s, the proper motion terms are in units of mas/yr, and the tidal radii r$_{\rm{tidal}}$ are listed in units of arcminutes.  The parameters N$_{\rm{A}}$, N$_{\rm{C}}$, N$_{\rm{P}}$, and N$_{\rm{aNN}}$ denote the number of matches in each cluster with a reported [Fe/H] measurement from that particular pipeline. Similarly for N$_{\rm{B}}$, for which the sample inclusion criteria are different, and described by \citet{2018arXiv181208817M}.} 
\begin{tabular}{|ll| cccccccc | rrrrr|}
\hline
Name & Alt. Name & [Fe/H] & $V_{r}$ & $\sigma_{Vr}$ & $\mu_{\alpha}\cos{\delta}$ &  $\mu_{\delta}$  & $\sigma_{\mu}$ &   r$_{\rm{tidal}}$ & $\log_{10}(M_{\rm{GC}}/M_{\odot})$ & N$_{\rm{A}}$ & N$_{\rm{C}}$ & N$_{\rm{P}}$ & N$_{\rm{aNN}}$ & N$_{\rm{B}}$ \\ 
\hline
NGC 7078 & M15 & $-$2.37 & $-$106.50 & 12.90 & $-$0.63 & $-$3.80 & 0.15 & 27.3 & 5.70 & 80 & 120 & 5 & 134 &  138  \\
NGC 6341 & M92 & $-$2.31 & $-$120.70 & 8.00 & $-$4.93 & $-$0.57 & 0.11 & 12.4 & 5.49 & 40 & 70 & 0 & 73 &  72  \\
NGC 5053 & -- & $-$2.27 & 42.50 & 1.60 & $-$0.37 & $-$1.26 & -- & 11.4 & 4.75 & 7 & 9 & 0 & 9 &  --  \\
NGC 5024 & M53 & $-$2.10 & $-$63.10 & 5.90 & $-$0.11 & $-$1.35 & 0.05 & 18.4 & 5.54 & 24 & 36 & 0 & 37 &  40  \\
NGC 5466 & -- & $-$1.98 & 106.90 & 1.60 & $-$5.41 & $-$0.79 & -- & 15.7 & 4.65 & 11 & 12 & 1 & 12 &  --  \\
NGC 5634 & -- & $-$1.88 & $-$16.20 & 5.30 & $-$1.67 & $-$1.55 & -- & 10.6 & 5.33 & 1 & 2 & 0 & 2 &  --  \\
NGC 4147 & -- & $-$1.80 & 179.10 & 3.10 & $-$1.71 & $-$2.10 & -- & 6.1 & 4.50 & 3 & 3 & 0 & 3 &  --  \\
NGC 7089 & M2 & $-$1.65 & $-$3.60 & 10.60 & 3.51 & $-$2.16 & 0.18 & 12.4 & 5.71 & 26 & 22 & 14 & 26 &  26  \\
NGC 6205 & M13 & $-$1.53 & $-$244.40 & 9.20 & $-$3.18 & $-$2.56 & 0.20 & 21.0 & 5.66 & 126 & 125 & 97 & 136 &  135  \\
NGC 5272 & M3 & $-$1.50 & $-$147.20 & 8.10 & $-$0.14 & $-$2.64 & 0.15 & 28.7 & 5.58 & 129 & 119 & 110 & 143 &  145  \\
NGC 6715 & M54 & $-$1.49 & 142.30 & 16.20 & $-$2.73 & $-$1.38 & 0.08 & 9.9 & 6.20 & 6 & 8 & 8 & 8 &  --  \\
NGC 6229 & -- & $-$1.47 & $-$138.30 & 7.10 & $-$1.19 & $-$0.46 & -- & 3.8 & 5.47 & 6 & 6 & 6 & 6 &  --  \\
Pal 5 & -- & $-$1.41 & $-$58.40 & 0.60 & $-$2.77 & $-$2.67 & -- & 7.6 & 4.23 & 4 & 4 & 4 & 4 &  --  \\
NGC 6544 & -- & $-$1.40 & $-$36.40 & 6.40 & $-$2.34 & $-$18.66 & 0.49 & 2.1 & 5.06 & 2 & 2 & 2 & 2 &  --  \\
NGC 6218 & M12 & $-$1.37 & $-$41.20 & 4.50 & $-$0.15 & $-$6.77 & 0.18 & 17.3 & 4.91 & 57 & 54 & 61 & 63 &  --  \\
NGC 5904 & M5 & $-$1.29 & 53.80 & 7.70 & 4.06 & $-$9.89 & 0.18 & 23.6 & 5.56 & 202 & 187 & 214 & 217 &  218  \\
NGC 6517 & -- & $-$1.23 & $-$39.60 & 15.00 & $-$1.49 & $-$4.23 & 0.17 & 4.0 & 5.56 & 0 & 1 & 1 & 1 &  --  \\
NGC 6171 & M107 & $-$1.02 & $-$34.70 & 4.30 & $-$1.93 & $-$5.98 & 0.14 & 19.0 & 4.95 & 57 & 63 & 66 & 66 &  67  \\
NGC 6522 & -- & $-$1.00 & $-$21.10 & 15.00 & 2.62 & $-$6.40 & 0.17 & 7.2 & 5.56 & 5 & 5 & 5 & 5 &  --  \\
Pal 6 & -- & $-$0.91 & 181.00 & 15.00 & $-$9.17 & $-$5.26 & 0.19 & 8.3 & 5.13 & 3 & 3 & 3 & 3 &  --  \\
NGC 6838 & M71 & $-$0.78 & $-$22.50 & 3.30 & $-$3.41 & $-$2.61 & 0.17 & 8.9 & 4.73 & 27 & 26 & 29 & 29 &  28  \\
Pal 1 & -- & $-$0.65 & $-$82.80 & 15.00 & $-$0.17 & 0.03 & -- & 3.7 & 3.25 & 1 & 1 & 2 & 2 &  --  \\
NGC 6539 & -- & $-$0.63 & 35.60 & 5.90 & $-$6.82 & $-$3.48 & 0.14 & 20.9 & 5.40 & 0 & 1 & 1 & 1 &  --  \\
Terzan 12 & -- & $-$0.50 & 94.10 & 15.00 & $-$6.07 & $-$2.63 & -- & 3.1 & 3.13 & 1 & 1 & 1 & 1 &  --  \\
NGC 6760 & -- & $-$0.40 & $-$1.60 & 7.20 & $-$1.11 & $-$3.59 & 0.17 & 15.2 & 5.43 & 8 & 8 & 9 & 9 &  --  \\
Terzan 5 & -- & $-$0.23 & $-$82.30 & 19.00 & $-$1.71 & $-$4.64 & 0.48 & 23.8 & 5.59 & 7 & 8 & 9 & 9 &  --  \\
NGC 6553 & -- & $-$0.18 & 0.50 & 8.50 & 0.30 & $-$0.41 & 0.21 & 7.7 & 5.52 & 9 & 9 & 9 & 9 &  --  \\
NGC 6528 & -- & $-$0.11 & 211.00 & 6.40 & $-$2.17 & $-$5.52 & 0.14 & 4.1 & 4.97 & 2 & 2 & 2 & 2 &  --  \\
\hline
Total (DR14) & -- & --     & -- & -- & --     & --    & -- & -- & -- & 844 & 907 & 659 & 1012 & 885 \\
\hline
NGC 1904 & M79 & $-$1.60 & 205.60 & 6.50 & 2.5 & $-$1.59 & -- & 8.02 & 5.23 & -- & -- & 27 & -- & --\\
NGC 6254 & M10 & $-$1.56 & 74.00 & 6.20 & $-$4.7 & $-$6.54 & 0.24 & 18.47 & 5.27 & -- & -- & 88 & -- & --\\
NGC 6752 & -- & $-$1.54 & $-$26.20 & 8.30 & $-$3.2 & $-$4.01 & 0.29 & 53.76 & 5.36 & -- & -- & 156 & -- & --\\
NGC 3201 & -- & $-$1.46 & 494.30 & 4.50 & 8.3 & $-$2.00 & 0.20 & 25.35 & 5.13 & -- & -- & 142 & -- & --\\
NGC 6218 & M12 & $-$1.37 & $-$41.20 & 4.50 & $-$0.1 & $-$6.77 & 0.18 & 17.28 & 4.91 & -- & -- & 48 & -- & --\\
NGC 362 & -- & $-$1.26 & 223.50 & 8.80 & 6.7 & $-$2.51 & 0.14 & 10.36 & 5.52 & -- & -- & 45 & -- & --\\
NGC 1851 & -- & $-$1.18 & 320.20 & 10.20 & 2.1 & $-$0.63 & 0.11 & 6.52 & 5.45 & -- & -- & 56 & -- & --\\
NGC 6121 & M4 & $-$1.16 & 71.00 & 4.60 & $-$12.5 & $-$18.99 & 0.49 & 51.82 & 4.96 & -- & -- & 154 & -- & --\\
NGC 2808 & -- & $-$1.14 & 103.70 & 14.40 & 1.0 & 0.28 & 0.20 & 9.08 & 5.91 & -- & -- & 81 & -- & --\\
NGC 104 & 47 Tuc & $-$0.72 & $-$17.20 & 12.20 & 5.2 & $-$2.53 & 0.44 & 42.30 & 5.88 & -- & -- & 106 & -- & --\\
NGC 6388 & -- & $-$0.55 & 83.40 & 18.20 & $-$1.3 & $-$2.68 & 0.21 & 6.75 & 6.02 & -- & -- & 36 & -- & --\\
\hline 
Total (DR16) & -- & --     & --    & -- & --     & --    & -- & -- & --  & --  & -- & 939 & -- & -- \\
\hline
\end{tabular}
\label{tab:GCparams}
\end{table*}


The criteria for matching stars to globular clusters are mostly similar to those used by \citet{2017MNRAS.466.1010S}, though we benefit from and employ the additional information provided by Gaia Data Release 2 \citep{2018A&A...616A...1G}. For stars with measured proper motions in right ascension and declination $\mu_{[\alpha,\delta],\ast}$, that have reported uncertainties $\sigma_{\mu,[\alpha,\delta],\ast}$ smaller than 0.30 mas/year, which are being associated with globular clusters for which \citet{2019MNRAS.482.5138B} reported measurements of both mean proper motion $\mu_{[\alpha,\delta],GC}$ and central proper motion dispersion $\sigma_{\mu,GC}$, we define the quantity:
\begin{equation}
    X^2_{\mu} = \biggl(  \frac{(\mu_{\alpha,\ast}-\mu_{\alpha,GC})^2}{\sigma_{\mu,GC}^2+ \sigma_{\mu,\alpha,\ast}^2} + \frac{(\mu_{\delta,\ast}-\mu_{\delta,GC})^2}{\sigma_{\mu,GC}^2+ \sigma_{\mu,\delta,\ast}^2} \biggl)
\end{equation}
This quantity is available for $\sim$60\% of our eventual candidate globular cluster stars. It cannot be used uniformly due to the incompleteness of Gaia Data Release 2 in crowded fields. Nevertheless, it is useful to quantify the rate of false positives and false negatives that would result purely from the criteria of \citet{2017MNRAS.466.1010S}, and to then apply those corrections to that 60\% of the sample. Following consultation with the AAS statistics consultant, we denoted this quantity as $X^2_{\mu}$ rather than  $\chi^2_{\mu}$ as it has not actually been shown (by theorem) to be asymptotically $\chi^2$ distributed. In particular, the sources of ``error" are heterogeneous and somewhat uncertain.  

Stars are classified as part of a globular cluster if they can satisfy the following three conditions:
\begin{enumerate}
    \item They have an APOGEE targeting flag classifying them as a cluster target (``Apogee\_Target2 $=$ 10") and do not have the flag classifying them as an open cluster member (``Apogee\_Target1 $=$ 9"). These stars are then associated with the nearest cluster, and constitute $\sim$80\% of our final sample. 
    \item They have a position within one tidal radius of the globular cluster, a metallicity within 0.30 dex of that of the cluster, and a radial velocity that differs from that of the cluster by less than the radial velocity dispersion of the cluster. We found that the Cannon and astroNN might be overestimating metallicities for metal-poor stars, and thus we shifted the metallicity requirement to 0.45 dex when [Fe/H]$_{\rm{GC}}\leq -2.0$ when the metallicity measurement is from the Cannon or astroNN. When the Payne's metallicity measurement is used, we only consider measurements from clusters with [Fe/H]$_{\rm{Harris}} \geq -1.55$. The metallicity criterion is waived for the globular cluster Terzan 5, because it is known to have a broad metallicity distribution function, which spans the range $-0.80 \lesssim \rm{[Fe/H]} \lesssim +0.70$ \citep{2014ApJ...795...22M}.
    \item If a star has proper motion measurements with precision better than 0.30 mas/yr in both axes, and if it is being associated with a cluster for which \citet{2019MNRAS.482.5138B} reported measurements of mean proper motion and dispersion, we require that $X^2_{\mu} \leq 12$. This criterion is waived if proper motion measurements are not available. 
\end{enumerate}
We have verified that none of the globular clusters with matches are overlapping with one another in the spaces of metallicity, radial velocity, and line of sight. There is thus no significant risk of stars belonging to one cluster being misidentified as a member of another. The proper motion criterion identifies a scant 15 stars that would be misdiagnosed as false positives if we were to rely on the selection criteria of \citet{2017MNRAS.466.1010S}. We also identify candidate globular cluster stars with criteria that are half-as-strict in  metallicity, radial velocity, and position on the sky. We add 26 such stars to our sample, which have proper motion measurements satisfying $X^2_{\mu} \leq 12$. 

The full sample of 1012 candidate globular cluster stars from APOGEE DR14 that meet our selection criteria  are listed in Table \ref{tab:GCcandidates}.

\begin{table*}
\centering
\caption{The APOGEE ID's, RA (degrees), DEC (degrees), signal-to-noise ratio, and a relevant selection of relevant atmospheric parameters and abundances from the Payne for the 1010 stars with spectra from APOGEE DR14 that we associate with Galactic globular clusters. Stars for which the Payne did not converge have parameters listed as "nan". A full version of this Table is available in the online version.  }
\renewcommand{\arraystretch}{0.95}
\begin{tabular}{|lllll|ll|rrrrr|}
\hline
APOGEE ID & 	 RA  & 	 DEC  & 	Cluster   & 	 SNR  & 	Teff  & 	logg  & 	 [Fe/H]  & 	[C/H]  & 	[N/H]  & 	[O/H]  & 	[Al/H] \\
\hline
M03332183+7935382  &  53.3409680  &  79.5939710  &  Pal1   &   54  &  4908.4  &  2.64  &  -0.55  &   -0.66  &  -0.52  &   -0.46  &  -0.34 \\
2M13415631+2825565  &  205.4846620  &  28.4323880  &  NGC 5272  &   129  &  4479.7  &  1.63  &  -1.24  &   -1.54  &  -1.02  &   -0.80  &  -1.47 \\
2M15181418+0201222  &  229.5591010  &  2.0228560  &  NGC 5904   &   122  &  5247.4  &  2.33  &  -1.32  &   -1.72  &  -0.98  &   -1.29  &  -1.14 \\
2M21333520-0046089  &  323.3967030  &  -0.7691410  &  NGC 7089   &   279  &  4332.5  &  1.46  &  -1.43  &   -1.80  &  -0.83  &   -1.04  &  -1.39 \\
\hline
\end{tabular}
\label{tab:GCcandidates}
\end{table*}

The first choice of stellar [Fe/H] used to compare to the globular cluster literature value is that of ASPCAP. If no ASPCAP metallicity is derived for a star, we use that from the Cannon. If a star has neither a Cannon nor an ASPCAP metallicity, we try that from the Payne, and finally we use that from astroNN. If none of the available pipelines has a metallicity estimate for a star, we discard the star from our analysis. 

With this approach, we identify 1012 stars in the APOGEE DR14 catalogue that we consider to be likely globular cluster members, of which 832 were deliberately selected as APOGEE calibration targets (criterion \#1 above). They are associated with 28 different globular clusters.  We list the 28 globular clusters, the literature values of their relevant physical properties, and the number of matches in APOGEE, in Table \ref{tab:GCparams}. 
We also ran our search on a sample of southern fields observed as part of APOGEE DR16, on which we ran the Payne pipeline. From these, we identified an additional 939 stars that we associate to 11 different globular clusters, one of which (NGC 6218) also had measurements in APOGEE DR14.  

\begin{figure}
\includegraphics[width=0.50\textwidth]{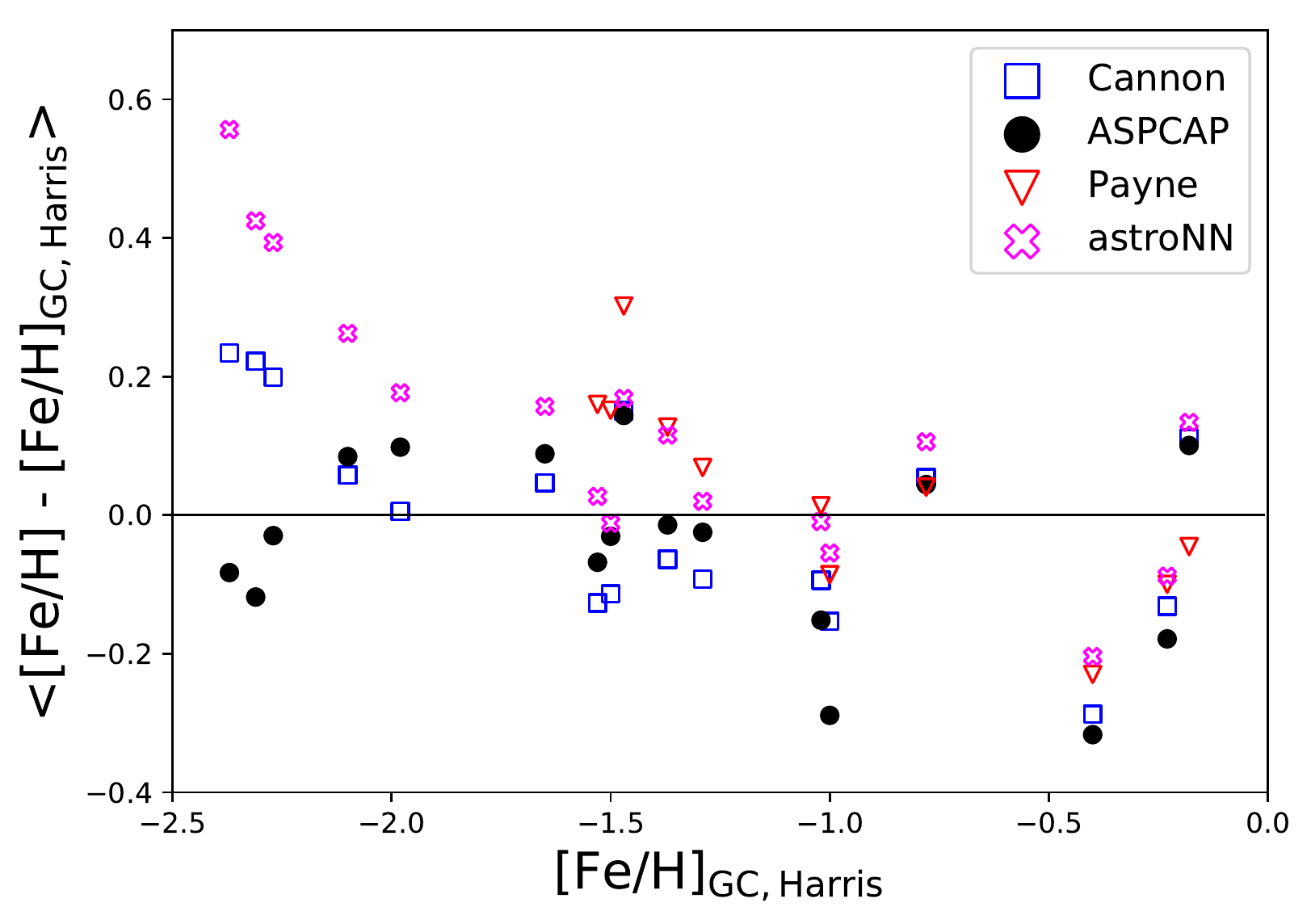}
\caption{The difference between the derived and literature cluster metallicities as a function of literature cluster metallicity, for the four APOGEE pipelines, for the 17 globular clusters with at least 5 associated members. Metallicities derived from ASPCAP are similar to those from the Harris catalogue across the whole metallicity range. Those from the Cannon and astroNN overestimate metallicity for [Fe/H]$\lesssim -2.0$. Those from the Payne overestimate metallicity for [Fe/H]$\lesssim -1.40$. The thin horizontal black line denotes  the line of equality between the APOGEE-derived matallicity values and those from the Harris catalog. }
\label{fig:GCmetallicityprecision}
\end{figure}

\begin{figure}
\includegraphics[width=0.50\textwidth]{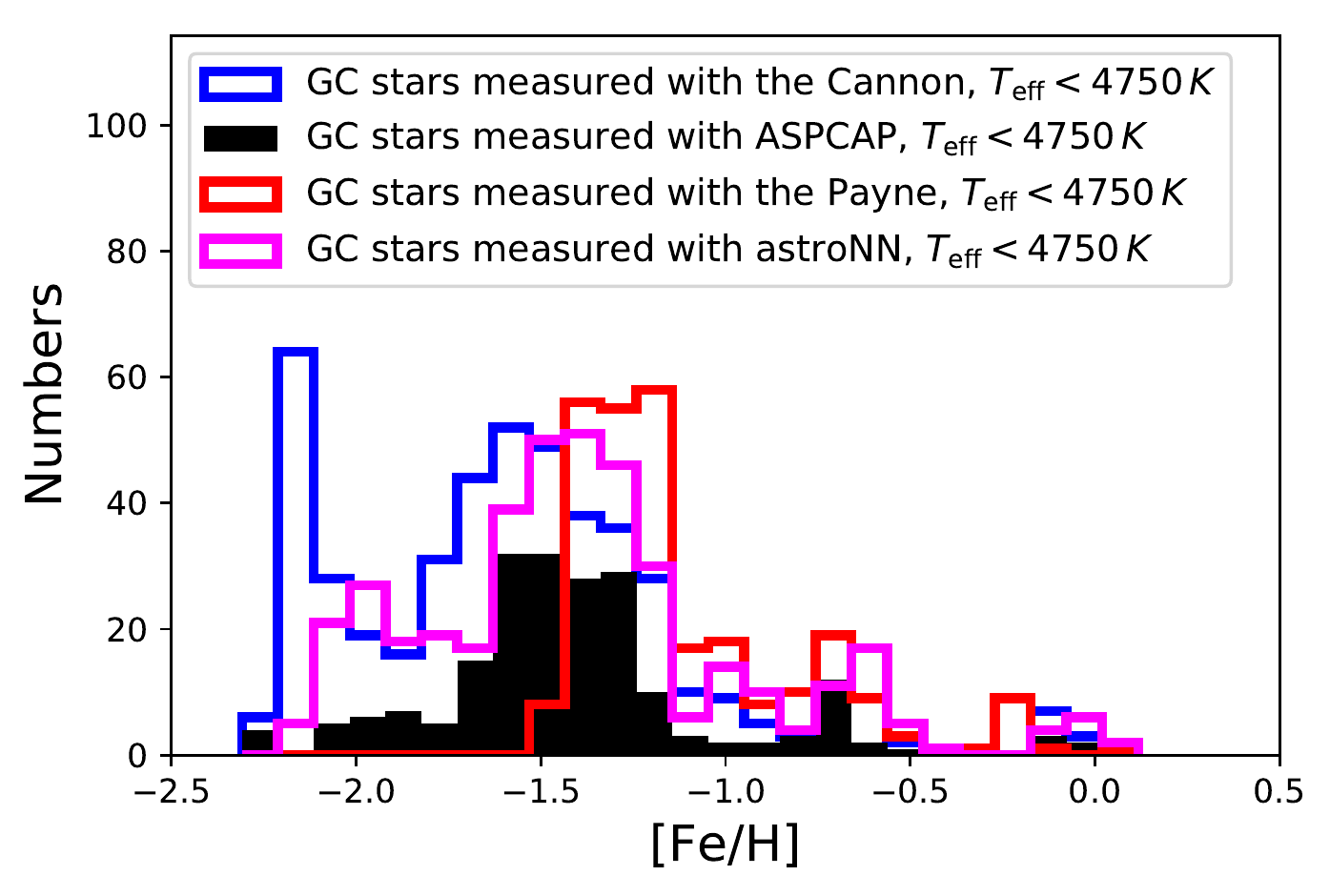}
\caption{The distribution of derived stellar [Fe/H] from ASPCAP (black), the Cannon (blue), the Payne (red), and astroNN (magenta) for globular cluster stars identified in APOGEE DR14. We restrict the comparison to those stars with $T_{\rm{eff}} \leq 4,750\,K$. }
\label{fig:FEH_Cannon_VS_ASPCAP_VS_PAYNE_VS_astroNN_GCs}
\end{figure}

\begin{figure}
\includegraphics[width=0.50\textwidth]{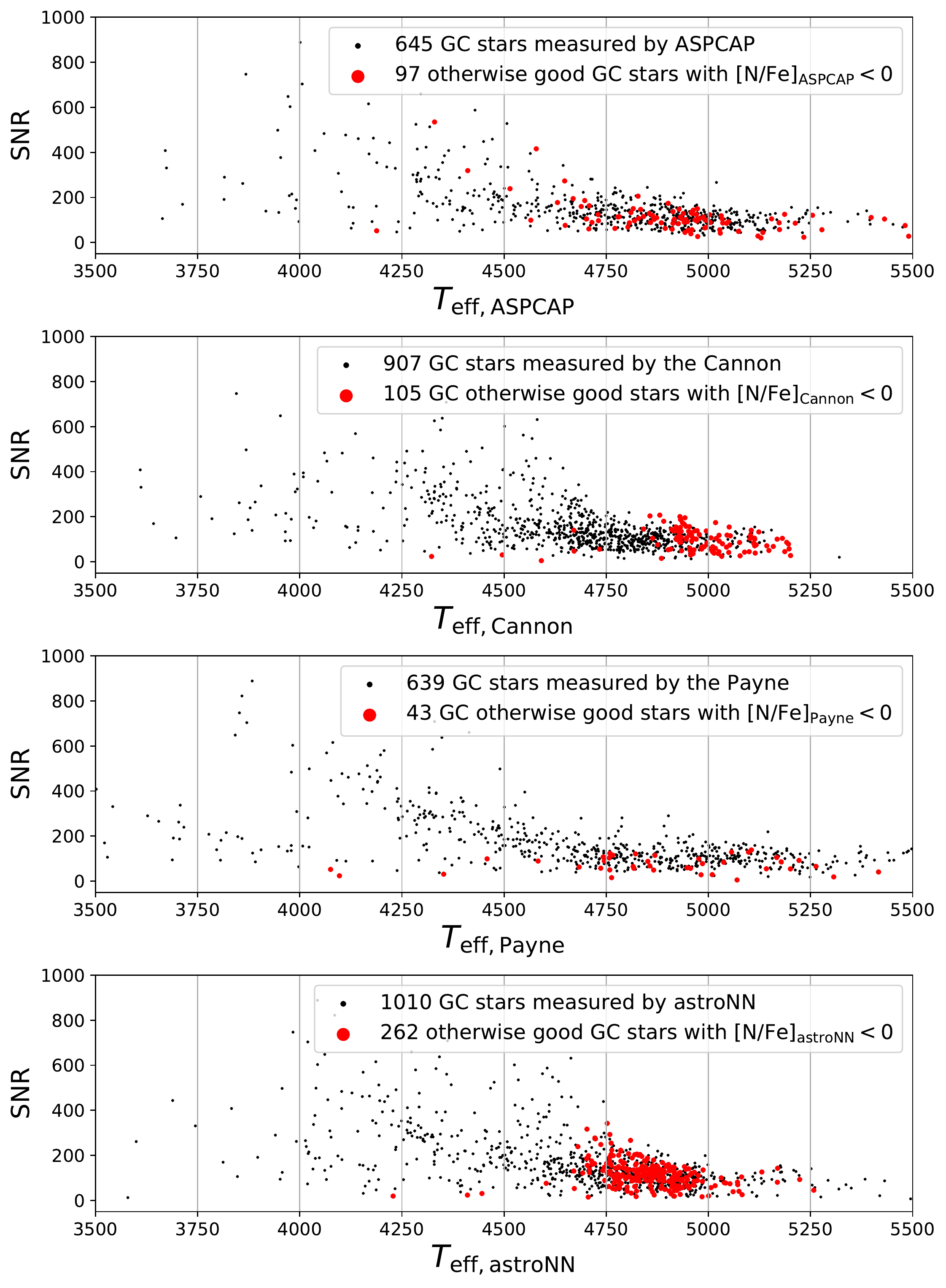}
\caption{The scatter of signal-to-noise ratio and effective temperature for globular cluster stars in the four APOGEE pipelines. Stars with [N/Fe] $\leq 0$ (red circles) are almost certainly indicative of errors in the pipeline. They are predominantly found at hotter temperatures, $T_{\rm{eff}}\geq 4,750\,K$, and are less frequently found by the Payne.}
\label{fig:Nitrogen_Sensitivity}
\end{figure}

The comparison between the derived mean [Fe/H] of these clusters by the four pipelines, and the literature values from \citet{1996AJ....112.1487H}, is shown in Figure \ref{fig:GCmetallicityprecision}. This is coarse diagnostic at best, but there is unfortunately no analog to the Gaia benchmark stars \citep{2014A&A...564A.133J} for globular clusters. Comparisons to BACCHUS-derived data are not included in Figures \ref{fig:GCmetallicityprecision}, \ref{fig:FEH_Cannon_VS_ASPCAP_VS_PAYNE_VS_astroNN_GCs}, and \ref{fig:Nitrogen_Sensitivity}, but the equivalent information can be found in \citet{2018arXiv181208817M}.


The mean of the differences between the ASPCAP-measured and literature values of globular cluster metallicities are close to zero across the full metallicity range of the Milky Way globular clusters. However, metallicities measured from both the Cannon and astroNN appear overestimated for [Fe/H]$_{\rm{Harris}}\lesssim -2.00$, and likewise with Payne-derived metallicities for [Fe/H]$_{\rm{Harris}}\lesssim -1.40$. The Payne also seems to consistently underestimate metallicities at [Fe/H]$_{\rm{Harris}} \gtrsim -0.50$, which is consistent with what \citet{2018arXiv180401530T} show in their Figure 14. The slight metallicity offset at the highest metallicities is plausibly due to the small metallicity-dependence of temperature bias by the Payne, see Figure 8 of \citet{2018arXiv180401530T}.  Some of these differences may be partly or fully due to the heterogeneities of the sample of \citet{1996AJ....112.1487H}. 

The metallicity distribution function of the globular cluster sample for each of the four pipelines is shown in Figure \ref{fig:FEH_Cannon_VS_ASPCAP_VS_PAYNE_VS_astroNN_GCs}. These distribution functions are restricted to stars with $T_{\rm{eff}} \leq 4,750\,K$, for reasons justified later in this work. The different relative effectiveness of the pipelines at reporting [Fe/H] values is a function of [Fe/H]$_{\rm{Harris}}$.  This contributes to the argument that the choice of pipeline, at least when using the DR14-related releases, will affect the diagnostic potential of the globular cluster sample.

AstroNN is the most effective pipeline at merely reporting [Fe/H] measurements for globular cluster stars, at all metallicities. Of the other three pipelines, the Cannon is the most effective at yielding [Fe/H] measurements at the metal-poor end, predominantly stars from NGC 7078 (M15) and NGC 6341 (M92), which are respectively listed at metallicities,  [Fe/H]$=-$2.37 and $-$2.31 in the Harris catalog. However, this seeming advantage of astroNN and the Cannon is in fact a limited one, as they do not perform as well at the task of measuring relative abundances [X$_{i}$/Fe]. That is discussed in greater detail in Sections \ref{subsec:GCsample2} and \ref{subsec:GCsample3}.

The Payne matches the yield of astroNN at higher metallicities, but the termination of its parameter space at [Fe/H]$=-1.50$ is, at least for this study, a severely limiting factor. 

We do not show error bars in our abundance plots. As we will see in the subsequent sections, the true errors are likely dominated by systematic issues, rather than signal-to-noise limitations that are more easily computed and generally available.

\subsection{Delineating an APOGEE Globular Cluster Sample Suitable for Multiple Populations Studies} \label{subsec:GCsample2}

\begin{figure}
\includegraphics[width=0.48\textwidth]{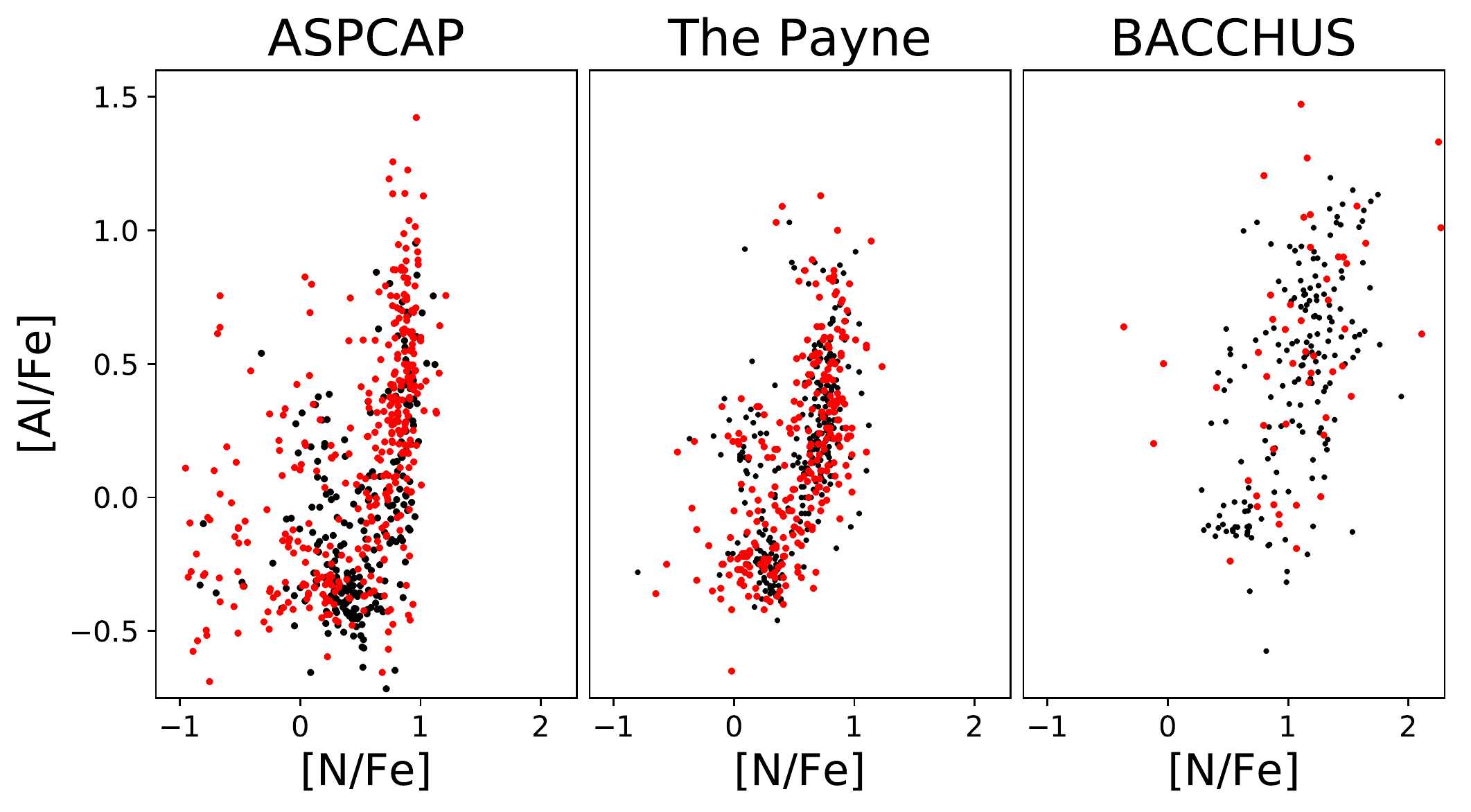}
\caption{Of the three pipelines shown here, the Payne is the most consistent at deriving an [Al/Fe]-[N/Fe] relation that is the nearly identical for both stars with $T_{\rm{eff}} \leq 4,750\, K$  (black points) and stars with $5,250\, K \geq T_{\rm{eff}} > 4,750\, K$ and signal-to-noise ratios greater than 50 (red points). } 
\label{fig:APOGEE_GC_NAL_Teff}
\end{figure}

\begin{figure*}
\centering
\includegraphics[width=1.00\textwidth]{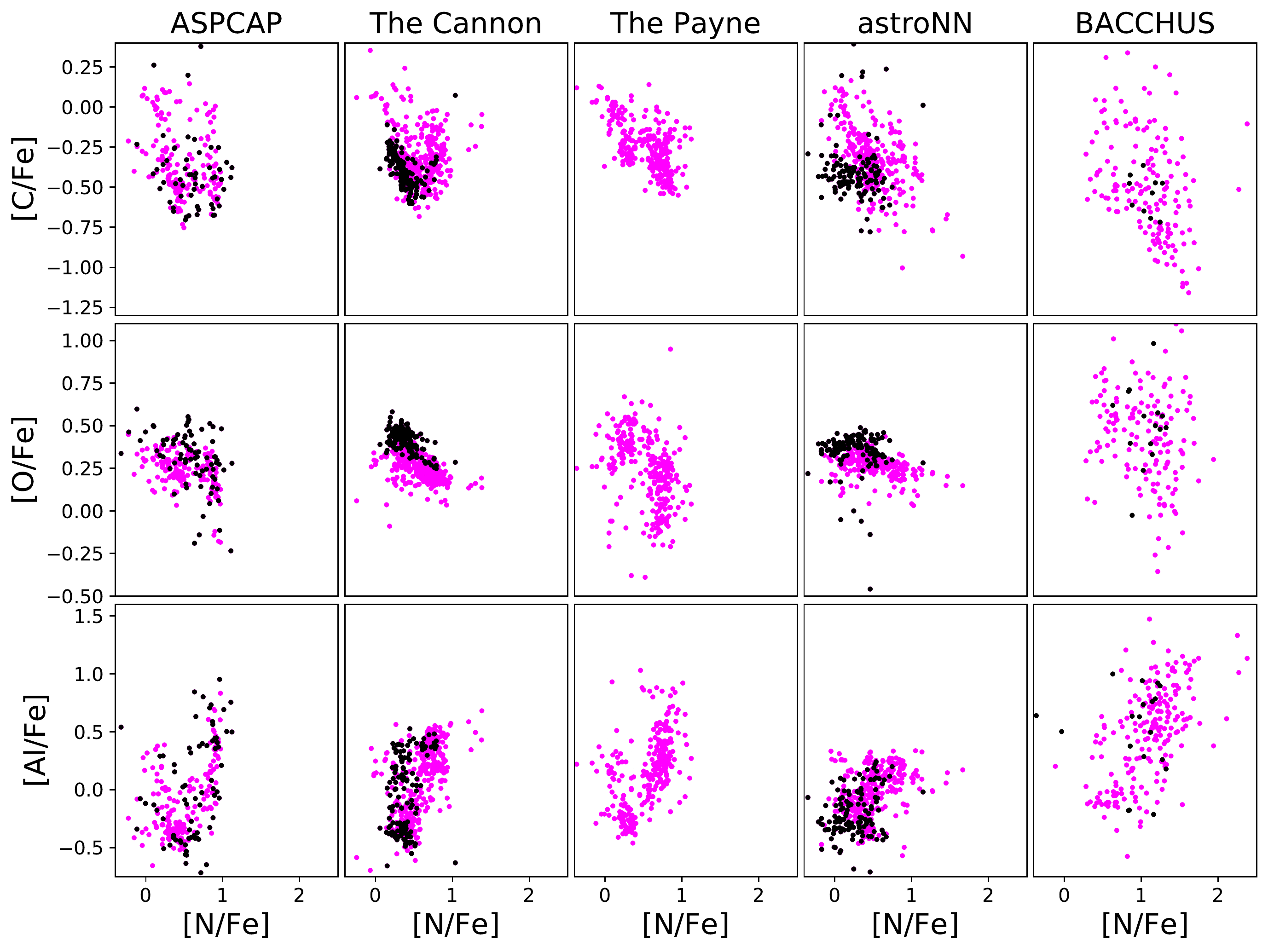}
\caption{The relative abundance diagrams for globular cluster stars in APOGEE with $T_{\rm{eff}} \leq 4,750 \,K$ from the five abundance pipelines. The abundances of stars from globular clusters with [Fe/H]$\leq -1.55$ are plotted in black, the remainder are plotted in magenta. The Payne and BACCHUS are the only pipelines that can recover convincing C-N anti-correlations, and for which the NO anti-correlations span a convincingly broad range in [O/Fe]. AstroNN's recovery of the [Al/Fe]-[N/Fe] correlation is, for reasons unknown to us, truncated at [Al/Fe]$\approx +0.25$. }
\label{fig:APOGEE_GC_AbundanceScatter}
\end{figure*}

There are 24 elemental abundances reported by ASPCAP, 18 reported by the Cannon, 18 reported by astroNN, and 15 reported by the Payne. The latter did measure sodium, phosphorus, vanadium, cobalt, and germanium, but the values were not reported, as it was deemed that further work is needed to understand the features contributing to those measurements. Of the elements reported, only some of these are useful for the analysis of multiple populations in globular clusters, and only some of these will be be precisely measured at the relevant range of temperatures, gravities, and metallicities of this sample. 
	
The abundances of carbon, nitrogen, and oxygen, to which APOGEE's $H$-band spectra are sensitive, have previously been established as excellent tracers of multiple populations in globular clusters  \citep{2009ApJ...695L..62Y,2012ApJ...746...14M,2015AJ....149..153M,2018ApJ...860...70C,2018arXiv181208817M}. 

However, the distributions of CNO abundances include a substantial noise source. Four of the five pipeline reports many values of [N/Fe]$< 0$, even approaching [N/Fe]$=-1.0$. Such measurements are not expected from previous literature studies, and further, the truncation at [N/Fe]$=-1.0$ is intrinsically unconvincing as it is at the edge of parameter spaces of two of the pipelines, ASPCAP and the Payne. Regardless of the pipeline used, almost all of the measurements with [N/Fe]$< 0$ occur for stars with $T_{\rm{eff}} > 4,750\, K$, as can be seen from the distribution of red points in Figure \ref{fig:Nitrogen_Sensitivity}. This is due to a steep temperature dependence of the molecular features responsible for the measurability of nitrogen in $H$-band spectra. 


It is not surprising that a threshold effective temperature exists, in this case  $T_{\rm{eff}} \approx 4,750\, K$. APOGEE's sensitivity to carbon, nitrogen, and oxygen is predominantly due to OH, CN, and CO, for which the rates of molecular disassociation are sensitive to temperature. This issue has been previously reported and investigated by \citet{2015AJ....149..153M}. In their analysis, which used a different methodology (such as using temperatures derived from photometry), the cutoff was set at $T_{\rm{eff}} = 4,500\, K$. The fraction of candidate globular cluster stars within APOGEE with $T_{\rm{eff}} \leq 4,750\, K$ is approximately 40\%.  We note that the stars with  $T_{\rm{eff}} \leq 4,750\, K$ have a median signal-to-noise ratio of approximately 180, and only $\sim$3\% of them have signal-to-noise ratios of less than 50. We found that the frequency of stars with [N/Fe] $< 0$ is lowest in the Payne. This remained true even as we experimented with additional cuts in metallicity and signal-to-noise ratio.

This motivated us to see if the the parameter space could be reliably expanded for the investigation of the trends in the [N/Fe]-[Al/Fe] plane, which will be the most important to our work. In Figure \ref{fig:APOGEE_GC_NAL_Teff}, we show the different trends in the relative abundances for each of ASPCAP, the Payne, and BACCHUS for the stars with $T_{\rm{eff}} \leq 4,750\, K$ and $5,250\, K \geq T_{\rm{eff}} > 4,750\, K$. The Payne is able to recover an indistinguishable distribution as long as restrict the sample of hotter stars to those with measurement signal-to-noise ratios greater than 50. The same is not true of the ASPCAP results. For the hotter stars, [N/Fe] measurements extend to much lower values, and the [Al/Fe] extend to higher values, both of which are unphysical. There is no temperature dependence to the BACCHUS measurements of [N/Fe], but there is to the measurements of [Al/Fe]. The Payne thus provides the largest pool of consistent measurements in the [Al/Fe]-[N/Fe] abundance place, with 454 in the APOGEE DR14 sample, compared to 244 with ASPCAP and 256 with BACCHUS.

\begin{figure}
\includegraphics[width=0.45\textwidth]{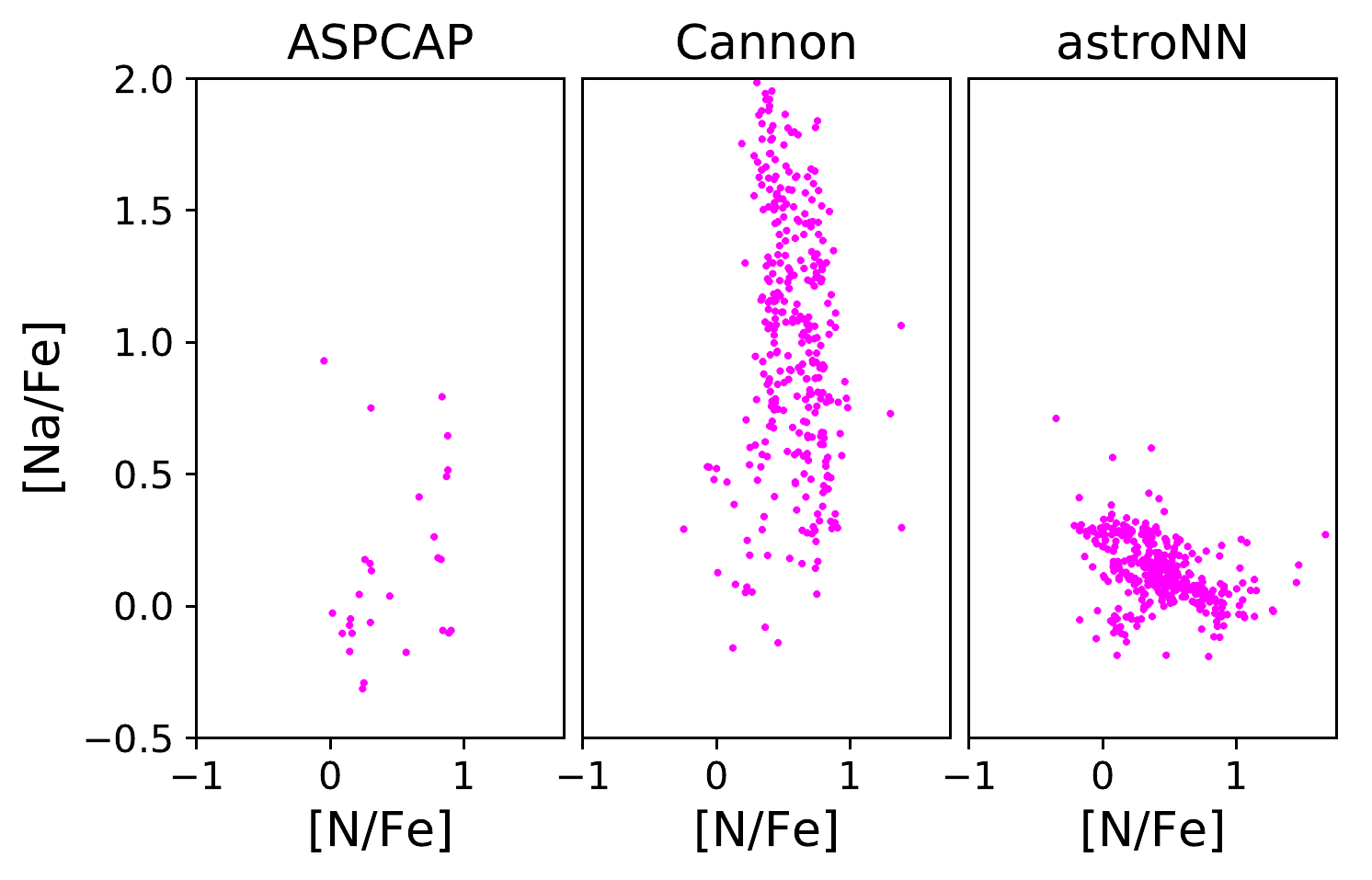}
\caption{None of the APOGEE pipelines consistently report reliable sodium abundances for globular cluster stars. The [Na/Fe]-[N/Fe] scatter is positively correlated in ASPCAP measurements, but only for a small sample of stars; it is a pure scatter extending to very high values of [Na/Fe] in the Cannon measurements; and it shows up as an anti-correlation in the astroNN measurements. All three are inconsistent with literature measurements documenting a significant and positive correlation \citep{2008ApJ...684.1159Y,2016MNRAS.459..610M}.}
\label{fig:APOGEE_GC_NitrogenSodium}
\end{figure}

Concerning elements other than CNO whose measurements have yielded the greatest empirical footprint on the literature, aluminum abundance variations are present and measurable from APOGEE spectra, whereas sodium abundance variations are present but typically not reliably measured. 

We show the relative abundance plots for C,N,O,Al, for all five pipelines, in Figure \ref{fig:APOGEE_GC_AbundanceScatter}.  All of the pipelines successfully recover the N-Al correlation and the N-O anticorrelation, though the latter is most extended when measured by the Payne or BACCHUS. The C-N anticorrelation is not recovered by ASPCAP and the Cannon,  though a previous data release of ASPCAP abundances (DR12), did recover the C-N anti-correlation \citep{2017MNRAS.466.1010S,2017MNRAS.465..501S}. None of the pipelines do well in the metal-poor ([Fe/H]$_{\rm{Harris}} \leq -1.55$) regime, for which the measurements are plotted in black. The scatter of the abundances are not distributed uniformly, but show hints of bimodality. That is discussed in Section \ref{subsec:GCsample7}. 

In contrast to aluminum, the sodium lines are very weak, with a shift of 0.05 dex in [Na/H] yielding flux changes of $\lesssim 1\%$ \citep{2018arXiv180401530T}. Sodium abundances were measured, but not reported, by the Payne. The Cannon and astroNN can derive sodium measurements for nearly all stars, but the values are not physically meaningful for globular cluster stars. ASPCAP derives sodium abundances for a small subset of globular cluster stars, which does better over that small sample. We show the reported [Na/Fe]-[N/Fe] scatters in Figure \ref{fig:APOGEE_GC_NitrogenSodium}. The correlation that is expected between [Na/Fe] and [N/Fe] \citep{2008ApJ...684.1159Y,2016MNRAS.459..610M} shows up as a pure scatter in the Cannon  measurements, and is an anti-correlation in the astroNN measurements. 

\subsection{Comparison to the Measurements of  \citet{2009A&A...505..117C}: Most APOGEE-Derived Variations in [O/Fe] Are Underestimated by $\gtrsim$30+\%}
\label{subsec:GCsample3}

\citet{2009A&A...505..117C} measured abundances of [Fe/H], [Na/Fe], and [O/Fe] for a sample of 1,409 spectra of red giant stars from 15 globular clusters.  There was also an analysis of spectra of red giants in NGC 6218,  done by \citet{2007A&A...464..939C} using the same methodology as \citet{2009A&A...505..117C}. The following five clusters have also been probed by APOGEE DR14: NGC 7078, NGC 6218, NGC 5904, NGC 6171, and NGC 6838. The overlap between the samples allows us to compare the standard deviation of [O/Fe] for each globular cluster. Given that \citet{2009A&A...505..117C} actually report upper bounds on [O/Fe] for many of the most oxygen-deficient stars, the standard deviations for those data are actually a lower bound since the stars with upper bounds on their [O/Fe] relative abundances are not included in the calculation. The comparison between the result of \citet{2009A&A...505..117C} and that derived by  ASPCAP, the Cannon, the Payne, astroNN, BACCHUS, and by \citet{2015AJ....149..153M} is listed in Table \ref{tab:GCcomparison}

 The dispersion in [O/Fe] measured by ASPCAP, the Cannon, and astroNN are typically far smaller than the literature values. The Payne's measured scatter in [O/Fe] is more consistent with the measurements of \citet{2009A&A...505..117C}, with the dispersion being a more modest $\sim$30\% lower. The dispersion measured by the BACCHUS pipeline, and by \citet{2015AJ....149..153M}, are consistent with the literature values. 

\begin{table*}
\centering
\caption{The standard deviation of [O/Fe], $\sigma_{\rm{[O/Fe]}}$, for clusters probed by both the investigation of \citet{2009A&A...505..117C} and APOGEE. The $\sigma_{\rm{[O/Fe]}}$ derived by ASPCAP, the Cannon, and astroNN  are typically much smaller than those reported by \citet{2009A&A...505..117C},  those derived by the Payne are typically slightly smaller, and those derived by the BACCHUS pipeline and by \citet{2015AJ....149..153M} are comparable. }
\renewcommand{\arraystretch}{0.9}
\begin{tabular}{|lll| ccccc | c |}
\hline
Name & [Fe/H]$_{\rm{Harris}}$ & ${\sigma}_{\rm{[O/Fe],C09}}$ &  ${\sigma}_{\rm{[O/Fe],A}}$   & ${\sigma}_{\rm{[O/Fe],C}}$  &   ${\sigma}_{\rm{[O/Fe],P}}$ &   ${\sigma}_{\rm{[O/Fe],N}}$ &   ${\sigma}_{\rm{[O/Fe],B}}$ &   ${\sigma}_{\rm{[O/Fe],M2015}}$ \\ 
\hline
NGC 7078 / M15 & $-$2.37 & 0.18 & 0.22  & 0.05  & --  & 0.17  & 0.16 &  0.19 \\
NGC 6218 / M12 & $-$1.37 & 0.28 & 0.05  & 0.06  & 0.16  & 0.05  & -- &  -- \\
NGC 5904 / M5 & $-$1.29 & 0.29 & 0.11  & 0.12  & 0.19  & 0.54  & 0.28 &  0.27 \\
NGC 6171 / M107 & $-$1.02 & 0.19 & 0.04  & 0.06  & 0.13  & 0.02  & 0.12 &  0.15 \\
NGC 6838 / M71 & $-$0.78 & 0.10 & 0.03  & 0.06  & 0.12  & 0.03  & 0.12 &  0.09 \\
\hline
\end{tabular}
\label{tab:GCcomparison}
\end{table*}

\subsection{Most APOGEE-Derived Variations in [N/Fe] and [C/Fe] Are Likely Underestimated by $\sim$50+\%}
\label{subsec:APOGEEoffset}

APOGEE-derived variations in [N/Fe] and [C/Fe] are at least 50\% lower than literature estimates for ASPCAP, the Cannon, the Payne, astroNN, but not for BACCHUS. As we will discuss, there are known sources of systematic error in most analyses of APOGEE spectra, and consistent evidence for a necessary re-scaling of [N/Fe] and [C/Fe] variations are found with several independent sources used as a comparison.  Given these factors, we conclude that the variations in [N/Fe] and [C/Fe] are likely being underestimated by four of the five APOGEE-based analyses. 

 APOGEE's CNO abundance determinations are predominantly derived from the absorption of three sets of molecular lines (OH, CN, and CO) and thus a mistaken assumption for one of the three abundances may propagate as an error to the determination of the other two abundances. ASPCAP, for example, fits for spectra with model grids for which the relative abundances extend no higher than [X/Fe] $= +1.0$. Given that [N/Fe] frequently extends to much higher abundances in globular clusters (shown below), this imposes a systematic error. For those stars, the models may be compensating for underestimated [N/Fe], which would deepen CN lines, by fitting a higher abundance of carbon. Some of these issues should be resolved in future data releases, as the model grid of stellar atmospheres from which ASPCAP abundances are derived is being expanded. 
 
 There is also a degeneracy between the  of $T_{\rm{eff}}$, [O/H], [C/H], and [N/H] determinations derived from H-band spectra, which is discussed in detail in Section 3 of \citet{2018arXiv181208817M}. Further research into understanding and potentially breaking or at least better constraining  this degeneracy is ongoing. 

\begin{figure}
\centering
\includegraphics[width=0.45\textwidth]{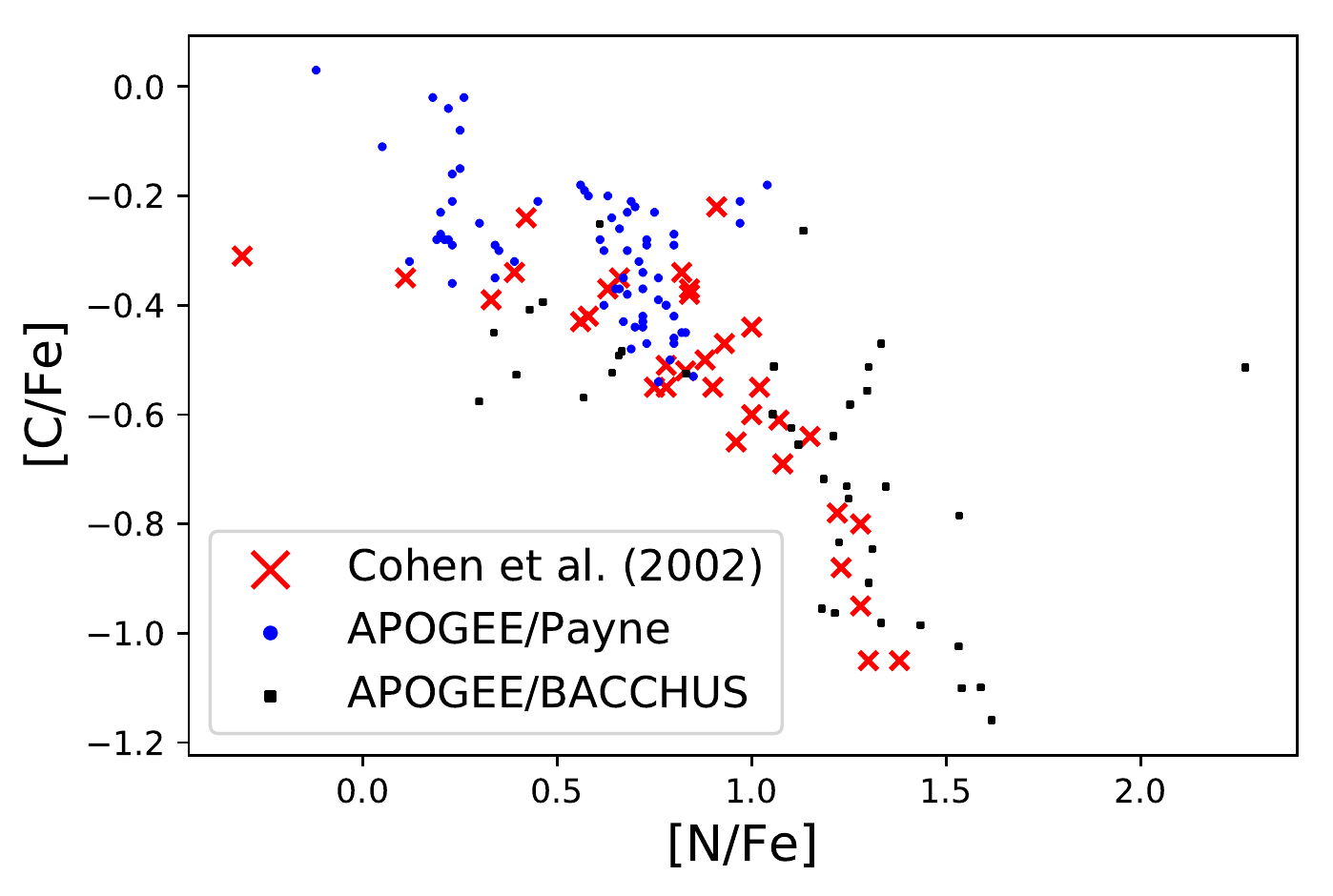}
\caption{The relation of [N/Fe] and [C/Fe] for the cluster NGC 5904 (M5) measured by \citet{2002AJ....123.2525C} is consistent with that measured with the BACCHUS pipeline, but is more extended than that measured with the Payne. The measurements of \citet{2002AJ....123.2525C} are derived off of ultraviolet absorption lines from stars at the base of the red giant branch. }
\label{fig:CohenVsAPOGEE}
\end{figure}

We show a comparison of the Payne's and BACCHUS's measured [C/Fe]-[N/Fe] abundance trend from APOGEE spectra to that of \citet{2002AJ....123.2525C}, for the globular cluster NGC 5904  (M5), in Figure \ref{fig:CohenVsAPOGEE}. \citet{2002AJ....123.2525C} measured [C/Fe] and [N/Fe] for stars in and near the base of the red giant branch of the cluster, using spectra taken with the the Low Resolution Imaging Spectrometer (LRIS) on Keck over the wavelength range 3600 \AA $\lesssim \lambda \lesssim$ 4800 \AA. They report a relative abundance trend extending to much higher [N/Fe] values and much lower [C/Fe] values than the values derived by the Payne, but much more consistent with that derived with the BACCHUS pipeline. 


Separately, \citet{2016MNRAS.459..610M} and \citet{2008ApJ...684.1159Y} have respectively shown that the [N/Fe] variations in 47 Tuc and NGC 6752 are approximately twice as large as the [Na/Fe] variations. That is in contrast to a result that we derive later in this work, in Section \ref{subsec:fullsample}, that the Payne-derived dispersion in [N/Fe] in globular clusters is of a similar size to the literature-based dispersion in [Na/Fe]. As above, this suggests a necessary factor of 2 rescaling, with the Payne-derived dispersion in [N/Fe] being $\sim$50\% smaller than it should be, and similarly for ASPCAP, the Cannon, and astroNN.

Given that there is likely a scaling factor between four of the five APOGEE-derived variations in [N/Fe] and [C/Fe], the linear trends derived later in this work are likely qualitatively valid but not quantitatively valid. Where they may be most useful, is as a self-consistent flag to identify and interpret field stars with abundances similar to those of second generation globular cluster stars \citep{2010A&A...519A..14M,2017MNRAS.465..501S}. These results are also useful for informing future abundance analyses of the APOGEE spectra. 

\subsection{Testing for the Robustness of APOGEE-Derived Dispersions in [Al/Fe]} \label{subsec:GCsample5}

An important assumption of our investigation is that spectroscopic analyses of the APOGEE data have similar sensitivity to [Al/Fe] variations as prior literature studies. We define two  criteria for this task. 

The first criterion is that for clusters satisfying [Fe/H]$_{\rm{Harris}} \geq -1.55$, the mean value of [Al/Fe] for stars with $5,250\,K \geq T_{\rm{eff}} \geq 4,750\,K$ is consistent with that for stars with $T_{\rm{eff}} < 4,750\,K$, as long as there are at least five stars in both groups when the ASPCAP-derived temperatures are used. This is meant to show that the presence or absence of CNO absorption features is not yielding an error that is covariant with an error in [Al/Fe] determinations, as that would be catastrophic for the study of multiple populations in globular clusters. The results of this comparison are shown in Table \ref{tab:ALFEcomparison}. The mean difference in [Al/Fe] between the relatively hot and cold samples, ${\delta}\rm{[Al/Fe]}=\rm{[Al/Fe]}_{\rm{cold}}-\rm{[Al/Fe]}_{\rm{warm}}$ is consistent with zero for the Payne, astroNN, and BACCHUS.  They are often non-zero, by a significant amount, for ASPCAP, the Cannon, and the BACCHUS pipeline.

\begin{table*}
\centering
\caption{The difference between the mean value of [Al/Fe] for stars with $5,250\,K \geq T_{\rm{eff}} \geq 4,750\,K$  and the mean value of [Al/Fe] for stars with $T_{\rm{eff}} < 4,750\,K$, ${\delta}\rm{[Al/Fe]}=\rm{[Al/Fe]}_{\rm{cold}}-\rm{[Al/Fe]}_{\rm{warm}}$, for each of the five pipelines (ASPCAP, the Cannon, the Payne, astroNN, and BACCHUS). We list the names of the clusters, their metallicity from the Harris catalog, and the five mean differences and the sample error in the mean differences,  for those clusters for which the cold and the warm ASPCAP samples both include at least 5 stars. The differences are statistically consistent with zero for the Payne and for astroNN, but not for ASPCAP, the Cannon, and BACCHUS. }
\renewcommand{\arraystretch}{0.9}
\begin{tabular}{|ll|rrrrr|}
\hline
Name & [Fe/H]$_{\rm{Harris}}$ & $\delta\rm{[Al/Fe]}_{\rm{A}}$ & $\delta\rm{[Al/Fe]}_{\rm{C}}$ & $\delta\rm{[Al/Fe]}_{\rm{P}}$ & $\delta\rm{[Al/Fe]}_{\rm{N}}$ & $\delta\rm{[Al/Fe]}_{\rm{B}}$ \\
\hline
NGC5904 & -1.29 & -0.27 $\pm$ 0.05 & -0.00 $\pm$ 0.04 & 0.04 $\pm$ 0.05 & 0.01 $\pm$ 0.03 & 0.10 $\pm$ 0.05 \\
16 30
NGC6171 & -1.02 & -0.12 $\pm$ 0.08 & -0.03 $\pm$ 0.04 & 0.00 $\pm$ 0.03 & -0.03 $\pm$ 0.03 & 0.06 $\pm$ 0.05 \\
9 12
NGC6838 & -0.78 & -0.01 $\pm$ 0.06 & 0.04 $\pm$ 0.06 & 0.04 $\pm$ 0.02 & -0.05 $\pm$ 0.04 & -0.08 $\pm$ 0.09 \\
70 62
NGC5272 & -1.50 & -0.14 $\pm$ 0.07 & 0.10 $\pm$ 0.05 & 0.01 $\pm$ 0.07 & 0.03 $\pm$ 0.03 & 0.10 $\pm$ 0.07 \\
45 58
NGC6205 & -1.53 & -0.43 $\pm$ 0.10 & -0.04 $\pm$ 0.05 & -0.03 $\pm$ 0.09 & 0.05 $\pm$ 0.04 & -0.15 $\pm$ 0.10 \\
\hline
\end{tabular}
\label{tab:ALFEcomparison}
\end{table*}

\begin{figure}
\centering
\includegraphics[width=0.50\textwidth]{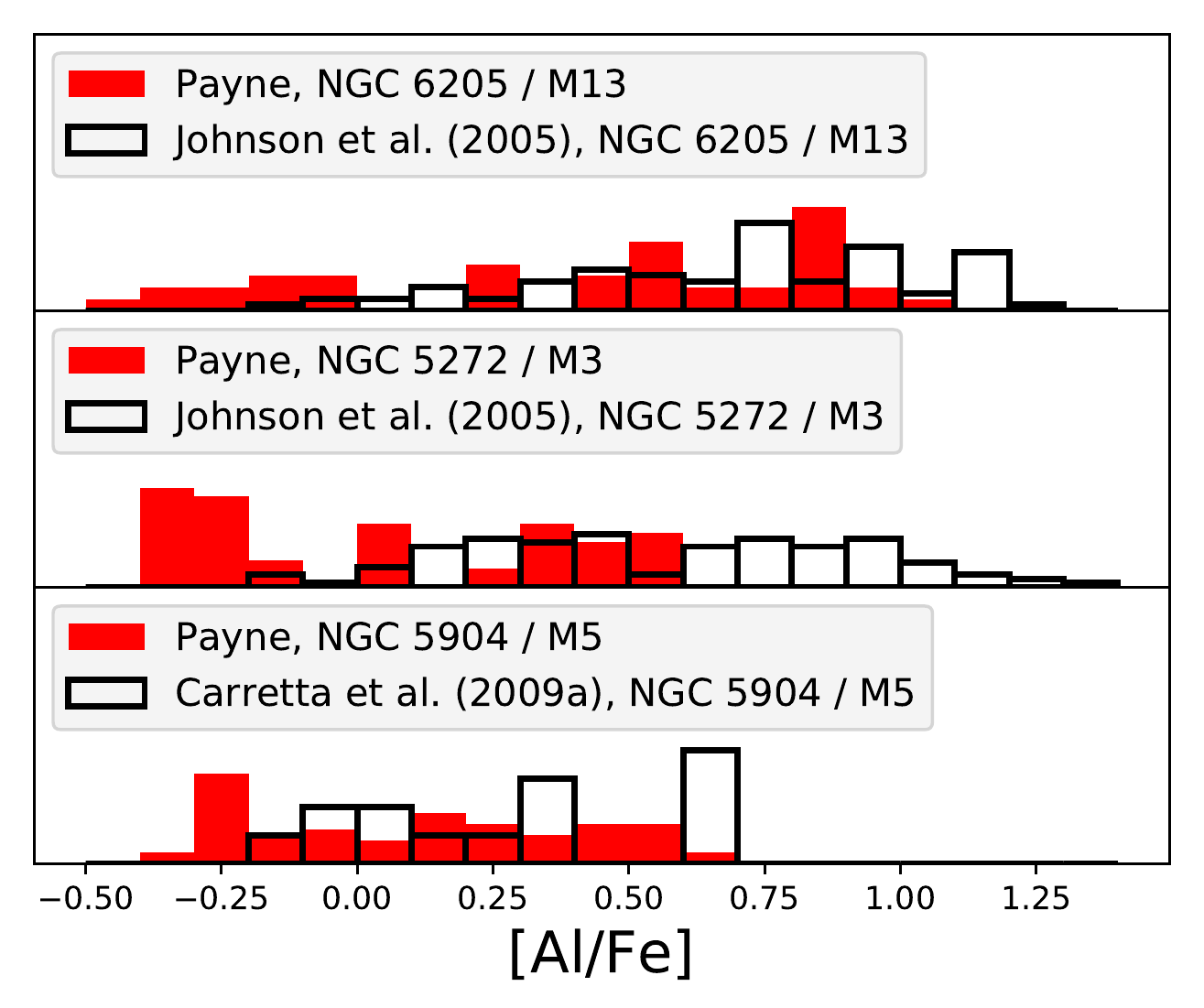}
\caption{Payne-derived values of the dispersion of [Al/Fe] in globular clusters are consistent with that measured by \citet{2005PASP..117.1308J} for NGC 5272 (M3) and NGC 6205 (M13) (top two panels), and with that measured by \citet{2009A&A...505..139C} for NGC 5904 (M5) (bottom panel). The mean values of [Al/Fe], however, are lower when measured by the Payne than when measured by the prior literature data.}
\label{fig:AluminumLiteratureComparison}
\end{figure}

The second criterion is a comparison of the dispersion in [Al/Fe] for stars with $T_{\rm{eff}} < 4,750\,K$ to that measured by prior literature studies. It is fortunate that we find three clusters that are well-sampled APOGEE whose [Al/Fe] distributions had already been well-sampled by other studies. These are the clusters NGC 6205 (M13) and NGC 5272 (M3), which were studied by \citet{2005PASP..117.1308J}, and NGC 5904 (M5), which was studied by \citet{2009A&A...505..139C}. We show comparisons of the literature data to that of the Payne in Figure \ref{fig:AluminumLiteratureComparison}. A comparison of the mean values of [Al/Fe] is not included within our criteria, as our investigation is concerned with \textit{differences} in abundances.

For NGC 6205, the literature value for the dispersion in [Al/Fe] is 0.34, compared to 0.38, 0.32, 0.44, 0.24, and 0.53 for ASPCAP, the Cannon, the Payne, astroNN, and BACCHUS. For NGC 5272, the literature value for the dispersion in [Al/Fe] is 0.38, compared to 0.32, 0.33, 0.34, 0.19, and 0.43 for ASPCAP, the Cannon, the Payne, astroNN, and BACCHUS.  For NGC 5904, the literature value for the dispersion in [Al/Fe] is 0.28, compared to 0.22, 0.31, 0.29, 0.19, and 0.34 for ASPCAP, the Cannon, the Payne, astroNN and BACCHUS.   The dispersion in [Al/Fe] from ASPCAP and the Payne are always within 0.10 dex of that measured by the prior literature studies. In contrast, those measured by the Cannon and astroNN are often lower. In particular, astroNN seems to have a ceiling in its derived [Al/Fe] values, which never go higher than [Al/Fe] $\approx +0.40$. Meanwhile, the dispersion in [Al/Fe] derived from the BACCHUS pipeline exceeds the literature value for the cluster NGC 6205.

The Payne is the only one of the five pipelines which meets both of our criteria for robust [Al/Fe] measurements.

\subsection{We Adopt Payne-Derived Abundances of the APOGEE Data for our Subsequent Analysis} \label{subsec:GCsample6}

\begin{figure}
\includegraphics[width=0.45\textwidth]{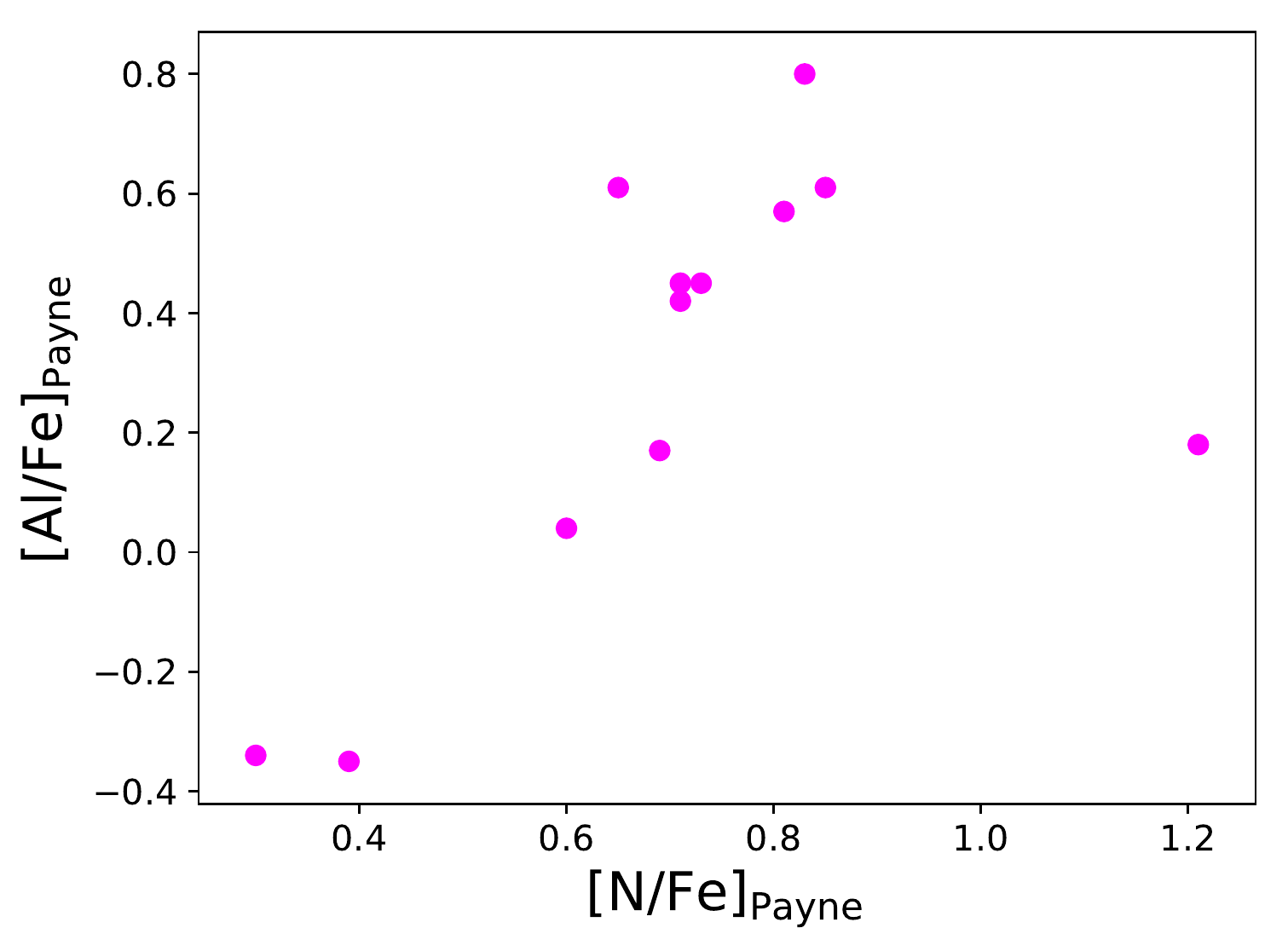}
\caption{The scatter of [Al/Fe] and [N/Fe] for stars in the metal-poor ([Fe/H]$-1.65$) globular cluster NGC 7089 (M2). We show stars with $T_{\rm{eff,Payne}} \leq 4,750\,K$, and stars with $5,250\, K \leq T_{\rm{eff}} \leq 4,750\,K$ and with measured signal-to-noise ratios greater than 50. As the metallicity of the cluster is less than the lower bound of the parameter space of the first Payne-derived data release, we cannot assume that the correlation is accurately recovered. However, its presence indicates that  future iterations of the Payne should be effective at lower metallicities.  }
\label{fig:Payne_NGC7089}
\end{figure}

Given the findings of Sections \ref{subsec:GCsample2}, \ref{subsec:GCsample3}, and \ref{subsec:GCsample5}, we adopt the Payne as our preferred option to study the relative abundances of globular cluster stars. Its relative strengths are:
\begin{enumerate}
\item More consistent determinations of [N/Fe], as evidenced by the decreased frequency of stars with [N/Fe] $< 0$, as well as the fact that this type of failure mode is more effectively suppressed by increased signal-to-noise ratio. The ability to include stars with $T_{\rm{eff}} > 4,750\,K$  and signal-to-noise ratios greater than 100 nearly doubles the sample available for study. 
\item Recovery of the C-N anticorrelation, which is not present in the DR14 releases of ASPCAP and the Cannon, and barely present in astroNN measurements. 
\item Dispersions of [O/Fe] that are nearly as large as those of \citet{2009A&A...505..117C}. 
\item The Payne is one of four pipelines to reliably recover the correlation between [N/Fe] and [Al/Fe]. 
\item The Payne is the only one of the five pipelines to both yield mean values of [Al/Fe] that are independent of the presence of molecular features as well as dispersions in [Al/Fe] consistent with prior literature values.
\item The Payne is one of four pipelines whose temperatures and gravity estimates are derived based off of the spectroscopy alone, and not dependent on literature estimates of reddening. 
\end{enumerate}
In contrast, the relative weaknesses of the Payne are not as significant:
\begin{enumerate}
\item The Payne does not report [Na/Fe]. This is normally an informative element when studying globular cluster stars, but the sodium lines within the APOGEE spectral window are very weak in the metallicity and temperature regimes typical of this work. 
\item The Payne, in its current implementation, has an effective abundance floor of [Fe/H]$=-1.50$. 
However, the reliability of the [X/Fe] abundance determinations in the metal-poor globular clusters, by ASPCAP, the Cannon, astroNN, and BACCHUS are uncertain. We show in Figure \ref{fig:Payne_NGC7089} that the Payne successfully recovers an [Al/Fe]-[N/Fe] correlation for stars in the metal-poor ([Fe/H]$=-1.65$) globular cluster NGC 7089  (M2). We do not include it within our sample as it is plausible that the correlation may be tilted due to the cluster's metallicity being outside the parameter space of the current implementation of the Payne, [Fe/H] $\geq -1.50$. Nevertheless, a correlation is recovered, and that is an indication that future implementations of the Payne could eventually perform effectively in globular clusters with a metallicity lower than its current floor of [Fe/H]$=-1.50$. 
\item A third weakness, one shared by four of the pipelines, is a ceiling of [X/Fe]$=+1.0$ on relative abundance determinations. That is sensible for studies of the Galactic field populations for which these pipelines were predominantly intended, it is not sensible for globular cluster stars. It is likely leading to underestimates of [N/Fe] and [Al/Fe] for the most chemically anomalous stars. 
\end{enumerate}

These limitations are unfortunate, but they are straightforward to modify by future iterations of the Payne and applications to APOGEE data. Ideally, the parameter spaces of [X/H] should be expanded down to $-$3.00 for [C/H] and [O/H]. The parameter spaces for [N/Fe] and possibly [Al/Fe] should be respectively expanded to $+2.0$ and $+$1.5. 

The current abundance floor will not effect the results presented in this work, as we limit our analysis to clusters with [Fe/H] $\geq -1.55$.

We note that the selection of the Payne is partly motivated by the scientific priority of this investigation, the analysis presented in Section \ref{subsec:fullsample}. That is, the dependency of aluminum abundance variations on that of other light element abundance variations, and globular cluster properties. The Payne is the pipeline which yields the largest sample of consistently measured values of [Fe/H], [Al/Fe], and [N/Fe]. However, a different investigation might be making a different choice. For example, the results of the BACCHUS pipeline are the most suitable for study of absolute abundance variations of carbon, oxygen, and nitrogen. 

\subsection{Comparison of Payne-Derived Abundances to those of \citet{2015AJ....149..153M}} \label{subsec:GCsample7}

\citet{2015AJ....149..153M} measured abundances for 428 red giants in 10 globular clusters with data from APOGEE. Their investigation differs from ours in a few ways. Among these, they used an earlier APOGEE data release which thus had a smaller sample of stars, and they used photometric rather than spectroscopic information to set their temperature and gravity scales. It is worthwhile to see if their derived [N/Fe] and [Al/Fe] are consistent with those derived by the Payne. 

There are 97 stars for which [N/Fe] was measured by both of our samples. The [N/Fe] values are  consistent in their trend, but the zero point of the [N/Fe] scale of \citet{2015AJ....149..153M} is shifted upwards by $\sim$0.20 dex. For the aluminum abundances, the abundances of \citet{2015AJ....149..153M} are shifted from those derived by the Payne by both a zero-point and a small rescaling. The relations are as follows:
\begin{equation}
    \begin{split}
       \rm{[N/Fe]}_{\rm{Meszaros15}} \approx \rm{[N/Fe]}_{\rm{Payne}}+0.20 \\
       \rm{[Al/Fe]}_{\rm{Meszaros15}} \approx 1.15 \rm{[Al/Fe]}_{\rm{Payne}}+0.30 \\ \\
    \end{split}
    \label{EQ:BestFits}
\end{equation}



\subsection{Abundance Correlations Measured by the Payne} \label{subsec:GCsample7}

\begin{figure*}
\includegraphics[width=1.00\textwidth]{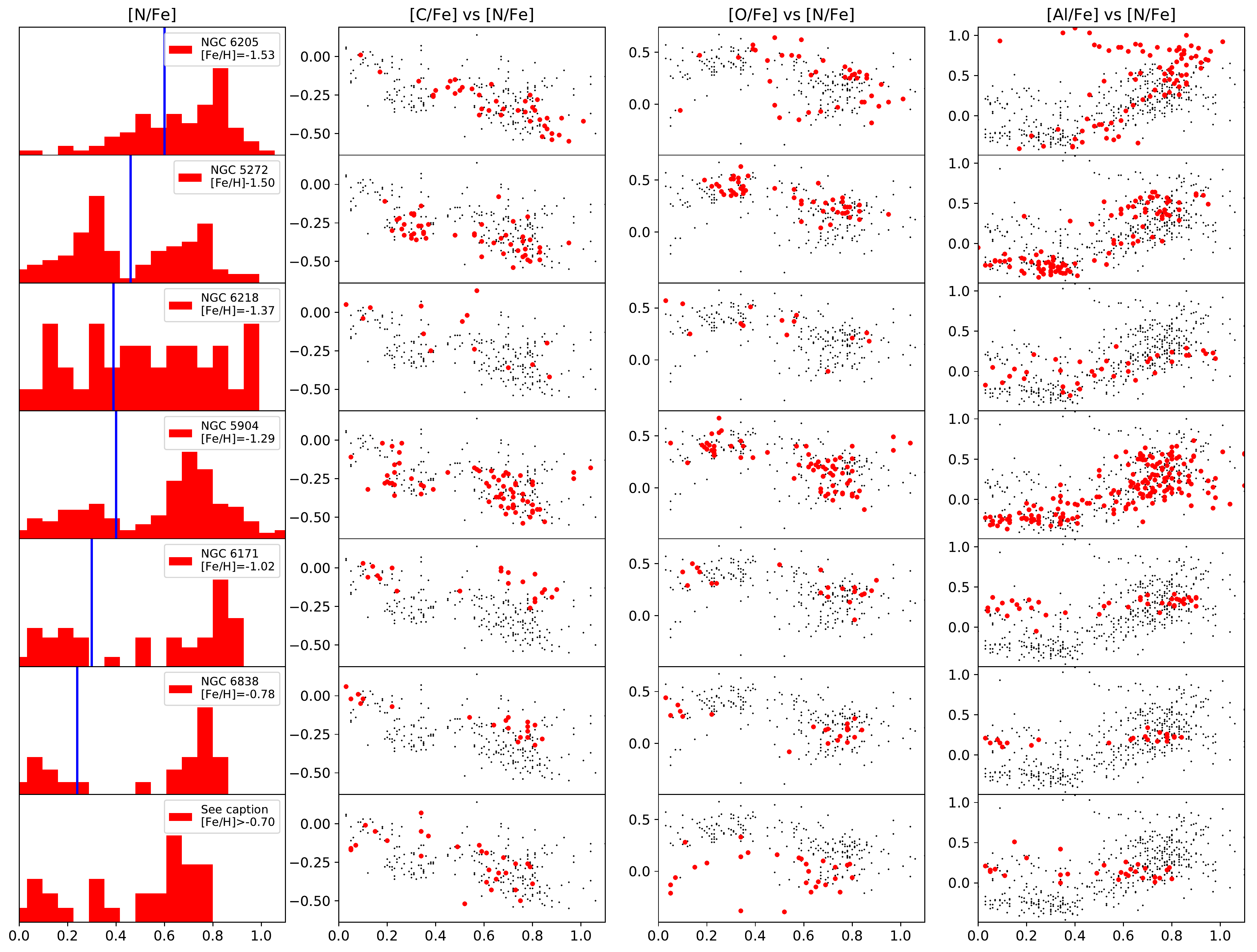}
\caption{LEFT panels: Histograms of [N/Fe] in each of six well-sampled clusters and an aggregate of 30 stars in 8 more metal-rich clusters (Pal1, NGC 6539, Terzan 12, NGC 6316, NGC 6760, Terzan 5, NGC 6553, and NGC 6528) in the bottom panel. The vertical blue lines denote the separation between the first and second generation stars, which is justified later in this work. RIGHT three panels: C-N-O-Al abundance correlations for all globular cluster stars in APOGEE DR14 are shown as the black points, and those of the specific clusters corresponding to each row are shown as the red points. The C-N-O abundance correlations are seen across the full metallicity range, whereas the correlation of [Al/Fe] with [N/Fe] appears only in the lower metallicity clusters.  [C/Fe] and [O/Fe] values are only shown for stars with  for stars with $T_{\rm{eff}} \leq 4,750\,K$, whereas the [N/Fe] and [Al/Fe] values are also shown for those stars and stars with $5,250\, K \leq T_{\rm{eff}} \leq 4,750\,K$ and with measured signal-to-noise ratios greater than 50.}
\label{fig:Payne_APOGEE_GC_CNOAl_AbundanceScatter}
\end{figure*}

The histogram of [N/Fe] abundances, and the CNO abundance correlations, are shown in Figure \ref{fig:Payne_APOGEE_GC_CNOAl_AbundanceScatter} for six globular clusters with measurements for at least ten stars, as well as an aggregate of 8 metal-rich globular clusters (Pal1, NGC 6539, Terzan 12, NGC 6316, NGC 6760, Terzan 5, NGC 6553, and NGC 6528) with [Fe/H] $\geq -0.70$, and a sample mean metallicity of [Fe/H]$=-0.40$ from 30 measurements. The CNO abundance correlations are present in all 7 groups and span a similar range. 

A striking feature of the [N/Fe] histograms is that the distributions distributions appear bimodal in five of the seven panels. That is consistent with photometric studies of globular clusters, which find that the multiple populations of globular clusters are distinct  \citep{2015MNRAS.451..312N,2017MNRAS.464.3636M,2018MNRAS.475.4088L}. In contrast, these distributions are not consistent with most other spectroscopic investigations, which typically find continuous sequences in abundance space. \citet{2015MNRAS.454.4197R} argues that the multiple populations of globular clusters are almost certainly distinct (and possibly discrete), and that the spectroscopic results are likely confounded by measurement error. However, APOGEE-derived Payne abundances show distinct populations, particularly for nitrogen and aluminum. It is clear, from Figure \ref{fig:APOGEE_GC_AbundanceScatter}, that distinct populations in the [N/Fe] distributions can also be identified by ASPCAP and the Cannon. That is an impressive achievement, and an argument for the continuing diagnostic potential of the APOGEE survey to study the stellar populations of Galactic globular clusters.

\begin{figure*}
\includegraphics[width=1.00\textwidth]{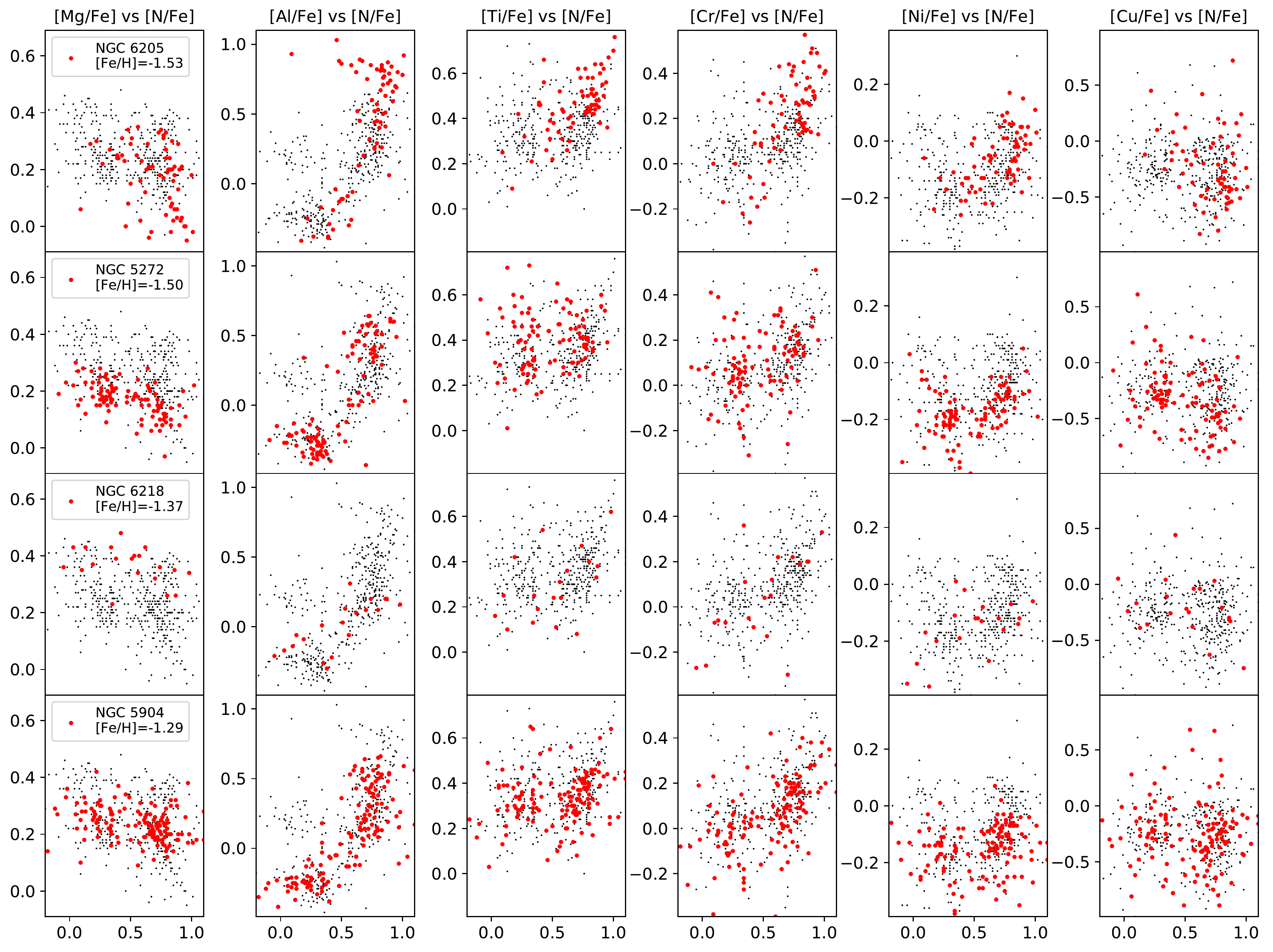}
\caption{The Payne-derived abundances from APOGEE spectra show that [Al/Fe], [Ti/Fe], [Cr/Fe], [Ni/Fe] are positively correlated with [N/Fe], and [Mg/Fe] and [Cu/Fe] are negatively correlated, with [N/Fe] in four well-sampled and metal-poor globular clusters. Points are as in Figure \ref{fig:Payne_APOGEE_GC_CNOAl_AbundanceScatter}, where we show the measurements for all stars with  $T_{\rm{eff}} \leq 4,750\,K$, and   all stars with $5,250\, K \leq T_{\rm{eff}} \leq 4,750\,K$ and with measured signal-to-noise ratios greater than 100. }
\label{fig:Payne_APOGEE_GC_MP_AbundanceScatter}
\end{figure*}

\begin{figure}
\includegraphics[width=0.48\textwidth]{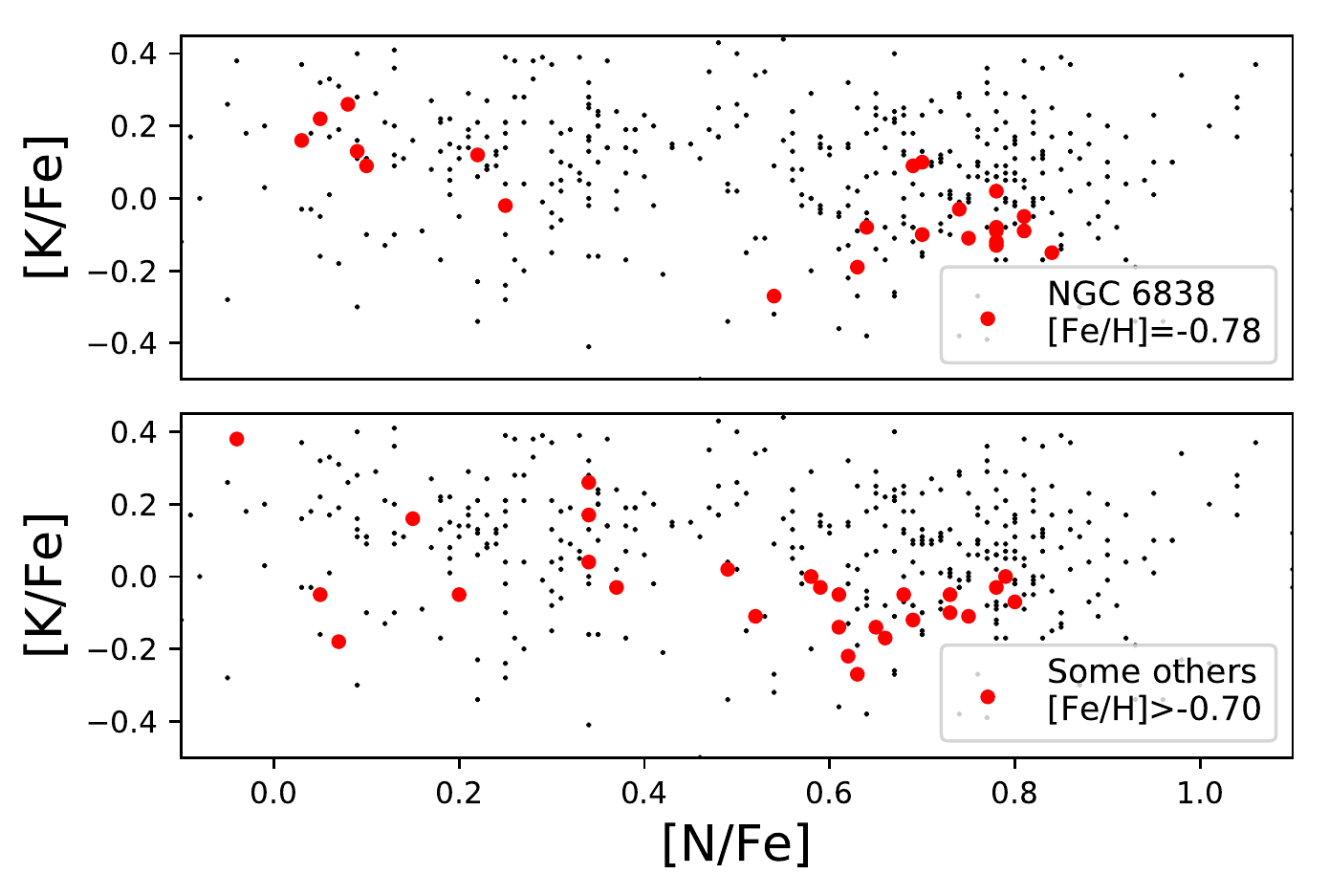}
\caption{For NGC 6838 (top panel), and for a collection of 26 stars in 8 more metal-rich clusters (Pal1, NGC 6539, Terzan 12, NGC 6316, NGC 6760, Terzan 5, NGC 6553, and NGC 6528) in the bottom panel, Payne-derived abundances from APOGEE spectra show an anti-correlation between [K/Fe] and [N/Fe]. Points are as in Figure \ref{fig:Payne_APOGEE_GC_CNOAl_AbundanceScatter}, where we show the measurements for all stars with  $T_{\rm{eff}} \leq 4,750\,K$ or measured signal-to-noise ratios greater than 50. }
\label{fig:Payne_APOGEE_GC_MR_AbundanceScatter}
\end{figure}

The [N/Fe]-[Al/Fe] correlation is present for stars in clusters as or more metal-poor than NGC 5904 (M5, [Fe/H]$=-$1.29), and is null or negligible for stars in clusters as or more metal-rich than NGC 6171 (M107,  [Fe/H]$=-$1.02).  Prior literature measurements \citep{2006A&A...455..271G,2008MNRAS.388.1419O,2009A&A...505..139C,2011ApJ...726L..20O,2013A&A...550A..34C,2017A&A...601A.112P} are consistent with a picture whereby the abundance variations of [Al/Fe] are reduced in more metal-rich clusters. As we show in the subsequent sections, the correlation between nitrogen and aluminum enrichment correlates not only with the globular cluster metallicity, but also with the present-day stellar mass of the cluster. 

There is no significant correlation between [N/Fe] and [Fe/H]. Such a correlation is not expected for most globular clusters \citep{2009A&A...505..117C}, but it could nonetheless result if there are issues with either the data or the reduction and analysis thereof. For the six well-sampled globular clusters shown in Figure \ref{fig:Payne_APOGEE_GC_CNOAl_AbundanceScatter}, we find that the mean increases of [Fe/H] in second generation stars relative to first generation stars are ${\Delta}$[Fe/H]$= 0.03, 0.01, 0.06, 0.01, -0.03, -0.01$. The average value is a negligible  $<{\Delta}$[Fe/H]$>= 0.01$.

We also show, in Figures \ref{fig:Payne_APOGEE_GC_MP_AbundanceScatter} and \ref{fig:Payne_APOGEE_GC_MR_AbundanceScatter}, the abundance correlations for other elements for the metal-poor and metal-rich clusters, respectively. The Payne abundances suggest that for the metal-poor clusters, nitrogen enrichment is positively correlated with enrichment in aluminum, titanium, chromium, and nickel; and anti-correlated with enrichment in magnesium and copper. 

Among the metal-rich clusters, we find only an anti-correlation with potassium. The decreased level of [K/Fe] in the nitrogen-enriched stars of metal-rich globular cluster is distinct from the primary literature finding in this area, which is that the second generation stars of metal-poor globular clusters typically have enhanced levels of [K/Fe]. For example, \citet{2017A&A...600A.104M} measured small increases in [K/Fe] in the three clusters NGC 104, NGC 6752, and NGC 6809, which have respective metallicities of [Fe/H]$_{\rm{Harris}}=-$0.72, $-$1.54, and $-$1.81. This followed the work of \citet{2015ApJ...801...68M}, who measured a similar increase of [K/Fe] in the second generation stars of NGC 2808, which has  [Fe/H]$_{\rm{Harris}}=-$1.14. \citet{2018MNRAS.tmp.1822K} have also found Mg-depleted, and K-enhanced, field stars in the LAMOST sample \citep{2015RAA....15.1095L} with spectroscopic abundances derived by \citet{2017ApJ...841...40H}. We do not know the origin of this discrepancy. It may be that the potassium line in the APOGEE spectra is too weak for precise measurements in metal-poor stars, and that the literature has simply not adequately studied potassium variations in more metal-rich clusters.

We do not discuss these abundance correlations in detail, as our investigation is primarily focused on the subjects of nitrogen and aluminum enrichment. However, we included these figures as they may be of interest to some readers. 

Some readers may be concerned by our choice to frame our discussion in terms of aluminum enrichment, rather than the aluminum-magnesium anti-correlation, as is more standard (e.g. \citealt{2017A&A...601A.112P}). The latter approach is almost certainly more physically correct, as the two abundance variations are likely linked by the ${}^{24}$Mg(p,$\gamma$)${}^{25}$Al nuclear reaction. However, the measurement precision of [Al/Fe] \textit{variations} are greater than that of the corresponding variations of [Mg/Fe]. Though the ${}^{24}$Mg(p,$\gamma$)${}^{25}$Al nuclear reaction conserves the total number of these two nuclei, the cosmic abundance of magnesium is approximately 13 times that of aluminum \citep{1993oee..conf...15G}. The combination of these two factors, with the fact that the measurement precision of [Al/Fe] and [Mg/Fe] are comparable, results in a different signal-to-noise ratio, which can be discerned from the two leftmost panels of Figure \ref{fig:Payne_APOGEE_GC_MP_AbundanceScatter}.

\subsection{A Note on Palomar 6}  \label{subsec:Pal6}

\citet{2016A&A...590A...9D} constructed a new metallicity scale for 51 globular clusters using medium-resolution spectra of $\sim$800 red giant stars. For the globular cluster Palomar 6, they reported $v_{helio}=177$ km/s, and [Fe/H]$=-0.85$, which were at odds with some of the prior literature measurements. We associate three of Palomar 6's stars with the APOGEE catalog, for which we find [Fe/H]=$-$0.95, $-$0.85, and $-$0.79. The third star is enhanced by $\sim$0.50 dex in [N/Fe] relative to the other two with correspondingly reduced values of [C/Fe] and [O/Fe]. This validates a globular cluster membership for these three stars, and by extension, these values of $v_{helio}$ and [Fe/H].

\section{Analysis} \label{sec:Analysis}

\subsection{Delineating First and Second Generation Stars} \label{subsec:Generations}

\begin{figure*}
\includegraphics[width=1.00\textwidth]{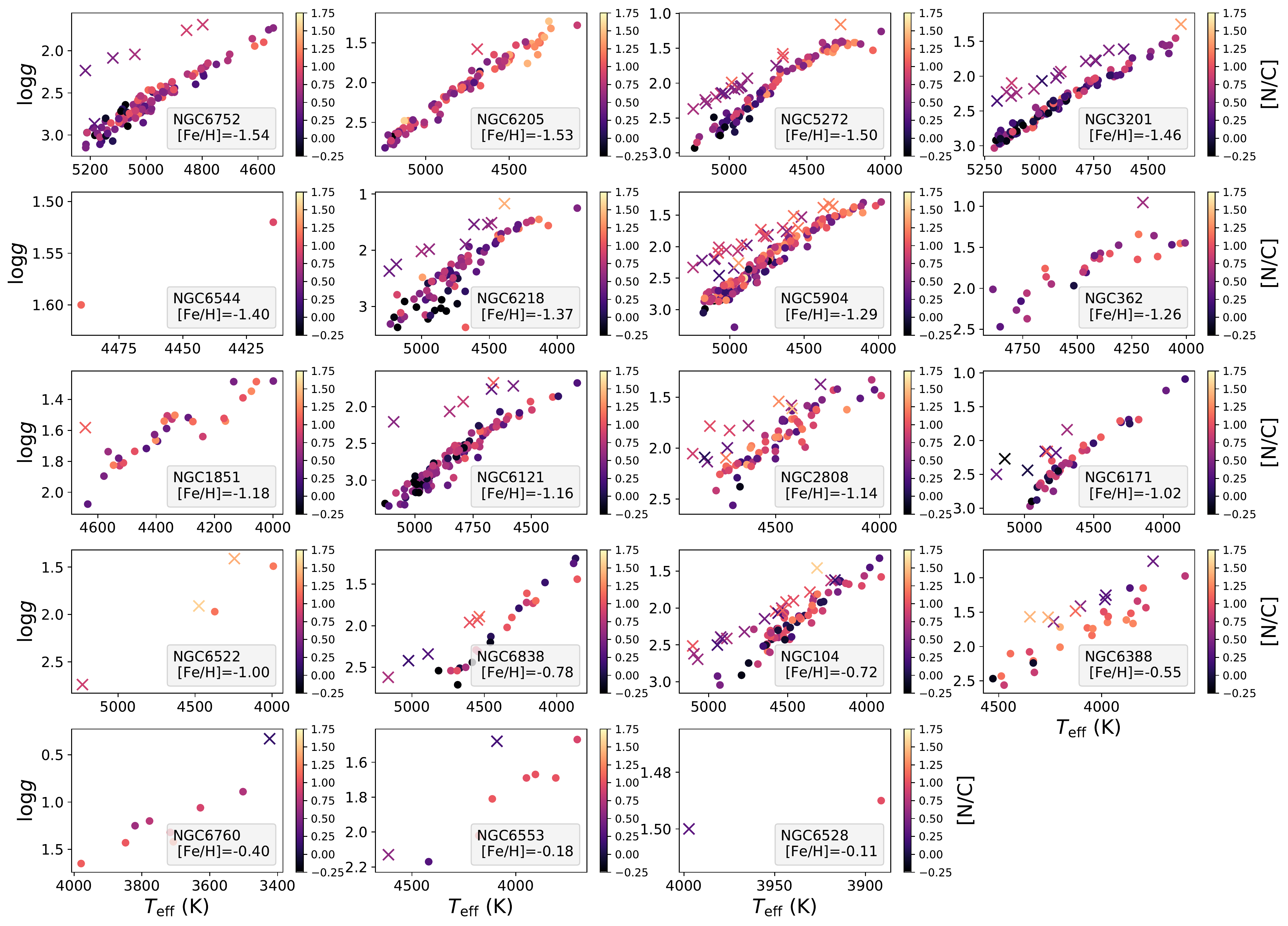}
\caption{For each of the 19 globular clusters with APOGEE data used in this section, we plot the spectroscopically determined effective temperature and gravity of each star, color-coded by their [N/C] ratio. The Payne provides sufficiently reliable parameter estimates to enable a coarse separation of first-ascent red giant branch (circles) and asymptotic giant branch stars (x's). }
\label{RGBselection}
\end{figure*}

We use the term ``first-generation" to refer to those stars with chemical compositions similar to those of the Halo, Bulge, and Thick Disk at their metallicity.  We use the term ``second generation" to refer to those stars showing some combination of enhanced nitrogen, sodium, and aluminum, as well as deficient carbon, oxygen, and magnesium. We acknowledge that this terminology is problematic in multiple ways. The implicit assumption, that the chemically anomalous stars were born \textit{after} the chemically mundane stars and partially \textit{from} their enriched ejecta, has not actually been demonstrated. Further, the ``second" generation may include a ``third" (or greater) generation.

The separation between the first and second generations in the compilation of \citet{2009A&A...505..139C}, which they respectively label as the primordial and intermediate/extreme populations, is stated in their Table 11. We adopt their delineation for all of their program clusters, and similarly for other works which identified their own demarcation lines. For the sample of  \citet{2009A&A...505..139C}, the first generation abundance of [Na/Fe] is assumed to be the midpoint of their specified range, and the first generation abundance of [Al/Fe] is the mean [Al/Fe] of stars with a first generation abundance of [Na/Fe]. For the other clusters in the literature sample, we estimate the cutoff between the first and second generations by eye. The first generation abundance of a cluster is then assumed to be the mean [Na/Fe] and [Al/Fe] of all first generation stars for that cluster, except for NGC 6656 (M22) \citep{2011A&A...532A...8M} where we use the median, and NGC 6139 \citep{2015A&A...583A..69B} where we use the mean weighted by the inverse square of the reported measurement errors. We require that there be at least two stars of the first generation with measured [Al/Fe] abundances for the cluster to be included in the literature compilation. 

For the globular cluster stars measured with APOGEE data, we first do a selection for red giant branch stars, at the expense of asymptotic giant branch stars, using the spectroscopic determinations of $T_{\rm{eff}}$ and $\log{g}$. The selection is shown in Figure \ref{RGBselection}, is approximate, and is done to reduce the risk of dredge-up among asymptotic giant branch stars \citep{2007A&A...463..251U,2018arXiv181207434U} contributing added variation to the abundance trends. Following this, we first use the guess that second generation stars are those with [N/Fe]  $-0.20 \leq$ [N/Fe]$_{\rm{Bulge}}$([Fe/H]), where the latter is the [N/Fe] of Bulge stars at the [Fe/H] value of that cluster, derived in a manner described below. We then iterate from the guess once, taking the median [N/Fe] abundance of those stars in each cluster that also satisfy [Al/Fe] $\leq +0.60$, and again using a cutoff of ${\Delta}$[N/Fe]$=$0.20 dex. The first generation [N/Fe] and [Al/Fe] values are the median values of the stars. If there are two or fewer first-generation stars, as is the case for NGC 6544, 6522, 6760, 6553, and 6528, we simply adopt the [N/Fe] and [Al/Fe] from the Bulge trend line. 

To construct a Bulge sample, we use the distance estimates from the StarHorse pipeline \citep{2018MNRAS.476.2556Q}, and require that the separation of stars from the Galactic plane $|Z_{g}|$ be no greater than 1.5 kpc, and that the separation of each star from the Galactic centre projected onto the Galactic plane, $R_{g}$, be no greater than 3.5 kpc. Posterior estimates of the distances to stars are derived by forward modeling stellar isochrones \citep{2012MNRAS.427..127B} and Galactic structure priors to derive probabilities of observing the measured \textit{Gaia} parallaxes \citep{2018A&A...616A...1G} and stellar atmosphere parameters. Unless stated otherwise, we use the median distance estimate from the posterior distribution of each star. The Halo and thick disk are not included, as that would require a more complex selection criteria, and regardless, their low-metallicity abundance trends are similar to those of the Bulge  \citep{2019ApJ...870..138Z}. A comprehensive comparison of these abundance trends will be presented by Queiroz et al. (2019, in preparation). 

\begin{figure}
\centering
\includegraphics[width=0.47\textwidth]{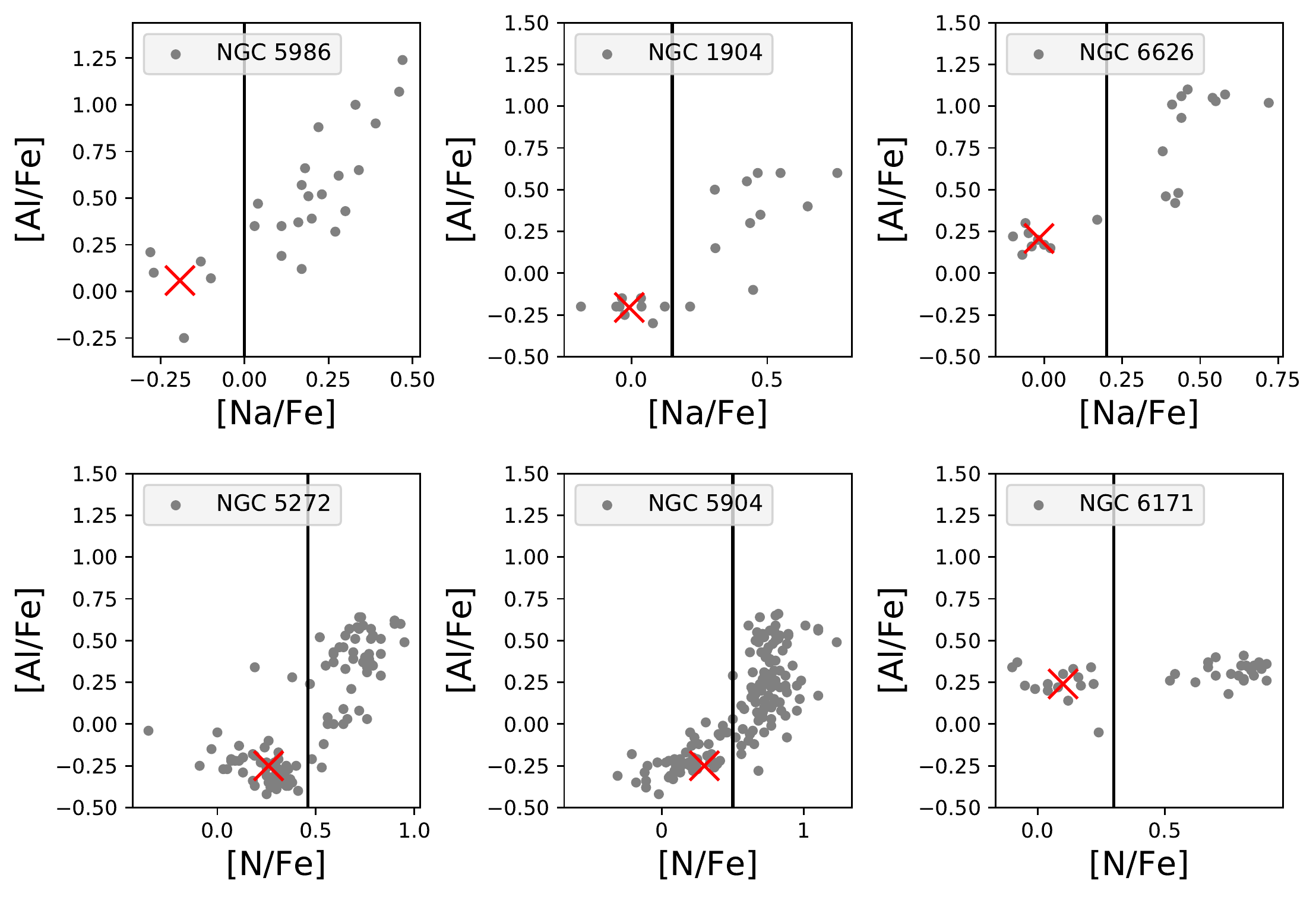}
\caption{The abundance correlations between [Na/Fe] or [N/Fe] and [Al/Fe] are shown for six globular clusters. The measurements are denoted by the grey points, the dividing line between the first and second generations by the vertical black lines, and the mean values of the first generation abundances by the red X's. }
\label{fig:Gen1Gen2Mosaic}
\end{figure}

We show the demarcation between the first and second generations of six globular clusters in Figure \ref{fig:Gen1Gen2Mosaic}, from which it is clear the error in separating the first and second generations is manifestly small. Our cutoffs between the first and second generations for the 32 clusters in the literature compilation are listed in Table \ref{tab:Generations}, and for the 24 clusters with data from APOGEE are listed in Table \ref{tab:GenerationsAPOGEE}. We thus compile data for 45 separate clusters, including 11 clusters which show up in both samples. 

\begin{figure}
\includegraphics[width=0.48\textwidth]{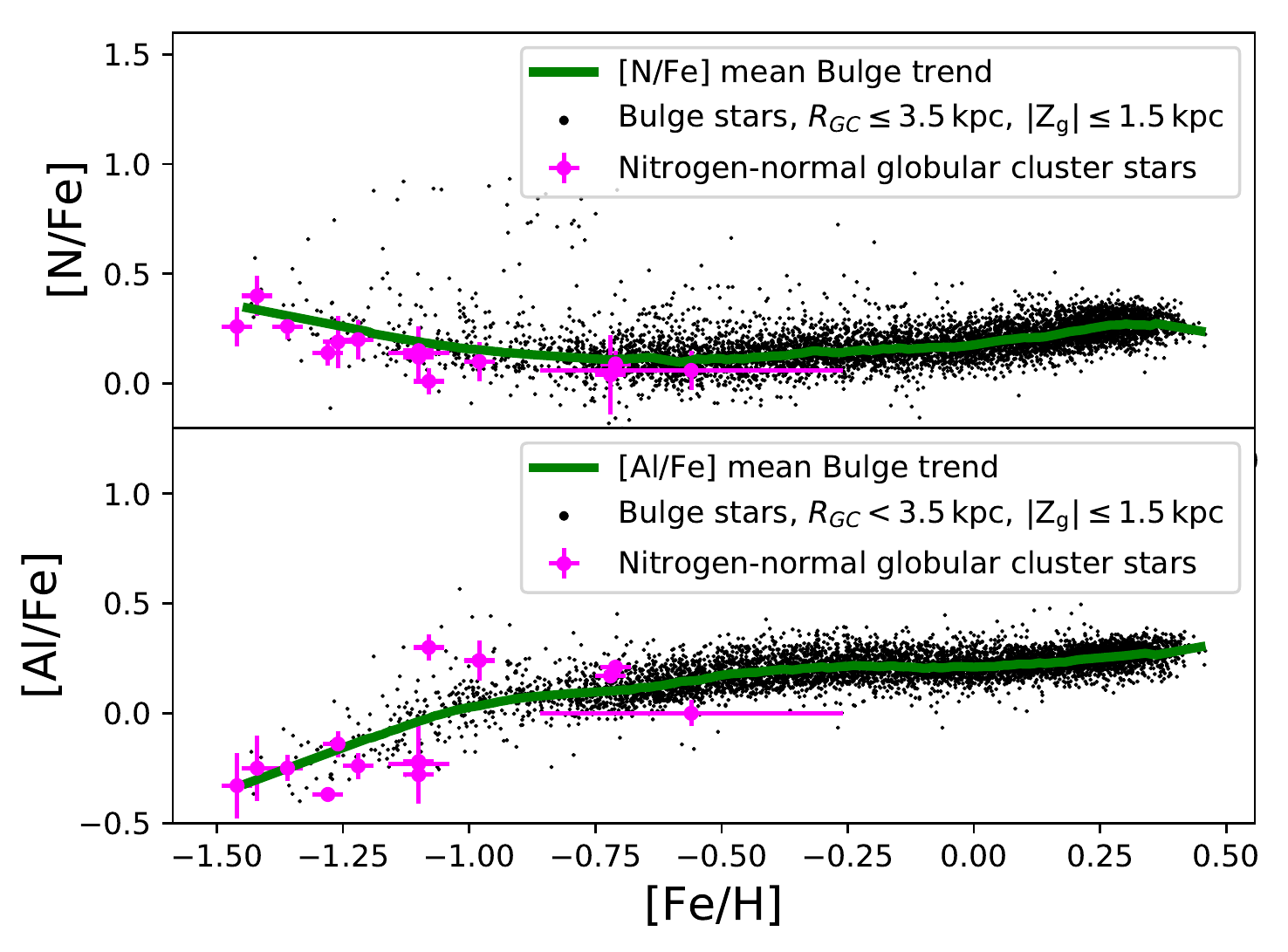}
\caption{The mean first generation globular cluster abundances (magenta points) follow a similar [N/Fe]-[Fe/H] mean trend (green line) as the Bulge field population (black points, selected based on estimated Galactic position as described in the caption), whereas they are often slightly offset from the [Al/Fe]-[Fe/H] relation. These abundances are Payne determinations derived from APOGEE spectra for globular clusters with at least three first generation stars estimated to be on the first-ascent red giant branch, in order of increasing metallicity: NGC 6752, NGC 6205 (M13), NGC 5272 (M3), NGC 3201, NGC 6218 (M12), NGC 5904 (M5), NGC 362, NGC 1851, NGC 6121 (M4), NGC 2808, NGC 6171 (M107), and NGC 6838 (M71), NGC 104 (47 Tuc), and NGC 6388.  The error bars represent the sample error in the mean, multiplied by three for clarity. Bulge stars are selected using StarHorse \citep{2018MNRAS.476.2556Q}. }
\label{Al_BulgeGCs}
\end{figure}

We show in Figure \ref{Al_BulgeGCs} that the mean [N/Fe]  of first-generation globular cluster stars is slightly lower than that of the bulge field population at their respective [Fe/H]$_{\rm{\rm{Payne}}}$, whereas the [Al/Fe] abundances are consistent in the mean, but often shifted to lower or higher values.

From this point on, our analysis treats second generation abundances in a differential manner. We focus on the abundances of N, Na, and Al in second generation stars, relative to the estimated mean value of those abundances in their first generation counterparts.  We thus define the quantity ${\Delta}\rm{[Al/Fe]}_{\rm{GenII,i}}=\rm{[Al/Fe]}_{\rm{GenII,i}}-\langle \rm{[Al/Fe]}\rangle_{\rm{GenI}}$, and similarly for nitrogen and sodium, as the relative abundance of the \textit{i}'th second generation star. 

There are several advantages to our differential approach, of which we mention two. The first is that it minimizes the instrumental and methodological zero-points that might vary between the different analyses. The second is that, relative chemical abundances need not have been homogeneous in time and space when the first generations of globular clusters were forming. 

\begin{table*}
\centering
\caption{The list of program clusters for which we found sufficient data to include in our investigation as part of our literature compilation, along with the cutoffs that we use to define their ``second-generation" stars, and the references for the data. The values sourced from \citet{2009A&A...505..139C} are derived by adding 0.15 dex to the [Na/Fe]$_{\rm{min}}$ listed in Table 11 of that work. } 
\begin{tabular}{lll}
\hline
Name & Second Generation Definition & Source of Data \\
\hline
NGC 104 & [Na/Fe] $\geq$ 0.45 & Carretta et al. (2009a) \\
NGC 288 & [Na/Fe] $\geq$ 0.20 & Carretta et al. (2009a) \\\
NGC 362 & [Na/Fe] $\geq$ 0.0 & Carretta et al. (2013) \& D'Orazi et al. (2015) \\
NGC 1851 & [Na/Fe] $\geq$ 0.00 & Carretta et al. (2012) \\
NGC 1904 & [Na/Fe] $\geq$ 0.15 & Carretta et al. (2009a) \& D'Orazi et al. (2015) \\
NGC 2808 & ``Group" $==$ 2,3,4,5 & Carretta (2015), Carretta et al. (2018) \\
NGC 3201 & [Na/Fe] $\geq$ 0.00 & Carretta et al. (2009a) \\
NGC 4147 & [Na/Fe] $\geq$ 0.20 & Villanova et al. (2016) \\
NGC 4833 & [Na/Fe] $\geq$ 0.30  &  Carretta et al. (2009a) \\
NGC 5897 & [Na/Fe] $\geq$ 0.30  & Koch \& McWilliam (2014) \\
NGC 5904 & [Na/Fe] $\geq$ 0.05 & Carretta et al. (2009a) \\
NGC 5927 & $\log{\epsilon_{\rm{Na}}} \geq$ 6.1 & Gilmore et al. (2012) \& Pancino et al. (2017) \\
NGC 5986 & [Na/Fe] $\geq$ 0.00 & Johnson et al. (2017a) \\
NGC 6093 & [Na/Fe] $\geq$ 0.00 & Carretta et al. (2015) \\
NGC 6121 &  [Na/Fe] $\geq$ 0.20 & Marino et al. (2008) \\
NGC 6139 & [Na/Fe] $\geq$ 0.20  &  Bragaglia et al. (2015) \\
NGC 6218 & [Na/Fe] $\geq$ 0.10 & Carretta et al. (2009a) \\
NGC 6229 & [Na/Fe] $\geq$ -0.05 & Johnson et al. (2017b) \\
NGC 6254 & [Na/Fe] $\geq$ 0.00 & Carretta et al. (2009a) \\
NGC 6266 & [Na/Fe] $\geq$ 0.30  & Lapenna et al. (2015) \\
NGC 6362 &  [Na/Fe] $\geq$ 0.25  & Mucciarelli et al. (2016) \& Massari et al. (2017)  \\
NGC 6388 & [Na/Fe] $\geq$ 0.00 & Carretta et al. (2018) \\
NGC 6397 & [Na/H] $\geq -$2.20 & MacLean et al. (2018) \\
NGC 6440 & [Na/Fe] $\geq$ 0.30 & Mu{\~n}oz et al. (2017) \\
NGC 6528 & [Na/Fe] $\geq$ 0.40 & Mu{\~n}oz et al. (2018) \\
NGC 6584 & [Na/Fe] $\geq$ 0.00 & O'Malley et al. (2018) \\
NGC 6626 & [Na/Fe] $\geq$ 0.20 & Villanova et al. (2017) \\
NGC 6656  &  [Na/Fe] $\geq$ 0.20 & Marino et al. (2011) \\
NGC 6569 & [Na/Fe] $\geq$ 0.15 & Johnson et al. (2018) \\
NGC 6681 & [Na/Fe] $\geq$ 0.10 & O'Malley et al. (2017) \\
NGC 6752 & [Na/Fe] $\geq$ 0.15 & Carretta et al. (2009a) \\
NGC 6809 & [Na/Fe] $\geq$ -0.05 & Carretta et al. (2009a) \\
NGC 7078 & [Na/Fe] $\geq$ 0.25 & Carretta et al. (2009a) \\
NGC 7099 & [Na/Fe] $\geq$ 0.10 & Carretta et al. (2009a) \\
\hline
\hline
\end{tabular}
\label{tab:Generations}
\end{table*}

\begin{table*}
\centering
\caption{The list of program clusters for which we found sufficient data to include in our investigation as part of our APOGEE compilation, along with the cutoffs that we use to define their ``second-generation" stars, and the references for the data. The 11 clusters for which the data are taken from the unpublished DR16 catalog are listed below the dividing line, and are computed with the same method as presented by \citet{2018arXiv180401530T}. } 
\begin{tabular}{lll}
\hline
Name & Second Generation Definition & Source of Data \\
\hline
NGC5024 & ``Group" $==$ 2 &  Majewski et al. (2017) \& Meszaros et al. (2015) \\
NGC5466 & ``Group" $==$ 2 & Majewski et al. (2017) \& Meszaros et al. (2015) \\
NGC6341 & ``Group" $==$ 2 &  Majewski et al. (2017) \& Meszaros et al. (2015) \\
NGC7078 & ``Group" $==$ 2 &  Majewski et al. (2017) \& Meszaros et al. (2015) \\
NGC7089 & ``Group" $==$ 2 & Majewski et al. (2017) \& Meszaros et al. (2015) \\
NGC 5272 & [N/Fe] $\geq$ 0.46 &  Majewski et al. (2017) \& Ting et al. (2018) \\
NGC 5904 & [N/Fe] $\geq$ 0.40 & Majewski et al. (2017) \& Ting et al. (2018) \\
NGC 6171 & [N/Fe] $\geq$ 0.30 & Majewski et al. (2017) \& Ting et al. (2018) \\
NGC 6205 & [N/Fe] $\geq$  0.60 & Majewski et al. (2017) \& Ting et al. (2018) \\
NGC 6218 & [N/Fe] $\geq$ 0.39 & Majewski et al. (2017) \& Ting et al. (2018) \\
NGC 6553 & [N/Fe] $\geq$ 0.35 & Majewski et al. (2017) \& Ting et al. (2018) \\
NGC 6760 & [N/Fe] $\geq$ 0.32 & Majewski et al. (2017) \& Ting et al. (2018) \\
NGC 6838 & [N/Fe] $\geq$ 0.24 & Majewski et al. (2017) \& Ting et al. (2018) \\
\hline
NGC 104 & [N/Fe] $\geq$ 0.29 & Majewski et al. (2017)  \\
NGC 362 & [N/Fe] $\geq$ 0.34 & Majewski et al. (2017)  \\
NGC 1851 & [N/Fe] $\geq$ 0.35 & Majewski et al. (2017)  \\
NGC 2808 & [N/Fe] $\geq$ 0.32 & Majewski et al. (2017)  \\
NGC 3201 & [N/Fe] $\geq$ 0.34 &  Majewski et al. (2017)\\
NGC 6121 & [N/Fe] $\geq$ 0.21 & Majewski et al. (2017) \\
NGC 6388 & [N/Fe] $\geq$ 0.26 & Majewski et al. (2017) \\
NGC 6522 & [N/Fe] $\geq$ 0.38 & Majewski et al. (2017) \\
NGC 6528 & [N/Fe] $\geq$ 0.35 & Majewski et al. (2017) \\
NGC 6544 & [N/Fe] $\geq$ 0.50 & Majewski et al. (2017) \\
NGC 6752 & [N/Fe] $\geq$  0.46 & Majewski et al. (2017) \\
\hline
\hline
\end{tabular}
\label{tab:GenerationsAPOGEE}
\end{table*}

\subsection{The mean second generation aluminum enrichment in globular clusters} \label{subsec:MeanAluminum}

\begin{figure}
\centering
\includegraphics[width=0.50\textwidth]{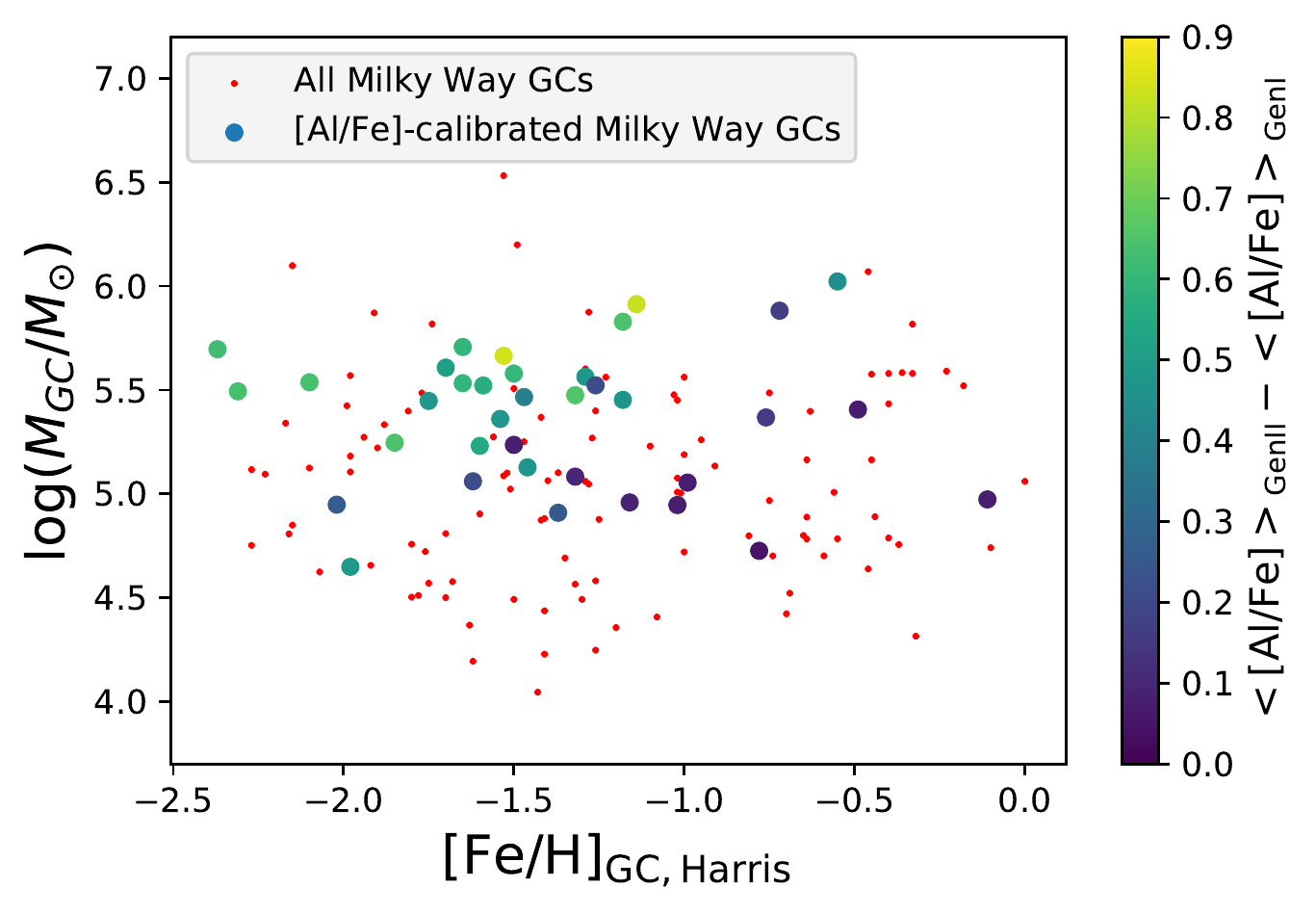}
\caption{Globular clusters with a lower [Fe/H] or a greater stellar mass tend to have a greater difference in the mean aluminum abundance of their multiple populations. The 36 clusters with measurements of mean aluminum enrichment are color coded between dark blue and yellow respectively corresponding to small and large mean differences in [Al/Fe]. The clusters without measurements are shown as the small red points.  }
\label{fig:AluminumDependency3}
\end{figure}

\begin{table*}
\centering
\caption{For the 19 globular clusters with APOGEE data whose chemical properties are estimated in this work, we list the name, literature estimate of the metallicity, the present-day stellar mass, the number of first generation stars assumed to be on the first-ascent red giant branch, their mean [N/Fe] and [Al/Fe] abundances, and likewise for the second generation. Note that the Payne-derived APOGEE abundances of [N/Fe] are expected to have variations approximately 50\% lower than the actual variations.  } 
\begin{tabular}{|lll|cccccc|}
\hline
Name & [Fe/H]$_{\rm{Harris}}$ & $\log{M_{GC}/M_{\odot}}$ & N$_{1}$ & $\langle \rm{[N/Fe]} \rangle_{1} $ &  $\langle \rm{[Al/Fe]} \rangle_{1} $  & N$_{2}$ & $\langle \rm{[N/Fe]} \rangle_{2} $ &  $\langle \rm{[Al/Fe]} \rangle_{2} $ \\
\hline
NGC 6752  & $-$1.54 & 5.36 & 26 & 0.26 & $-$0.33 & 75 & 0.73 & 0.51 \\
NGC 6205  & $-$1.53 & 5.66 & 15 & 0.40 & $-$0.25 & 57 & 0.80 & 0.59 \\
NGC 5272  & $-$1.50 & 5.58 & 48 & 0.26 & $-$0.25 & 42 & 0.71 & 0.35 \\
NGC 3201  & $-$1.46 & 5.13 & 56 & 0.14 & $-$0.37 & 38 & 0.65 & 0.18 \\
NGC 6544  & $-$1.40 & 5.06 & 0 & 0.30 & $-$0.24 & 2 & 0.73 & 0.16 \\
NGC 6218  & $-$1.37 & 4.91 & 33 & 0.19 & $-$0.14 & 48 & 0.72 & 0.11 \\
NGC 5904  & $-$1.29 & 5.56 & 53 & 0.20 & $-$0.24 & 113 & 0.74 & 0.24 \\
NGC 362  & $-$1.26 & 5.52 & 12 & 0.14 & $-$0.23 & 15 & 0.63 & 0.15 \\
NGC 1851  & $-$1.18 & 5.45 & 11 & 0.15 & $-$0.28 & 16 & 0.73 & 0.09 \\
NGC 6121  & $-$1.16 & 4.96 & 36 & 0.01 & 0.30 & 86 & 0.66 & 0.42 \\
NGC 2808  & $-$1.14 & 5.91 & 24 & 0.12 & $-$0.22 & 41 & 0.66 & 0.26 \\
NGC 6171  & $-$1.02 & 4.95 & 15 & 0.10 & 0.24 & 23 & 0.77 & 0.31 \\
NGC 6522  & $-$1.00 & 5.56 & 0 & 0.18 & $-$0.01 & 2 & 0.93 & 0.36 \\
NGC 6838  & $-$0.78 & 4.73 & 10 & 0.04 & 0.17 & 12 & 0.72 & 0.22 \\
NGC 104  & $-$0.72 & 5.88 & 22 & 0.09 & 0.21 & 43 & 0.70 & 0.27 \\
NGC 6388  & $-$0.55 & 6.02 & 3 & 0.06 & 0.00 & 21 & 0.73 & 0.14 \\
NGC 6760  & $-$0.40 & 5.43 & 0 & 0.12 & 0.12 & 8 & 0.68 & 0.10 \\
NGC 6553  & $-$0.18 & 5.52 & 1 & 0.15 & 0.21 & 6 & 0.62 & 0.18 \\
NGC 6528  & $-$0.11 & 4.97 & 0 & 0.15 & 0.21 & 1 & 0.66 & 0.23 \\
\hline
\hline
\end{tabular}
\label{tab:APOGEEclusters}
\end{table*}

In this work, we are primarily interested in investigating the trends between aluminum enrichment and either sodium or nitrogen enrichment among globular cluster stars. It is however interesting, as a first step, to simply assess the \textit{mean} value of aluminum enrichment in globular clusters as a function of globular cluster parameters. 

We compute the linear regression of the difference in the mean aluminum abundance of first and second generation globular cluster stars, ${\Delta}\langle \rm{[Al/Fe]} \rangle = (1/n_{GenII})\sum_{i} {\Delta}\rm{[Al/Fe]}_{\rm{GenII,i}}$, where $n_{\rm{GenII}}$ is the number of second generation stars of a particular cluster, as a function of globular cluster mass and metallicity, such that ${\Delta}\langle \rm{[Al/Fe]}  \rangle = a + b(\rm{[Fe/H]}+1.30) + c(\log{M_{GC}/M_{\odot}}-5.50)$. We include within our fits a step function, which is equal to 0 if the predicted ${\Delta}\langle \rm{[Al/Fe]}  \rangle$ is negative, and is equal to 1 otherwise. The step function is included on empirical grounds, as we do not see globular clusters for which [Al/Fe] decreases in second generation stars, and without the inclusion of the step function the coefficients $a,b,c$ may end up shifted to compensate. We acknowledge that in principle, the location of the step function may be at a small non-zero value, perhaps even at a small negative value. 

We restrict the fit to the 36 clusters for which there are at least three stars of both of the first and second generations. We include the measurements of \citet{2015AJ....149..153M} for the metal-poor clusters NGC 6341 / M92, NGC 5024 / M53, NGC 5466, and NGC 7089 / M2. Their aluminum abundance variations have been re-scaled by (1/1.15) to be consistent with the values determined by the Payne. For the clusters that have both prior literature measurements and APOGEE measurements, the APOGEE-derived values are given priority if they are from the DR14 sample. We obtain the relation:
\begin{equation}
\begin{split}
    {\Delta}\langle \rm{[Al/Fe]} \rangle = \rm{Max}\{0.43 -0.30([Fe/H]+1.30)  \\
    + 0.44(\log{M_{GC}/M_{\odot}}-5.50),0\}.
    \end{split}
    \label{EQ:meanAl}
\end{equation}
The three coefficients are measured with statistical significances of $15.5$, $5.2$, and $5.5$-$\sigma$. The sample's scatter to the relation is 0.14 dex. The value of the mass coefficient is $\sim$47\% higher than that of the metallicity coefficient, whereas the ratio was only 26\% higher in the work of \citet{2017A&A...601A.112P}. The inclusion of the step function has a very modest effect on both the best-fit parameters and their errors. 

We show the distribution of ${\Delta} \langle \rm{[Al/Fe]} \rangle $ as a function of globular cluster mass and metallicity in Figure \ref{fig:AluminumDependency3}. The clusters with the greatest aluminum enrichment between the first and second generation, those shown as green and yellow dots, tend to be of higher mass, lower metallicity, or both. Conversely, those with the lowest measured aluminum enrichment tend to be of lower mass, greater metallicity, or both. 



At high metallicity, [Fe/H]$_{\rm{GC}} \gtrsim   -0.50$, the one cluster with significant aluminum enrichment is NGC 6388, with  ${\Delta}\langle \rm{[Al/Fe]} \rangle $ = 0.45 dex. It is unique within our sample, but it is not unique within the Galaxy. Variations in [Al/Fe] have also been confirmed for the comparably massive and metal-rich globular clusters NGC 6440 \citep{2017A&A...605A..12M} and NGC 6441 \citep{2008MNRAS.388.1419O,2009A&A...505..139C}, but those samples are not large enough for us to confidently estimate the mean abundances of the first and second generations. Large samples of abundance measurements for NGC 6440 and NGC 6441 would be informative. 


At the request of the AAS statistics consultant, we estimated the impact of the assumption of linearity in Equation \ref{EQ:meanAl} is valid by also fitting the data using a local non-parametric regression. We computed a second-order bivariate spline regression at each point using the scipy.interpolate.SmoothBivariateSpline function in Python. We found that the differences between the two predictions are mostly small, regardless of the smoothing factor. 

The exception is at the low-metallicity end. The linear fit is found to overestimate the value of aluminum enrichment relative to the bivariate spline regression. This suggests that metallicity-dependence of aluminum enrichment may level off at low metallicity. Though it would be helpful to have more low-metallicity clusters with measurements in order to confirm this trend, it is consistent with what has been recently reported by \citet{2019A&A...622A.191M}. They reported a turnover in the magnesium-aluminum abundance ant-correlations at low metallicity, where a peak value of aluminum enrichment was reached. They suggested that it was due to leakage of nucleons into silicon production. 

The global parameters, and the derived chemical parameters for the 19 globular clusters with APOGEE data that we study in detail in this work, are listed in Table  \ref{tab:APOGEEclusters}.

\subsection{The Relationship Between NNaAl abundance variations, Globular Cluster Metallicity, and Globular Cluster Mass} \label{subsec:fullsample}


In practice, [Al/Fe] varies not just with the mass and metallicity of a globular cluster, but also with [N/Fe]. That is because the chemical abundances for many (but possibly not all) of the second generations of globular clusters are distributed as \textit{sequences}, rather than discrete points. We thus need to fit for [Al/Fe] abundances as a function of at least three parameters: globular cluster metallicity, mass, and the [N/Fe] of a particular star. For this task, we include 19 clusters with data from APOGEE with measurements for 649 second generation stars, and 34 clusters with prior literature data with measurements for 482 second generations. The total sample spans 42 clusters, including 11 that are in both of the literature and APOGEE samples. 



\begin{figure*}
\includegraphics[width=1.00\textwidth]{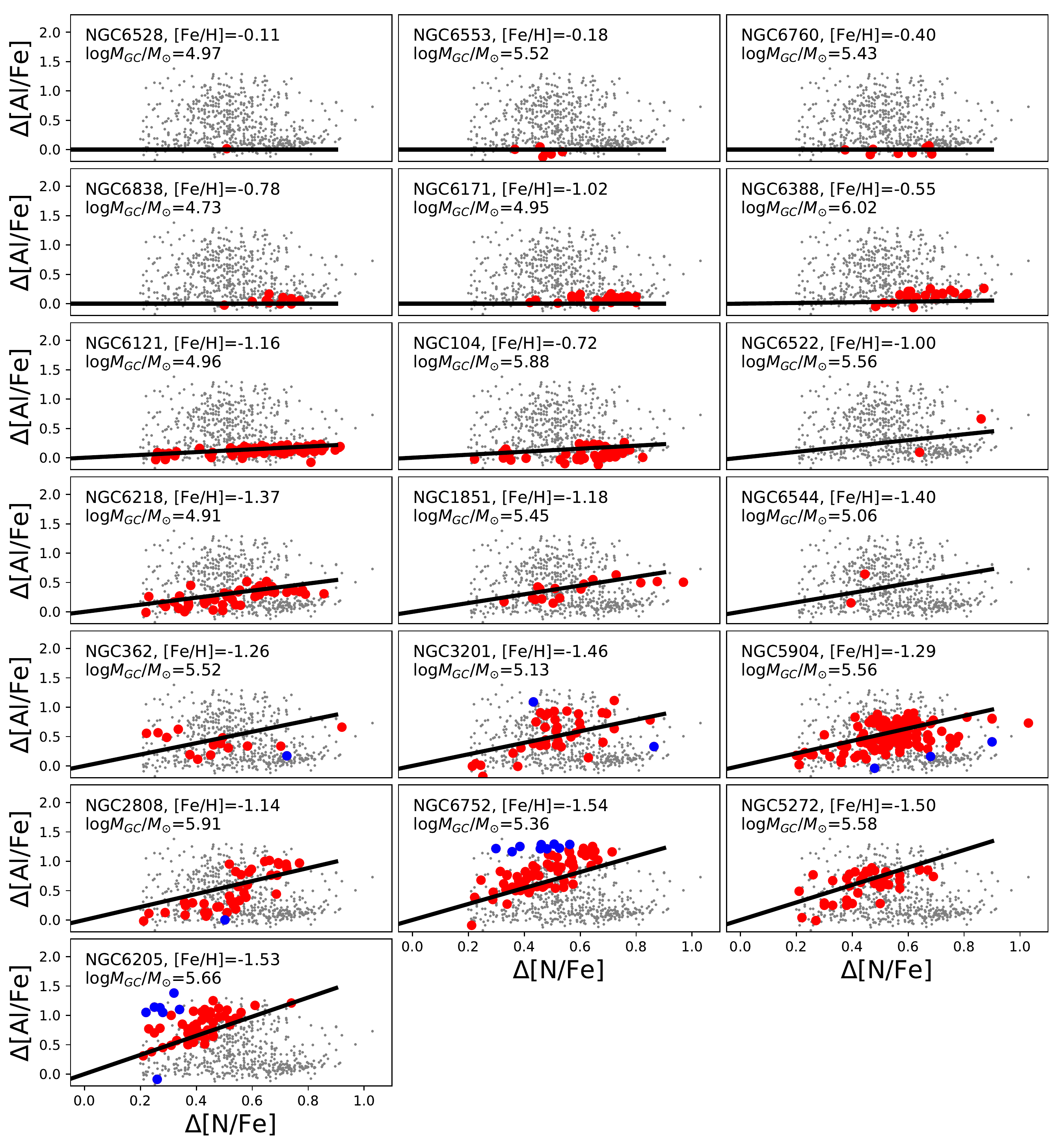}
\caption{The relationship between nitrogen and aluminum enrichment within our APOGEE-derived sample is correlated with cluster mass and metallicity. For each panel, we show the values of aluminum and nitrogen enrichment for all cluster stars (grey points), the stars in that cluster which contribute to the fit (red circles), the stars in that cluster that are outliers from the ${\Delta}$[Al/Fe] fit by 0.50 dex or more (blue points), and the predicted best-fit relation (black line) from Equation \ref{EQ:BestFits}. The clusters are ordered by increasing predicted value of  ${\partial \rm{[Al/Fe]}}/{\partial \rm{[N/Fe]}}$. The data for NGC 2808 are shown, but do not contribute to the fit.}
\label{fig:NAlplots}
\end{figure*}

\begin{figure*}
\includegraphics[width=1.00\textwidth]{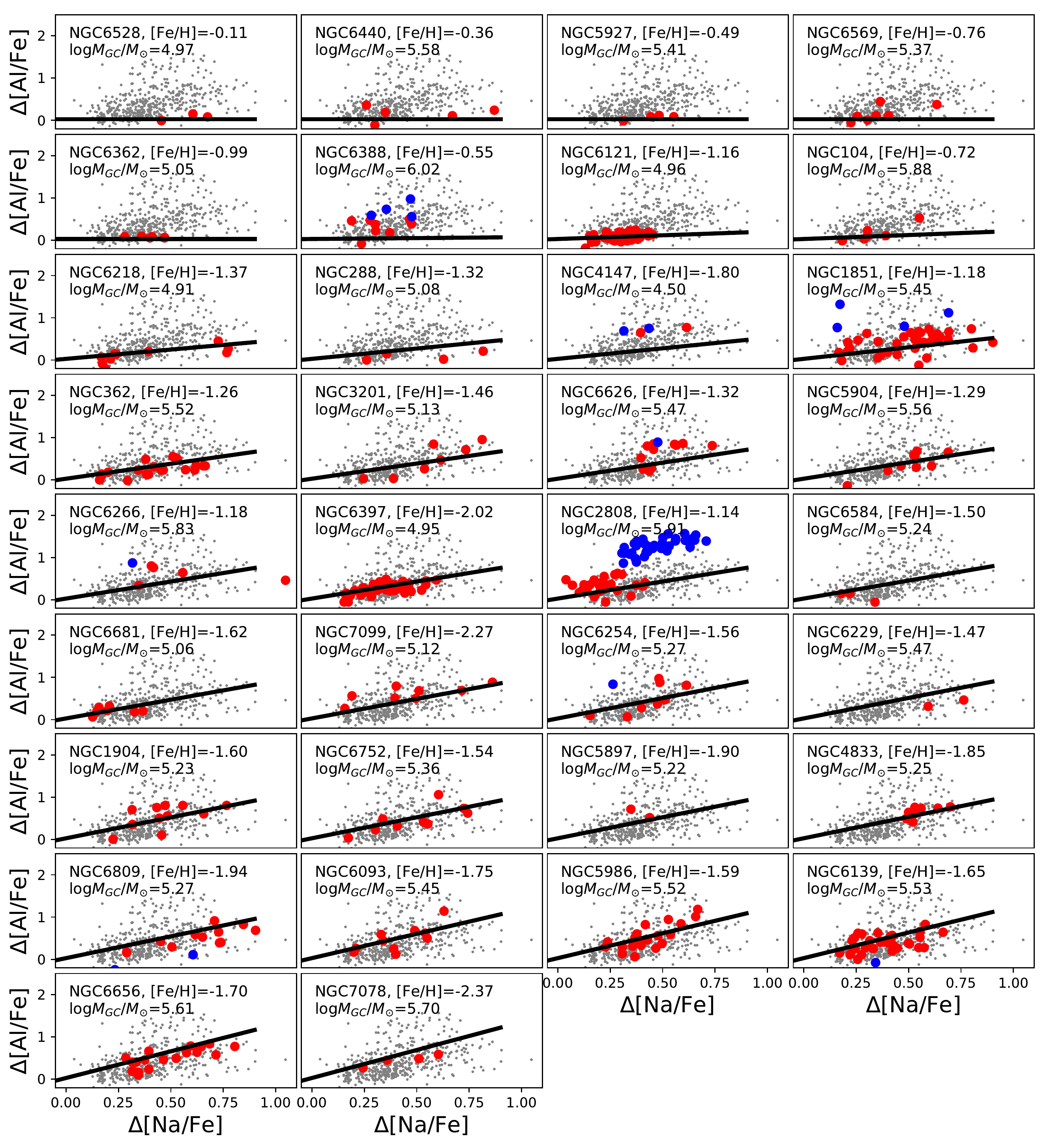}
\caption{The relationship between sodium and aluminum enrichment within the literature sample is correlated with cluster mass and metallicity. Points, lines, and ordering as in Figure \ref{fig:NAlplots}.}
\label{fig:NaAlplots}
\end{figure*}

We fit for linear relationships relating aluminum enrichment to sodium or nitrogen enrichment with globular cluster metallicity and mass by minimizing chi square using the Metropolis-Hastings implementation of Markov Chain Monte Carlo (MCMC, e.g. \citealt{2017ARA&A..55..213S,2018ApJS..236...11H}). The model, which is only applied to the abundances of second generation stars, can be written as:
\begin{equation}
    \begin{split}
    \Delta\rm{[Al/Fe]} = a_{1} +  Max  \{[ Min  \{ b_{1}([Fe/H]+1.30), b_{1,\rm{max}} \} + \\ c_{1}(\log{M_{GC}/M_{\odot}}-5.50) + d_{1}] *\Delta\rm{[Na/Fe]}, 0   \} , \\ 
    \Delta\rm{[Al/Fe]} = a_{2} +  Max  \{[ Min  \{ b_{2}([Fe/H]+1.30), b_{2,\rm{max}} \} + \\ c_{2}(\log{M_{GC}/M_{\odot}}-5.50) + d_{2}] *\Delta\rm{[N/Fe]}, 0   \} , 
    \end{split}
\label{EQ:NaBestFits}
\end{equation}
The x-intercepts of [Fe/H]$= - 1.30$ and $\log{M_{GC}/M_{\odot}}=5.50$ are chosen for purely heuristic reasons, as these values are approximately the mean of these parameters in our sample. The values of $a_{i}$ are y-intercepts, and thus should have best-fit values close to zero if we have properly estimated the first-generation abundances, and if these trends are actually linear in the manner that our model assumes. The terms $\rm{Min} \{ b_{i}\rm{[Fe/H]}+1.30), b_{i,\rm{max}}\} $ denote the metallicity dependence, where the term $b_{i,\rm{max}}$ parameterizes a step function, such that that the dependence of the ratio of aluminum abundance variations  to nitrogen abundance variations as a function of metallicity levels off at lower metallicity. That has recently been empirically supported by the analysis of \citet{2018arXiv181208817M}, who measured that the increase in [Al/Fe] in the most extreme second generation stars levels off, a feature they label a ``hook". 

The values of $c_{i}$ prameterizes the dependence on the present-day stellar mass of the globular clusters. The values of $d_{i}$ denote the predicted values of ${\partial \rm{[Al/Fe]}}/{\partial \rm{[Na/Fe]}}$ and ${\partial \rm{[Al/Fe]}}/{\partial \rm{[N/Fe]}}$ for clusters with [Fe/H]$= - 1.30$ and $\log{M_{GC}/M_{\odot}}=5.50$. Similarly, $b_{i}$ and $c_{i}$ can be thought of as second-order partial derivatives, relating the respective dependence of $\Delta\rm{[Al/Fe]}$ on both of $\Delta\rm{[N/Fe]}$ and then on [Fe/H] or $\log{M_{GC}/M_{\odot}}$. Finally, if the value of  $[ Min  \{ b_{i}(\rm{[Fe/H]}+1.30),0\} + c_{i}(\log{M_{GC}/M_{\odot}}-5.50) + d_{i}]$ is negative for a cluster, then it is replaced by zero. We assume that aluminum enrichment is either positively correlated with nitrogen and sodium enrichment, or null.

The following conditions are imposed on the fitting procedure:
\begin{itemize}
     \item  We do not know the true amplitude of the expected scatter, which should be a quadratic sum of the measurement errors and the actual intrinsic scatter in the relations. We instead assume that $\sigma_{\rm{[Al/Fe]}}=$ 0.20. This value  approximately corresponds to the measured scatter in the best-fit relations, and results in $\chi ^2 _{\rm{DoF}}$ being approximately rescaled to unity.
    \item To reduce the impact of outliers on the fit, we neglect points which are offset from the predicted best fit by 0.50 dex or more. Each such point imposes a penalty of $\Delta \chi^2 = +6.25$, to prevent the MCMC from exploring unphysical fits where all of the points are assumed to be outliers.
    \item We assign equal weight to every point from globular clusters with 10 or fewer measurements in a sample. When a globular cluster contains 10 or more measurements within the sample, the weight $w_{i}$ of each point is re-scaled as $w_{i} = 10/N_{2}$, where $N_{2}$ is the number of second generation stars in that sample and of that cluster.
    \item We impose the prior that the y-intercepts $a_{i}$ of the relations of Equations \ref{EQ:NaBestFits}, \ref{EQ:NBestFits}, and \ref{EQ:BestFits} (defined above, in Equation \ref{EQ:NaBestFits}) are close to zero, such that ${\Delta}\chi ^2 = 2(a_{i}/0.01)^2$. This is a small correction, as the y-intercepts would otherwise converge to values of $a_{i} \approx \pm 0.05$. 
    \item The data for NGC 2808 are assigned a weight of 0, due to inconsistent literature findings on that cluster. Each of the data from APOGEE, Gaia-ESO  \citep{2017A&A...601A.112P}, and  \citet{2015MNRAS.449.4038D} indicate a span of ${\Delta}$[Al/Fe]$ \approx $ 1.0 dex; the data of \citet{2015ApJ...801...68M} indicate a span of ${\Delta}$[Al/Fe] $\approx$ 1.2 dex; and the data of \citet{2018A&A...615A..17C} indicate a span in ${\Delta}$[Al/Fe] of $\approx$ 1.6 dex. Moreover, the cluster is known to host stars with exceptionally high initial abundances of helium reaching $Y \approx 0.38$ \citep{2014MNRAS.437.1609M,2015ApJ...808...51M}. This increases the odds of a different chemical evolution history, as the yields from helium-enriched asymptotic giant branch stars are expected to be different \citep{2014ApJ...784...32K,2015MNRAS.452.2804S} from those of helium-normal stars.  
\end{itemize}

As stated in our introduction, the three abundances being studied here emerge from different nuclear reactions, for which ${}^{23}$Na(p,$\alpha$)${}^{20}$Ne, ${}^{14}$N(p,$\gamma$)${}^{15}$O, and  ${}^{24}$Mg(p,$\gamma$)${}^{25}$Al are plausibly the predominant ones. As the nature and thus number of chemical polluters in globular clusters is unknown, it is also unknown whether one, two, or three required degrees of freedom are needed to jointly model sodium enrichment, nitrogen enrichment, and aluminum enrichment. We thus first fit for sodium and nitrogen separately.

Restricting the fit to the sodium (literature) sample, we obtain:
\begin{equation}
    \begin{split}
    \Delta\rm{[Al/Fe]} = 0.00 +  Max  \{[ Min  \{ -0.64([Fe/H]+1.30), 0.43 \} + \\ 0.58(\log{M_{GC}/M_{\odot}}-5.50) + 0.82] *\Delta\rm{[Na/Fe]}, 0   \}
    \end{split}
\label{EQ:NaBestFits}
\end{equation}
The scatter in [Al/Fe] to the best-fit relation is 0.179 dex, with only 17 of the 413 measurements not of NGC 2808 being 0.50+ dex outliers.

We repeat the exercise for the nitrogen (APOGEE) sample, though we shut off the step-function to the metallicity term as the APOGEE sample does not include any globular clusters with [Fe/H]$ \leq -1.54$. We obtain:
\begin{equation}
    \begin{split}
        \Delta\rm{[Al/Fe]} = 0.02 +  Max  \{[ -2.81([Fe/H]+1.30) + \\ 0.79(\log{M_{GC}/M_{\odot}}-5.50) + 0.97] *\Delta\rm{[N/Fe]}, 0   \}
    \end{split}
\label{EQ:NBestFits}
\end{equation}
The scatter in [Al/Fe] to the best-fit relation is 0.171 dex, with only 20 of the 608 measurements not of NGC 2808 being 0.50+ outliers. One major difference between the literature fit and the APOGEE-derived fit is that for the latter, the maximum slope to the metallicity coefficient, $b_{2,\rm{max}}$ is not constrained. That is because the APOGEE sample with Payne-derived parameters lacks a sample of lower-metallicity globular clusters, a challenge which can be overwhelmingly resolved by future implementations of the Payne. 


Both the literature sample and the APOGEE sample have a relative paucity of more metal-rich, more metal-poor, and lower-mass clusters, so jointly fitting them increases the statistical leverage where there is currently little. There is a useful physically-motivated constraint, in that if ${\partial \rm{[Al/Fe]}}/{\partial \rm{[Na/Fe]}} = 0$, then it is necessarily the case that ${\partial \rm{[Al/Fe]}}/{\partial \rm{[N/Fe]}} = 0$. Further, it is also the case that the two samples yield similar parameters values when fit for separately, in that the ratios of $c_{i}/b_{i}$ and $d_{i}/b_{i}$ are all of order unity. 

We thus actually impose a greater constraint, we assume the \textit{ansatz} that $\{b_{2},b_{2,\rm{max}},c_{2},d_{2}\} = C*\{b_{1},b_{1,\rm{max}},c_{1},d_{1}\}$, where  $C = {\partial \rm{[Na/Fe]}}/{\partial \rm{[N/Fe]}}$. In other words, we assume that the relation between aluminum and nitrogen is simply a re-scaled version of the relation between aluminum and sodium, or alternatively, that  ${\partial \rm{[Na/Fe]}}/{\partial \rm{[N/Fe]}}$ is not varying or only weakly varying with globular cluster stellar mass and metallicity. We will show further justification of this assumption later in this section, and in the next section.

We derive the following best-fit relations, when fitting both data sets together with the total weight of the literature and APOGEE samples fixed to be equal to one another:
\begin{equation}
    \begin{split}
    \Delta\rm{[Al/Fe]} = 0.03 +  Max  \{[ Min  \{ -1.42([Fe/H]+1.30), 0.44 \} + \\ 0.69(\log{M_{GC}/M_{\odot}}-5.50) + 0.75] *\Delta\rm{[Na/Fe]}, 0   \} \\
     \Delta\rm{[Al/Fe]} = 0.00 +  Max  \{[ Min  \{ -1.95([Fe/H]+1.30), 0.60 \} + \\ 0.95(\log{M_{GC}/M_{\odot}}-5.50) + 1.03] *\Delta\rm{[Na/Fe]}, 0   \}.
    \end{split}
    \label{EQ:BestFits}
\end{equation}
On the [Na/Fe]-[Al/Fe] plane, 17/413 stars not of NGC 2808 are offset from the prediction by ${\Delta}$[Al/Fe] $\geq 0.50$ dex, and the remaining stars have a scatter to the fit of 0.190 dex. On the [N/Fe]-[Al/Fe] plane, 24/608 stars not of NGC 2808 are offset from the prediction by ${\Delta}$[Al/Fe] $\geq 0.50$ dex, and the remaining stars have a scatter to the fit of 0.176 dex. Thus, this reduction in the number of degrees of freedom by two, leads to an increase in the total number of outliers from 37 to 41, and a small increase in the statistical scatters.

The measured and predicted relations in the $\Delta\rm{[Al/Fe]}$-$\Delta\rm{[N/Fe]}$ plane and $\Delta\rm{[Al/Fe]}$-$\Delta\rm{[Na/Fe]}$ are respectively plotted in Figures \ref{fig:NAlplots} and \ref{fig:NaAlplots} . The best-fit relations, shown as the black lines, are decent albeit imperfect matches to the data for each cluster, shown as the red points. It is not surprising that the data for some of the clusters (e.g. NGC 362) are offset from the fits, as there are various possibilities for deviation from the model. For example, the correlation with the present-day stellar mass of the cluster is plausibly due to a correlation with the initial stellar mass and star formation environment of the cluster. If that is the case, some clusters will be shifted from the fit if they have lost a different fraction of mass than typical of most of the other clusters in the sample. From Equation \ref{EQ:BestFits}, if a cluster has half as much stellar mass remaining as is typical of the other clusters in the sample, then its predicted slope ${\partial \rm{[Al/Fe]}}/{\partial \rm{[Na/Fe]}}$ will be shifted by 0.38. 

There are multiple options to validate, refute, or simply better constrain Equation \ref{EQ:BestFits}, of which we discuss three: \begin{enumerate}
    \item There is a paucity of measurements of aluminum abundance variations for globular clusters of low metallicity, of all stellar masses. This can be discerned from Figure \ref{fig:AluminumDependency3}. We considered including the metal-poor ([Fe/H]$_{\rm{Harris}}-1.91$) cluster NGC 5824, but the available spectroscopic data include mostly upper bounds on [Al/Fe] \citep{2018ApJ...859...75M}. Further measurements in that regime could inform if there is indeed a maximum value of ${\partial \rm{[Al/Fe]}}/{\partial \rm{[Na/Fe]}}$ at fixed mass, or if it is simply an artifact of our sample. 
    \item There are no measurements for clusters with $\log M_{GC}/M_{\odot} < 4.50$, which can also be discerned from Figure  \ref{fig:AluminumDependency3}. Adding a few such measurements at different metallicities should be helpful in constraining the nature of the polluters.
    \item NGC 6388 and NGC 6440 are currently the only  metal-rich clusters with a measured aluminum enrichment that contributes to the fit. Further measurements of massive, metal-rich clusters (e.g. NGC 6441, Liller 1), as well as a larger sample of data for NGC 6388 and NGC 6440, would constrain the validity of Equation \ref{EQ:BestFits} at the metal-rich end. 
    \item Our fit assumes that the relationship between sodium and nitrogen enrichment is independent or nearly independent of globular cluster mass and metallicity.  
    A sample that would include more clusters with both nitrogen and sodium abundance measurements could help one investigate the validity of the assumption that nitrogen and sodium vary together. The scarce available such measurements are discussed in the next subsection.
\end{enumerate}

\subsection{Possible Physical Interpretations of the Trends Between Light Element Abundances, Metallicity, and Globular Cluster Mass} \label{subsec:Physics}

The results of the prior two sections demonstrate two trends concerning aluminum enrichment in globular clusters. In Section 4.2, we showed that the mean difference in [Al/Fe] between the chemically mundane and chemically anomalous stars of globular clusters is positively correlated with present-day  globular cluster mass, and negatively correlated with globular cluster metallicity. That confirms and expands on the findings of \citet{2017A&A...601A.112P} and \citet{2009A&A...505..139C}. In Section \ref{subsec:fullsample}, we expanded the analysis to factor out the enrichment in nitrogen or sodium, and showed that the slopes of relative enrichment, ${\partial \rm{[Al/Fe]}}/{\partial \rm{[Na/Fe]}}$ and ${\partial \rm{[Al/Fe]}}/{\partial \rm{[N/Fe]}}$, were themselves linearly dependent on present-day globular cluster mass and metallicity.  

\begin{figure}
\includegraphics[width=0.47\textwidth]{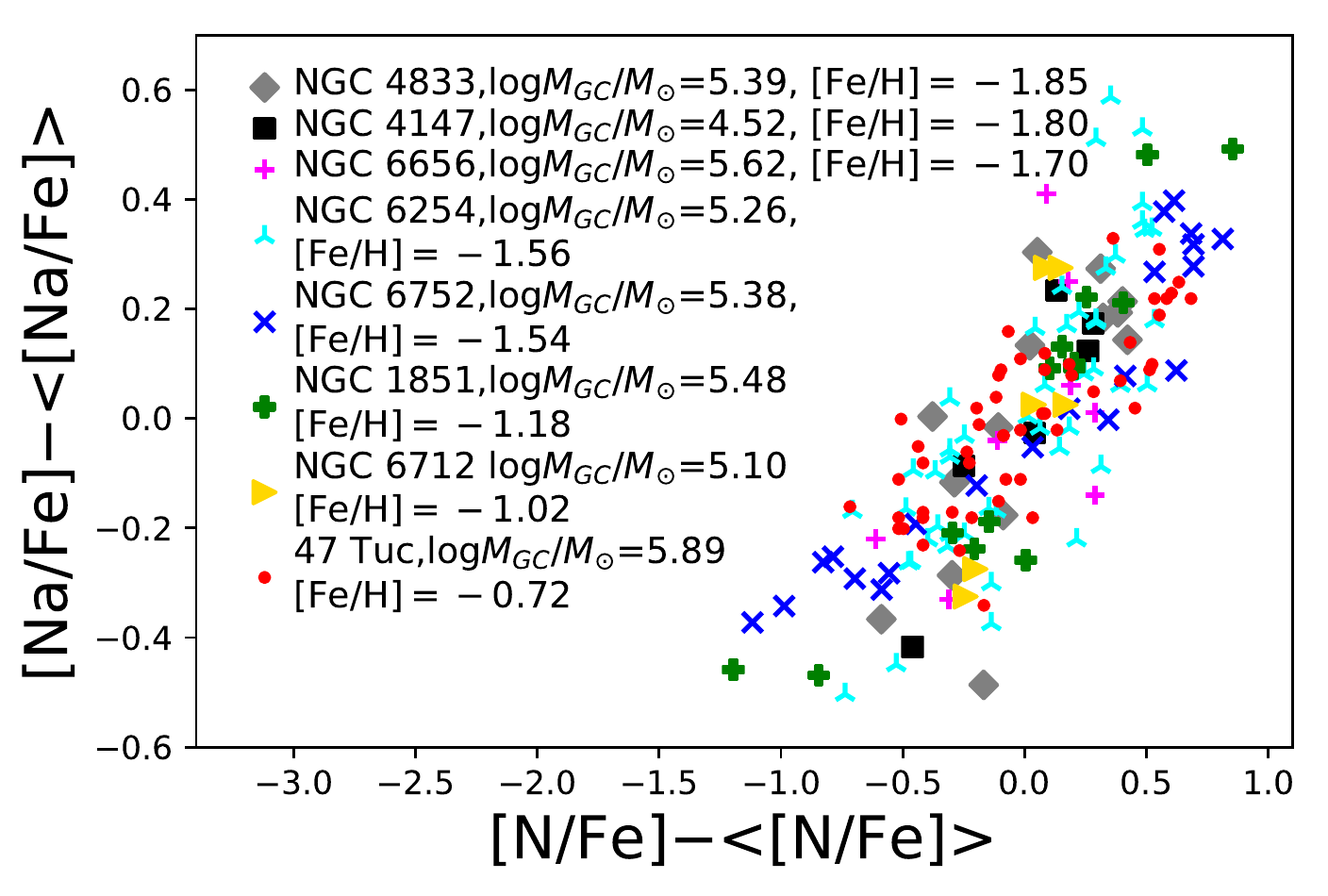}
\caption{Literature measurements suggest that the trend between [Na/Fe] variations and [N/Fe] variations is equal or nearly equal in different clusters, in contrast to the trends with [Al/Fe]  variations. We show the data for NGC 4833 \citep{2015MNRAS.449.3889R}, NGC 4147 \citep{2016MNRAS.460.2351V}, NGC 6656 / M22 \citep{2012A&A...540A...3A}, NGC 6254 / M10 \citep{2018AJ....156....6G},  NGC 6752 \citep{2008ApJ...684.1159Y},  NGC 1851 \citep{2015MNRAS.446.3319Y}, for NGC 6712 \citep{2008ApJ...689.1020Y}, and 47 Tuc / NGC 104 \citep{2016MNRAS.459..610M}. We subtracted the relevant mean values from each abundance set in order to emphasize the trends in differential abundances, which should be independent of zero-point calibrations. }
\label{fig:SodiumNitrogenLiteratureComparison}
\end{figure}

The latter is a new finding. It suggests that there may be two classes of non-supernovae chemical polluters that were active in the era of globular cluster formation, and that their relative contributions somehow scaled with globular cluster metallicity and present-day stellar mass. The first class of polluters is largely responsible for the ${}^{14}$N(p,$\gamma$)${}^{15}$O and the ${}^{23}$Na(p,$\alpha$)${}^{20}$Ne nuclear processing, and its contribution does not scale or does not significantly scale with globular cluster metallicity and mass. That is supported by our finding that a similar relation can be used to fit for both of the [Al/Fe]-[N/Fe] and the  [Al/Fe]-[Na/Fe] relations.

To further support this claim, we show, in Figure \ref{fig:SodiumNitrogenLiteratureComparison}, a comparison of the trends of [Na/Fe] versus [N/Fe] variations for eight globular clusters where the two relative abundances were measured in the same sample of stars. These clusters span a range of approximately 1.1 dex in metallicity and 1.5 dex in stellar mass, yet their [Na/Fe]-[N/Fe] relations are consistent with a slope,  ${\partial \rm{[Na/Fe]}}/{\partial \rm{[N/Fe]}} \approx 0.50$, whose dependencies on globular cluster mass and metallicity are null or negligible. 



Indeed, the contrast between the narrow scatter seen in Figure \ref{fig:SodiumNitrogenLiteratureComparison}, and that seen in Figures  \ref{fig:NAlplots} and \ref{fig:NaAlplots}, is large. For the latter two the relations with [Al/Fe] can vary by 1,000\% or more. Asymptotic giant branch stars, by themselves, can explain the trend with metallicity, as hot bottom burning is predicted to take place at higher temperatures in lower metallicity asymptotic giant branch stars \citep{2000MmSAI..71..737L}. \citet{2018MNRAS.475.3098D}, building on the work of \citet{2016ApJ...831L..17V}, have shown that the predicted chemical yields of asymptotic giant branch stars with initial masses in the range $4 M_{\odot} \leq M \leq 8 M_{\odot}$ can reproduce the abundance variations in  carbon, nitrogen, oxygen, magnesium, aluminum, and silicon, as measured in 9 globular clusters probed by APOGEE. Thus there is a straightforward explanation for the correlation between aluminum variations and globular cluster metallicity, but not that with present-day globular cluster stellar mass. We suggest that a second class of polluters is responsible for ${}^{24}$Mg(p,$\gamma$)${}^{25}$Al processing, in higher-mass globular clusters, that is separate from that which would be obtained purely from asymptotic giant branch stars. 

It is worth stating that the correlation with present-day stellar mass is a correlation by proxy. Present-day stellar mass cannot be the cause of these variations, as the clusters formed $\sim$12 Gyr ago. What may be responsible is a causal relation between the stellar and gas density, and the depth of the gravitational well, during the birth of these clusters, and the eventual stellar mass. 

For example, \citet{2013ApJ...775..134V}, who estimated helium abundance variations in a large sample of clusters with well-sampled photometry,  found that their inferred helium enrichment correlated with the present-day central escape velocity and surface mass density of these clusters. Their Section 6.2.1, and in particular their Figure 40, elaborates on these issues. We note that \citet{2018MNRAS.475.4088L} have also measured a correlation between inferred helium enrichment and the present-day stellar masses of clusters, by measuring brightness variations in the red giant branch bump, a completely independent tracer of helium abundance \citep{1997MNRAS.285..593C,2013ApJ...766...77N}. The correlation between helium abundance variations and globular cluster mass is convincing, but further research is needed to ascertain if that is due to the gas surface density, and depth of the gravitational well, during the birth of the globular clusters. 

If it is the case that most globular clusters have lost a lot of mass, and that the correlation found here is actually one with initial stellar mass, then there must be some regulatory process which has restricted globular cluster mass loss to some narrow fractional range. 

Some preliminary results of this investigation, including the correlation between aluminum enrichment and globular cluster mass, were presented at the ``\textit{Survival of Dense Star Clusters in the Milky Way System}" conference\footnote{http://www.mpia.de/~mwstreams/} held in Heidelberg, Germany, in November 2018. Following the presentation, Long Wang, of Peking University, suggested a model consistent with the above conjectures, whereby clusters more massive today were denser at birth, and thus had more mergers and mass transfers between massive stars, leading to a different  (``top-heavy") \textit{effective} mass function of polluters. Massive binaries were first suggested as the source of globular cluster abundance anomalies by \citet{2009A&A...507L...1D}. Massive binary mergers within globular clusters have also been linked to the formation of intermediate-mass \citep{2004Natur.428..724P} and supermassive black holes \citep{2018MNRAS.478.2461G}. It may be possible to investigate this theoretical framework further, given that the N-body simulations of globular clusters continue to improve (e.g. \citealt{2016MNRAS.458.1450W}). 

The second polluter may also be the winds of fast-rotating massive stars \citep{2006A&A...448L..37M,2007A&A...464.1029D,2016A&A...593A..36C}. It may be, for example, that a minimum gravitational well depth is required to hold on to some of their ejecta, and that this requirement ends up as a correlation between globular cluster mass and aluminum enrichment at the present day. 

As noted by the anonymous referee, an important constraint is to be found in the dependence of the abundance variations on one another.  There are several clusters with sodium abundance spreads and no measurable spread in aluminum, but there are no examples of the opposite. This suggests that while the polluters responsible for the aluminum spread are possibly not the same responsible for the sodium abundance variation, there is however a close relation between the two. For example, there may be one class of polluters that contributes extra sodium but not extra aluminum, and another polluter that contributes both sodium and aluminum. In the next subsection, we discuss evidence from the literature of two globular clusters that aluminum may sometimes vary in the absence of corresponding variations in sodium.

\subsection{Two Instructive Outliers: NGC 6121 (M4) and NGC 104 (47 Tuc)}  \label{subsec:TwoOutliers}

\begin{figure*}
\centering
\includegraphics[width=1.00\textwidth]{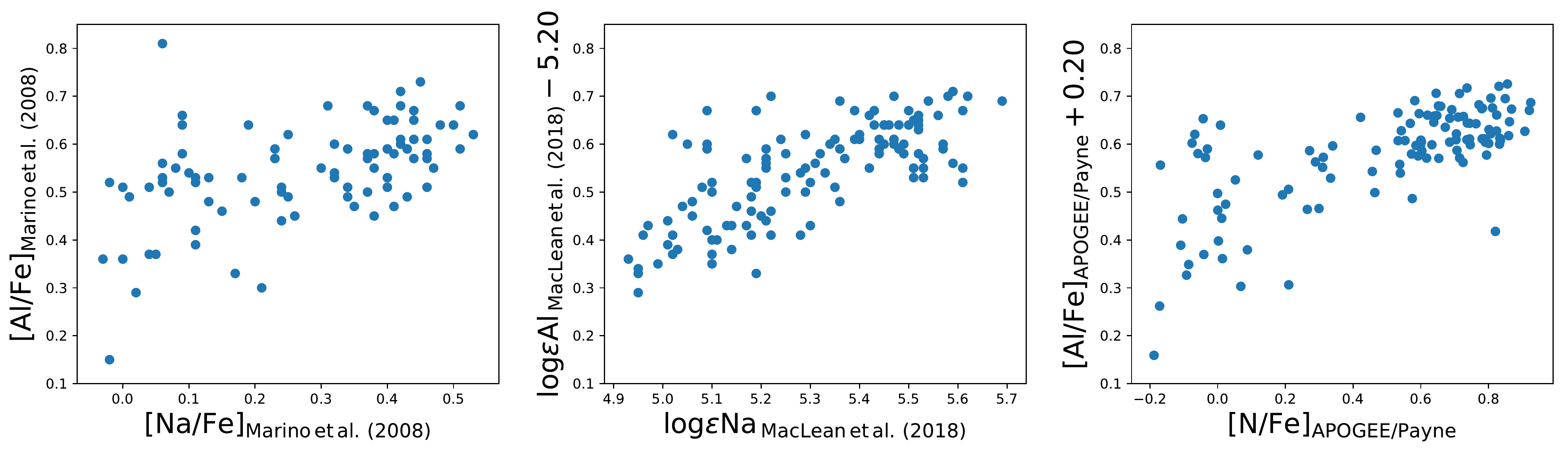}
\caption{The sodium-normal stars of NGC 6121 (M4, left panel) show a scatter in [Al/Fe] that is independent of [Na/Fe] in two independent sets of data (that of \citealt{2008A&A...490..625M} in the left panel, and that of \citealt{2018MNRAS.475..257M} in the middle panel), whereas the sodium-enhanced stars follow an [Al/Fe]-[Na/Fe] correlation. The same behavior is seen when nitrogen rather than sodium is used as the independent variable (right panel), with that data derived by the Payne from APOGEE spectra. We shift the zero-points of the ordinate in both of the middle and right panels, to align them with that of the left panel.  }
\label{fig:Outlier_6121}
\end{figure*}

\begin{figure}[H]
\centering
\includegraphics[width=0.47\textwidth]{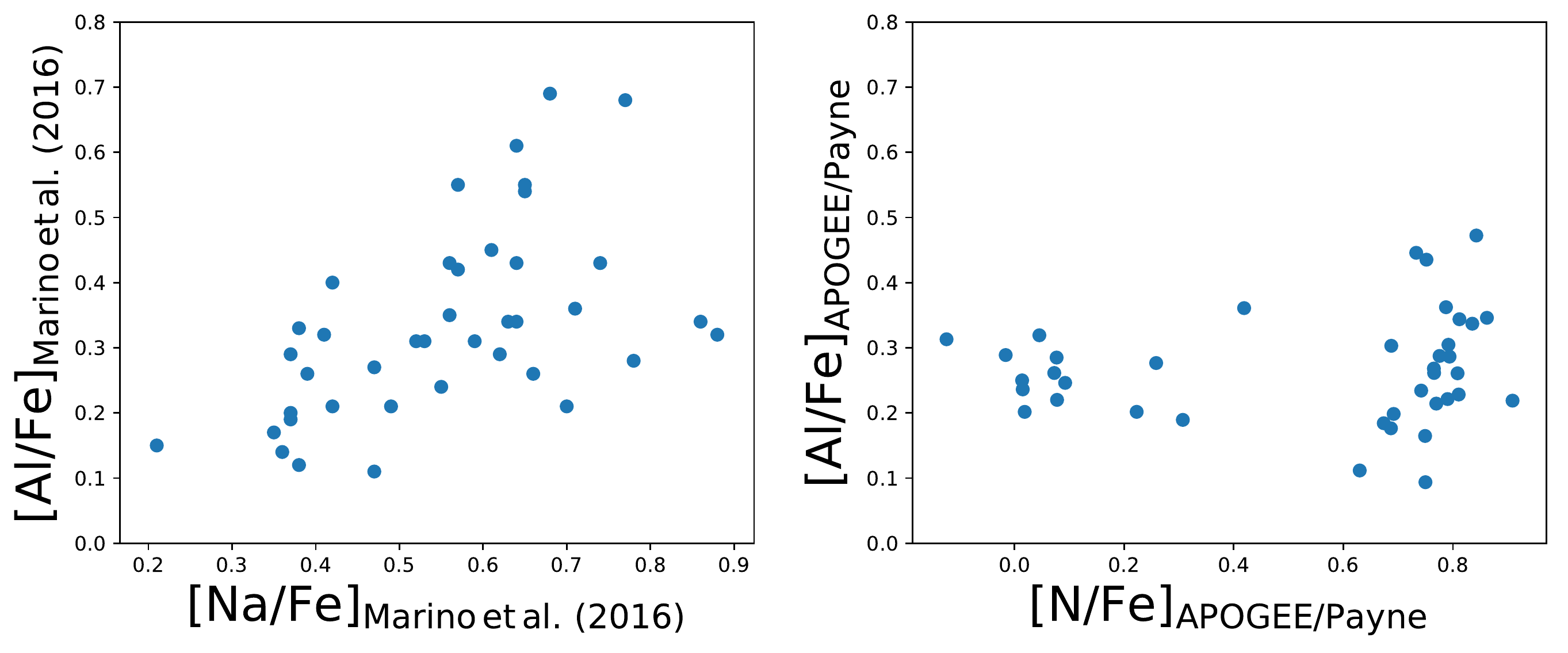}
\caption{The sodium-rich stars of NGC 104 (47 Tuc, left panel) show a scatter in [Al/Fe] that is independent of [Na/Fe], whereas the sodium-normal stars show little scatter in their [Na/Fe] abundances, data are from \citet{2016MNRAS.459..610M}. The same behavior is seen when nitrogen rather than sodium is used as the independent variable (right panel), with that data derived by the Payne from APOGEE spectra.   }
\label{fig:Outlier_104}
\end{figure}

In the course of our investigation we have noticed that the [Al/Fe]-[N/Fe] relations of two globular clusters, NGC 6121 (M4) and NGC 104 (47 Tuc), are clearly deviating from the main trends identified in this work. We show their respective [Al/Fe]-[N/Fe] distributions in Figures \ref{fig:Outlier_6121} and \ref{fig:Outlier_104}, where we plot both the APOGEE/Payne data and the data of \citet{2008A&A...490..625M} and of \citet{2018MNRAS.475..257M} for NGC 6121 and of \cite{2016MNRAS.459..610M} for NGC 104. In both cases, the  independent data sets show remarkably consistent distributions. 

In the case of NGC 6121, the [Al/Fe] of the nitrogen-enhanced stars does behave as expected from the results of the prior sections. However, for the nitrogen-normal and sodium-normal stars, [Al/Fe] has a significant scatter at fixed [N,Na/Fe], in fact the total scatter is larger than that of the nitrogen-enhanced stars. Such a distribution is not consistent with a picture whereby a single polluter is responsible for all of the light-element abundance variations in globular clusters. 

The converse holds for NGC 104. For that cluster, the stars with normal [N,Na/Fe] show little or no variation in [Al/Fe]. However, the stars with enhanced [N,Na/Fe] show a variable [Al/Fe] at fixed [N,Na/Fe]. This again suggests that different stars were responsible for the ${}^{23}$Na(p,$\alpha$)${}^{20}$Ne and ${}^{14}$N(p,$\gamma$)${}^{15}$O nuclear processing on the one hand, and the ${}^{24}$Mg(p,$\gamma$)${}^{25}$Al nuclear processing on the other hand. In the case of NGC 104, most of the second generation stars may have formed when only one of the polluter had contributed to the surrounding gas. This can explain why some studies find no [Al/Fe] variations at fixed [N,Na/Fe] (e.g. \citealt{2008AJ....135.1551K,2014ApJ...780...94C}), as the [Al/Fe]-enhanced stars are rare. It also explains why NGC 104 is an outlier to the relations identified in the previous sections: the gas from which its second generation stars formed was not well mixed. This result is also consistent with the discovery of \citet{2010MNRAS.408..999D}, that the ``second generation" of 47 Tuc is composed of two components. The second component of the second generation, which they called SGII, has a fainter subgiant branch, possibly due to having an enhanced sum of C+N+O, and constitutes $\sim$10\% of the population of the cluster. 

We have verified that the choice to include these two clusters in the fits of the prior sections has a negligible impact on the derived parameters.

\subsection{An Alternative Explanation to the Metal-Rich, Nitrogen-Rich, Aluminum-Rich, Magnesium-Poor Stars Found in the Field}  \label{subsec:FernTrin}

\citet{2017ApJ...846L...2F} searched for and identified nitrogen-rich stars in the field from the APOGEE spectroscopic database. As discussed in our Section \ref{subsec:Martell}, the origin of these stars is not currently understood. 

They found seven stars with [Fe/H] $\geq -1.0$, of which five had [Al/Fe] values that are high relative to the field trend, and [Mg/Fe] that are low relative to the field trend. \citet{2017ApJ...846L...2F} pointed out that the globular clusters with [Fe/H] $\gtrsim -1.0$ with stars that have been observed by the APOGEE survey do not show the [Al/Fe]-[Mg/Fe] anti-correlation \citep{2015AJ....149..153M}, and thus it did not seem likely that these stars dynamically evaporated from those kinds of globular clusters. They suggested two alternative possibilities. The first was that these stars may have originated from globular clusters of extragalactic origins, and that the interstellar medium of those galaxies had intrinsically lower relative abundances of magnesium and greater relative aluminum due to a different galactic chemical evolution. Their second suggested possibility is that the atmospheric abundances of these stars may have been polluted by gas transfer from a binary companion. 

However, what our findings suggest is that metal-rich globular clusters can host aluminum-rich stars, as long as the clusters are sufficiently massive. For example, the globular clusters NGC 6569 \citep{2018AJ....155...71J}, NGC 6440 \citep{2017A&A...605A..12M}, and NGC 6441 \citep{2008MNRAS.388.1419O}  all host [Al/Fe]-rich stars, and all have metallicities as high or higher than that typical of the sample of \citet{2017ApJ...846L...2F}, for which [Fe/H] $\approx -0.80$. In fact, NGC 6440, and NGC 6441 have substantially higher metallicities, with [Fe/H]$_{\rm{Harris}} \approx -0.40$. Admittedly, enriched [Al/Fe] only translates to an appreciable deficiency in [Mg/Fe] in the clusters with the most extreme chemically anomalous populations, but there are a few such cases at comparably high metallicities: the [Mg/Fe] abundances span approximately 0.60 dex in NGC 2808,  \citep{2018A&A...615A..17C} and no less than 0.25 dex in NGC 6441 \citep{2008MNRAS.388.1419O}. 

We thus suggest a third alternative explanation to the [Fe/H] $\approx -0.80$, [N/Fe]-rich, [Al/Fe]-rich, [Mg/Fe]-poor stars identified by \citet{2017ApJ...846L...2F}, that these stars formed in metal-rich globular clusters with present-day masses substantially greater than the approximate threshold for aluminum enrichment estimated in this work (Equation \ref{EQ:BestFits}), $\log{M_{GC}/M_{\odot}}  \approx 4.50+2.17(\rm{[Fe/H]}+1.30)$. It would be informative to measure sodium abundances for those stars. 


\section{Terzan 5: A true globular cluster after all?} \label{sec:Terzan5}

\begin{table*}
\centering
\small\addtolength{\tabcolsep}{-1.5pt}
\caption{Candidate Terzan 5 cluster members identified within APOGEE DR14. Listed are APOGEE IDs, the separation from the cluster center ${\Delta}\Psi$  in units of arcminutes, the Gaia-derived proper motion in RA and DEC in units of mas/yr where available, the APOGEE-derived heliocentric radial velocity in units of km/s, the signal-to-noise ratio of the spectrum, eight atmospheric parameters and relative abundances derived by the Payne, and the [Na/Fe] estimates derived by ASPCAP. The value of [$\alpha$/Fe] is the mean of the Payne-derived values of [Si,Ca,Ti,Mg/Fe]. For comparison, the adopted physical parameters for Terzan 5 are a tidal radius of 24 arcminutes, a heliocentric velocity and velocity dispersion of $-$81.4 and 19 km/s, mean proper motions of $-$1.71 and $-$4.64 mas/yr, and a proper motion dispersion of 0.48 mas/yr. } 
\begin{tabular}{|ccccc|cccc| rrrrrrr|}
\hline
APOGEE ID & ${\Delta}\Psi$ &	$\mu_{\alpha}$ &  	$\mu_{\delta}$ & $V_{\rm{Helio}}$ & SNR & [Fe/H] & $\log{g}$ & $T_{\rm{eff}}$ & [C/Fe] & [N/Fe] & [O/Fe] & [Al/Fe] & [K/Fe] & [$\alpha$/Fe] & [Na/Fe] \\
\hline
2M17472880$-$2423378 &  24 & $-$1.21 & $-$5.25 & $-$79 & 139 &  $-$0.66 &  1.13 &  3796 &  $-$0.08 & 0.09  & 0.22 & $-$0.02  & 0.12 & 0.22 & $-$0.03  \\
2M17480857$-$2446033 &  1 & $-$0.95 & $-$5.21 & $-$64 & 169 &  $-$0.65 &  0.61 &  3522 &  $-$0.05 & 0.34  & 0.14 & 0.10  & 0.04 & 0.19 & $-$0.18  \\
2M17480576$-$2445000 &  1 & $-$0.44 & $-$3.51 & $-$76 & 92 &  $-$0.56 &  1.32 &  3805 &  $-$0.19 & 0.61  & 0.07 & 0.11  & $-$0.05 & 0.18 & $-$0.09  \\
2M17480668$-$2447374 &  0 & $-$1.70 & $-$4.61 & $-$89 & 186 &  $-$0.50 &  1.43 &  3705 &  $-$0.43 & 0.73  & $-$0.07 & 0.01  & $-$0.10 & 0.14 & 0.31  \\
2M17480088$-$2447295 &  1 & $-$2.12 & $-$4.95 & $-$99 & 265 &  $-$0.46 &  1.58 &  3654 &  $-$0.50 & 0.75  & $-$0.20 & 0.06  & $-$0.11 & 0.19 & 0.75  \\
\hline
2M17482019$-$2446400 &  3 & $-$1.40 & $-$6.19 & $-$77 & 261 &  $-$0.03 &  0.45 &  3255 &  $-$0.16 & 0.05  & $-$0.13 & 0.16  & $-$0.02 & 0.10 & --  \\
2M17475169$-$2443153 &  4 & -- & -- & $-$75 & 94 &  0.07 &  1.21 &  3688 &  $-$0.17 & 0.05  & $-$0.21 & 0.14  & $-$0.05 & 0.05 & --  \\
2M17481414$-$2446299 &  2 & 0.22 & $-$4.12 & $-$76 & 106 &  0.12 &  0.98 &  3529 &  $-$0.14 & 0.07  & $-$0.06 & 0.17  & $-$0.18 & 0.08 & $-$0.31  \\
2M17473477$-$2429395 &  18 & $-$1.58 & $-$3.23 & $-$80 & 158 &  0.24 &  1.77 &  3979 &  $-$0.18 & 0.27  & $-$0.10 & 0.24  & $-$0.31 & 0.07 & 0.19  \\
\hline
\hline
\end{tabular}
\label{tab:Terzan 5}
\end{table*}

There are several independent and complementary arguments for Terzan 5 not being a ``true" globular cluster. These include the photometric evidence that there is a large spread in the metallicity and age  of its stars \citep{2009Natur.462..483F,2016ApJ...828...75F}, photometric evidence that it has a particularly large mass and low central concentration \citep{2010ApJ...717..653L}, and finally a spectroscopic confirmation of its large spread in iron abundance and the lack of an aluminum-oxygen anti-correlation among its stars \citep{2011ApJ...726L..20O}. 

 \citet{2011ApJ...726L..20O} measured chemical abundances for 33 red giant stars in Terzan 5, for which their main result was the identification of two main chemical groups, one with [Fe/H]$=-$0.25, [$\alpha$/Fe]$=+$0.34 and the other with [Fe/H]$=+$0.27, [$\alpha$/Fe]$=+$0.03. That is the largest measured metallicity spread of any stellar system classified as a Galactic globular cluster, and is the first of three arguments for the position that Terzan 5 is not a true globular cluster enumerated in the conclusion of  \citet{2011ApJ...726L..20O} . The metal-poor component is approximately twice as numerous \citep{2014ApJ...795...22M}. \citet{2011ApJ...726L..20O} also reported the absence of a measurable aluminum-oxygen anti-correlation in either group of stars, which they list as their second argument. This second argument followed the work of \citet{2010A&A...516A..55C}, who recommended that globular clusters be \textit{defined} as the stellar aggregates showing the sodium-oxygen anti-correlation. 

In this section we are not concerned with the merit of the definition proposed by \citet{2010A&A...516A..55C}. Rather, we are stating that the extension of the definition assumed by \citet{2011ApJ...726L..20O}, that globular clusters are the stellar aggregates showing an aluminum-oxygen abundance anti-correlation, might not apply. That is because the expected abundance scatter in both of [Al/Fe] and [O/Fe] for globular cluster stars as metal-rich as those of Terzan 5 is significantly lower than that found in most globular clusters. \citet{2011ApJ...726L..20O} did acknowledge this possibility, though at the time the available data were not as plentiful as they are now. 

We floated the predicted aluminum enrichment of Terzan 5 as a derived parameter in our Markov chain, assuming its estimated physical parameters of [Fe/H]$=-$0.23 \citep{1996AJ....112.1487H} and $\log{M_{GC}/M_{\odot}} =5.75$ \citep{2018MNRAS.478.1520B}. We derived the expected relation ${\partial \rm{[Al/Fe]}}/{\partial \rm{[Na/Fe]}} \approx 0$. In other words, it activates the step function for the slope to be set to zero, as it would otherwise be very negative.  Even this fiducial large negative value might be an overestimate, as the analysis of \citet{2017ApJ...845..148P}, which is based on long-term radio pulsar timing of 36 millisecond pulsars in the cluster, estimates a lower value for the cluster mass, $\log{M_{GC}/M_{\odot}} \approx 5.40$. Admittedly, there is also the issue that we have not calibrated how these relations might be shifted in clusters with large spreads in [Fe/H]. Regardless, the general trend that [Al/Fe] variations are reduced or eliminated in more metal-rich systems is likely a robust conclusion from our analysis (and also that of \citealt{2017A&A...601A.112P}). 

It is also the case that [O/Fe] variations are reduced in more metal-rich clusters, though we are unsure by how much. Within both our study (see Figure \ref{fig:Payne_APOGEE_GC_CNOAl_AbundanceScatter}) and that of \citet{2009A&A...505..117C}, the scatter in [O/Fe] is reduced (but not eliminated) in more metal-rich systems. Pertinently, \citet{2018A&A...620A..96M} measured that the stars in the metal-rich globular cluster NGC 6528 ([Fe/H]$_{\rm{Harris}}=-$0.11) show a scatter in [Na/Fe] without a corresponding scatter in [O/Fe]. This empirical trend is now supported by theoretical arguments. \citet{2018ApJ...869...35K} have shown that the predicted decrease in the [O/Fe] variations in more metal-rich globular clusters can be explained by a model of globular cluster chemical evolution that incorporates the predicted metallicity-dependent yields of both asymptotic giant branch stars and the winds of fast-rotating massive stars. 

Thus, [Al/Fe] and [O/Fe] variations are respectively expected to be negligible and small in metal-rich globular clusters, and thus the absence of an aluminum-oxygen abundance anti-correlation cannot be used as a diagnostic criterion to evaluate the nature of systems such as Terzan 5.

Separately, we have identified nine stars as candidate members of Terzan 5, using a combination of APOGEE DR14 and Gaia DR2 data, We list several of their parameters and best-fit chemical abundances in Table \ref{tab:Terzan 5}. These data confirm several of the findings of \citet{2011ApJ...726L..20O} and \citet{2014ApJ...795...22M}: Terzan 5 contains two metallicity groups; the group with sub-solar [Fe/H] has a higher  $\langle [\alpha/\rm{Fe}] \rangle  $ than the group with super-solar [Fe/H], and the lower metallicity group is at least as numerous as the higher metallicity group. 

\begin{figure}
\centering
\includegraphics[width=0.48\textwidth]{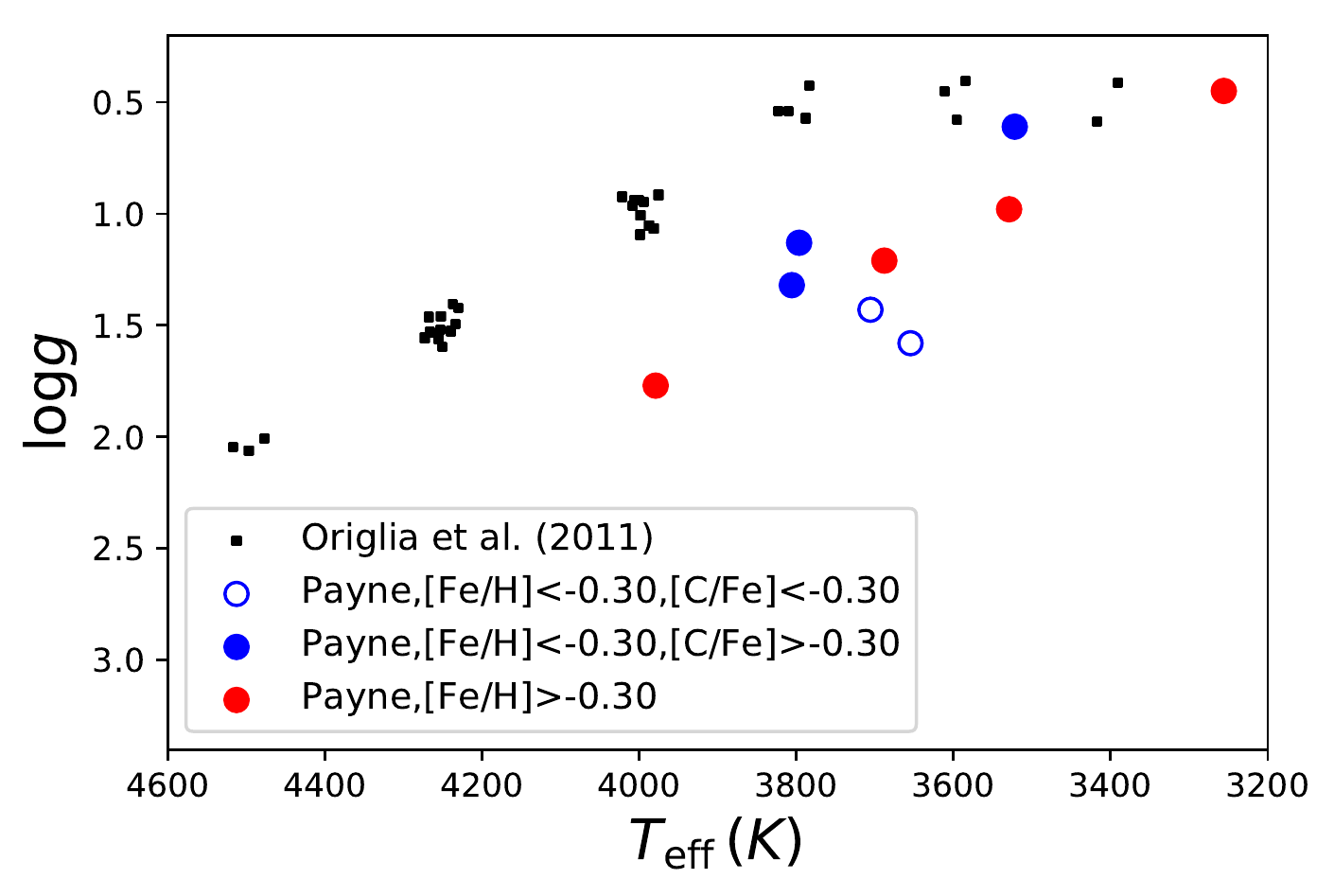}
\caption{A comparison of the temperature and gravity of red giants in Terzan 5 as measured by \citet{2011ApJ...726L..20O}, and by the Payne.  Relative to the measurements of \citet{2011ApJ...726L..20O}, the red giant branch is shifted to approximately 300 K colder temperatures by the Payne. The two stellar populations of Terzan 5 are color coded, to show that they have no obvious dependency on evolutionary state. For the metal-poor stars, the carbon-poor, nitrogen-rich stars are actually colder at fixed surface gravity, and thus the CNO variations are not due to the asymptotic giant branch phase. We have added 25 Kelvin and 0.10 dex of symmetrically distributed noise to the temperature and gravity measurements of \citet{2011ApJ...726L..20O} to improve the clarity of the figure. }
\label{fig:Terzan5Comparison}
\end{figure}

However, there are also some differences. Among these, the Payne-derived values of the mean chemistry are shifted. For the five most metal-poor stars, we measure mean values of [Fe/H]$=-$0.56, [$\alpha$/Fe]$=+$0.18, and for the four more metal-rich stars we measure  mean values of [Fe/H]$=+$0.10, [$\alpha$/Fe]$=+$0.07, where  the  [$\alpha$/Fe] is the arithmetic mean  of [Mg/Fe], [Ca/Fe], [Ti/Fe], and [Si/Fe]. This offset does not go away if one uses the ASPCAP or Cannon abundances rather than the Payne abundances. To investigate the discrepancy, we plotted both samples in Figure \ref{fig:Terzan5Comparison}. The metallicity offsets are clearly coupled to offsets in the estimates of effective temperature and/or surface gravity. The red giant branch derived from the Payne values is shifted to colder temperatures by approximately 300 Kelvin relative to that derived by \citet{2011ApJ...726L..20O}. \citet{2018arXiv180401530T} showed (see their Figure 8) that the Payne's temperatures for giants at the metallicity of Terzan 5 are approximately 100 Kelvin colder than photometric estimates using the infra-red flux method  of \citet{2009A&A...497..497G}, which can explain some of the offset.  The discrepancy between the temperature estimates of \citet{2011ApJ...726L..20O} and those derived from APOGEE spectra is reduced by one half if atmospheric parameters derived by ASPCAP are used.  Thus, it is likely that the temperature scale of the cluster, and by extension its mean chemistry, are intermediate between the Payne-derived values and those reported by  \citet{2011ApJ...726L..20O}. 

\begin{figure}
\centering
\includegraphics[width=0.47\textwidth]{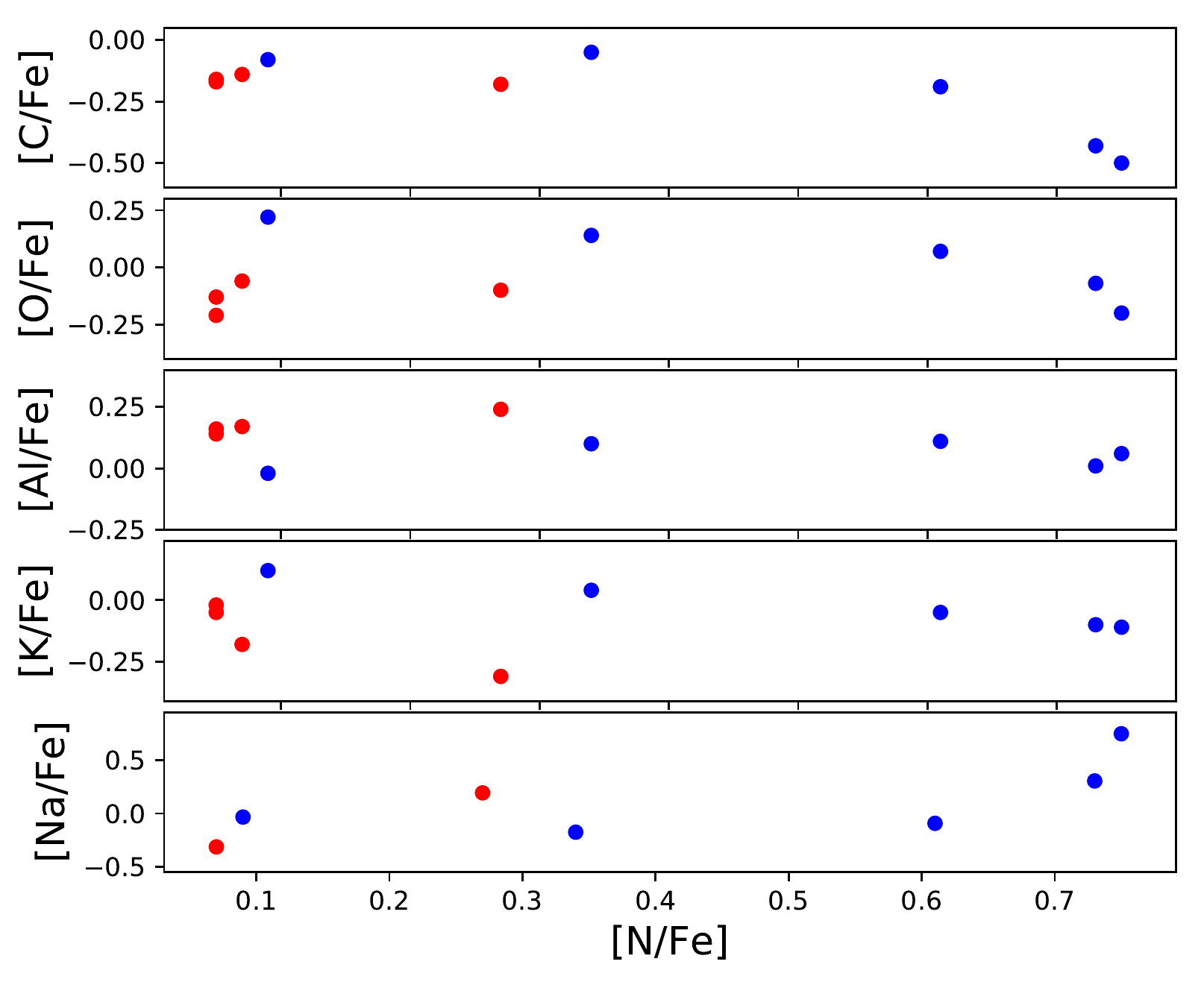}
\caption{APOGEE measurements of the metal-poor stars in Terzan 5 (blue) show the usual CNO abundance correlations expected of globular clusters, whereas those for the metal-rich stars (red) do not. The latter null result could be due to the small sample size. The increase of [Al/Fe] with [N/Fe] is null or negligible, consistent with our analysis. }
\label{fig:Terzan5Abundances}
\end{figure}

We show some of the abundance trends in Figure \ref{fig:Terzan5Abundances}. For the five more metal-poor stars, we find that each of [C/Fe], [O/Fe], and [K/Fe] are negatively correlated with [N/Fe], that [Na/Fe] is positively correlated with [N/Fe], and that [Al/Fe] is uncorrelated with [N/Fe]. The derived temperatures are all substantially colder than the previously estimated threshold temperature of $T_{\rm{eff}}\approx 4,750\,K$ at which CNO abundances become less reliable in APOGEE spectra.  The findings for carbon, oxygen, and aluminum are thus safely robust. More analysis would be needed to confirm the trends with sodium and potassium, which are less robust.  No abundance correlations are identified among the four metal-rich stars. We note that \citet{2017MNRAS.466.1010S} also reported abundance anti-correlations for the ASPCAP-derived APOGEE measurements of globular clusters in the inner Galaxy, though in their case they used an earlier data release, APOGEE DR12 \citep{2015ApJS..219...12A,2015AJ....150..148H}. 

The spread in CNO abundances is not an artifact of a low signal-to-noise ratio. For the five metal-poor stars, the signal-to-noise ratios are all greater than 92, and the two stars with the highest nitrogen abundances have signal-to-noise ratios of approximately 200. The [N/Fe] variation is also not likely to be due to mixing on the asymptotic giant branch. The nitrogen-rich, [Fe/H]$\leq -0.30$ stars are actually shifted to lower temperatures at fixed gravity relative to the nitrogen-poor, [Fe/H]$\leq -0.30$ stars, whereas they would be at higher temperatures if they were similar stars having evolved through the horizontal branch. 

\begin{figure}
\centering
\includegraphics[width=0.47\textwidth]{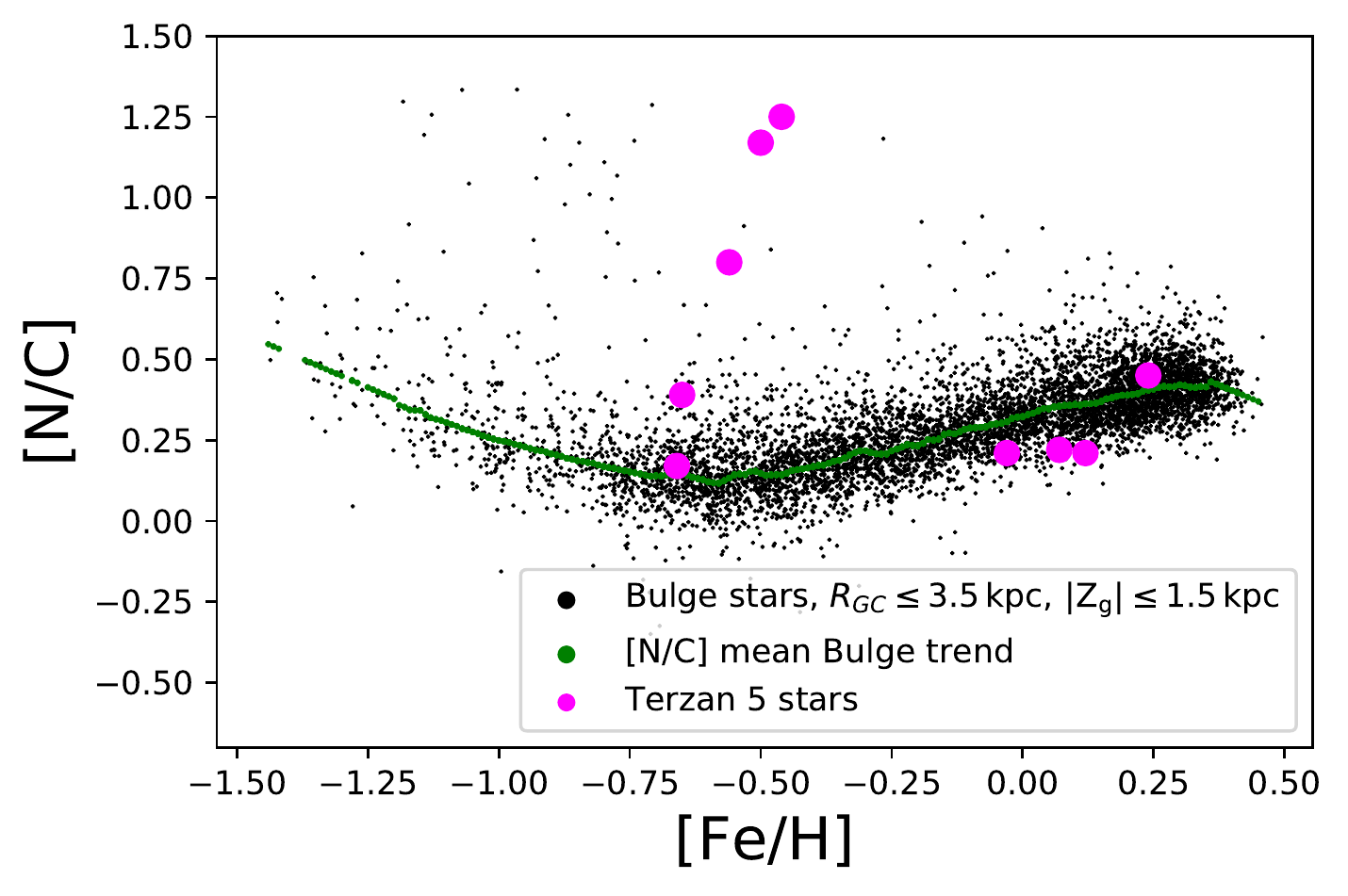}
\caption{A comparison of the [N/C] abundances of Terzan 5 stars (magenta points) to those of the Bulge field, green and black points are as in Figure \ref{Al_BulgeGCs}. The relatively metal-poor stars in Terzan 5 show [N/C] variations that are much larger than would be statistically expected from field stars of the same metallicity. Thus, the variations are more consistent with a globular cluster origin than with internal mixing processes. }
\label{fig:BulgeTerzan5_Comparison}
\end{figure}

Though small variations in light-element abundances are relatively common among [Fe/H] $\lesssim -1.0$ red giants in the field \citep{2000A&A...354..169G}, large enhancements in [N/Fe] or [N/C] are not that common among bulge red giants, which was recently shown by \citet{2017MNRAS.465..501S}. We also show this in Figure \ref{fig:BulgeTerzan5_Comparison}. Thus, something as common as mixing along the asymptotic giant branch phase cannot explain the large abundance shifts seen, of ${\Delta}\rm{([C/Fe],[N/Fe],[O/Fe])} \approx (-0.40,+0.65,-0.35)$. We can thus confidently argue that the metal-poor component in Terzan 5 includes stars showing the abundance variations expected for a globular cluster at its metallicity. We reiterate that the CNO abundance variations derived by the Payne from APOGEE spectra are qualitatively robust, but may be off by a factor of two. Regardless, the Bulge field stars are analyzed in the same way from the same kinds of spectra, they are thus an ideal control sample, and they do not exhibit such variations.

Thus, the findings of this investigation are consistent with Terzan 5 being a globular cluster. First, we explain the absence of an aluminum spread in Terzan 5 as being expected, due to its high metallicity. Second, we do find tentative evidence that the carbon, nitrogen, and oxygen abundances vary among the metal-poor stars in Terzan 5, with the potassium and sodium abundances possibly varying as well. Obviously, it would be better to have a larger sample of stars for both the metal-poor group and the metal-rich group. The statistical probability of the abundance correlations measured in the metal-poor group are negligible, however, there is the possibility of some undiagnosed and correlated systematic error. It would be advantageous for there to be another large, and independently studied sample. 

We ran the BACCHUS pipeline on these stars in order to gauge if they had s-process variations, specifically in the elements neodymium and cerium, using the same method as \citet{2017ApJ...846L...2F} and \citet{2018arXiv180107136F}. We did not measure any statistically significant variations in neodymium and cerium.

Aside from the findings previously mentioned in this Section, Terzan 5 is a complex system in several other ways. As noted by \citet{2011ApJ...726L..20O}, the stars in Terzan 5 have [C/Fe] $< 0 $, which also holds for all 9 of the members for which APOGEE measured spectra. Terzan 5  also hosts two trace subcomponents, with 5\% each of its total stellar mass, with metallicities of [Fe/H]$\approx -0.80$, and [Fe/H]$\approx +0.70$ \citep{2014ApJ...795...22M}.  \citet{2018arXiv181204325O} have also spectroscopically confirmed the  membership of three RR Lyrae variables and one Mira  variable as part of Terzan 5, for which the combination of spectroscopic abundances and pulsational properties can be used to provide constraints on the age-helium-metallicity relationship of the cluster.
Finally, there is an approximate spread of 0.70 magnitudes in the luminosity of the turnoff stars in Terzan 5  \citep{2016ApJ...828...75F}. 

Thus, we are not suggesting that Terzan 5 is a mundane stellar system. Quite the contrary, we believe that our findings validate its status as one of the most interesting stellar systems in the Galaxy, and confirm that further investigation is needed. In particular, a larger sample of APOGEE spectra, and separately, a large sample of precise sodium abundances, would each be very informative. 

\section{Discussion and Conclusion} \label{sec:Conclusion}

We investigated aluminum abundance variations in the stellar populations of globular clusters by conducting a  meta-analysis of the APOGEE data and the largest literature sample that we could assemble. We showed that aluminum enrichment operates independently of the CNONa abundance variations, and that it is reduced in more metal-rich and in lower-mass globular clusters, consistent with a prior analysis of globular clusters studied by the Gaia-ESO survey \citep{2017A&A...601A.112P}, and in an analysis of northern clusters analyzed with APOGEE data \citep{2018arXiv181208817M}.

We then derived, in Equations \ref{EQ:NaBestFits}, \ref{EQ:NBestFits}, and \ref{EQ:BestFits}, that the ratio of aluminum enrichment to sodium and/or nitrogen enrichment correlates with globular cluster metallicity and present-day stellar mass. The predicted relationships for 41 globular clusters are plotted along with the corresponding data in Figures \ref{fig:NAlplots} and \ref{fig:NaAlplots}. The data are consistent with the relative variations in [N/Fe] and [Na/Fe] being uncorrelated or weakly correlated with the mass and metallicity of globular clusters. That is also consistent with the sparse measurements available measurements of sodium and nitrogen being measured in the same samples of stars, as shown in Figure \ref{fig:SodiumNitrogenLiteratureComparison}. 

These relations can constrain the formation scenarios of chemically anomalous populations in globular clusters. For example, \citet{2016A&A...593A..36C} show that the [N/Al] ratio in the winds from fast rotating massive stars is extremely sensitive to the treatment of material injection from the helium-burning zone to the hydrogen-burning zone. If one supposes that rotating massive star winds will be proportionately more important in the more massive clusters, which is fair given that the globular clusters likely had deeper gravitational potentials at birth, then the finding of a lesser [N/Al] ratio in the chemically anomalous stars of the more massive clusters tightly constrains the treatment of injection. 

As a necessary component of our work, we investigated APOGEE's potential as a diagnostic of multiple populations in globular clusters. We first conducted a census of likely globular cluster stars within APOGEE DR14. We found 1012 stars which are associated with 28 globular clusters, consisting of 832 which were deliberately targeted as calibration targets, and 180 which were serendipitously observed due to globular clusters being quasi-randomly distributed in the field. We showed that most APOGEE-derived carbon, nitrogen, and oxygen abundances are not meaningful for stars with T$_{\rm{eff}} \geq 4,750\,K$, but that the Payne can still reliably measure [N/Fe] as long as the signal-to-noise ratio of the spectra is at least 50. We evaluated the suitability of the five currently-available pipelines applied to the APOGEE spectra for the study of multiple populations in globular clusters.  None of the pipelines are ideal, but the Payne performs best for our purposes: it tightly recovers the direction of the expected trends in the CNO abundance planes (though not the amplitudes), it does so for the largest sample of stars, and it has the most reliable [Al/Fe] determinations. 



One of the best prospects to improving APOGEE's diagnostic potential for globular clusters is expanding the parameter space explored by the Payne. We recommend lower bounds on the effective parameter space of [C/H],[O/H]$=-3.0$ and at least [X$_{i}$/H]$=-2.50$ for the other elements, as well as upper bounds of [N/Fe]$=2.00$ and [Al/Fe]$=1.50$.  APOGEE DR14 and DR16 incorporate data for at least ten more metal-poor ([Fe/H]$_{\rm{Harris}} \leq -1.50$) globular clusters, with more serendipitous inclusions being likely.  



Another argument for the diagnostic potential of APOGEE can be discerned from Figures  \ref{fig:Payne_APOGEE_GC_CNOAl_AbundanceScatter}, \ref{fig:Payne_APOGEE_GC_MP_AbundanceScatter}, and \ref{fig:Payne_APOGEE_GC_MR_AbundanceScatter}. The chemical abundances of the multiple populations appear as distinct sets, with discontinuities in their sequences. As discussed in the review of \citet{2015MNRAS.454.4197R}, the multiple populations of globular clusters are almost certainly distinct (and possibly discrete), rather than continuous sequences, as they appear that way in the more densely-sampled, and more precise, \textit{HST} color-magnitude diagrams \citep{2017MNRAS.464.3636M}. These populations normally form continuous sequences in spectroscopic samples, which may be due to greater relative measurement errors. APOGEE-derived Payne abundances show distinct populations, particularly for nitrogen and aluminum. That is an impressive achievement.

We have presented two lines of evidence against the notion that Terzan 5 is not a true globular cluster. First, we predict that the aluminum enrichment in that cluster is expected to follow a slope of ${\partial \rm{[Al/Fe]}}/{\partial \rm{[Na/Fe]}} \approx 0$.  Thus, the absence of an aluminum abundance scatter in Terzan 5 does not qualify it as an anomalous globular cluster. Second,  we show in Figure \ref{fig:Terzan5Abundances} that the five metal-poor stars in Terzan 5 measured by APOGEE follow the CNO abundance variation, with variations with sodium and potassium possibly detected as well. We find no scatter in the four metal-rich stars, but there are only three of them. A small sample can be sufficient to confidently demonstrate a scatter, but not to confidently negate a scatter. 

There are numerous options for follow-up. First, the GALAH survey \citep{2015MNRAS.449.2604D,2018MNRAS.478.4513B} is likely to include many serendipitously targeted globular cluster stars, as we found in APOGEE. Separately, though the seven globular clusters deliberately targeted by GALAH have all been studied by prior spectroscopic investigations, there is likely a lot to be learned from the numerous other abundances that the GALAH survey is measuring. There are also likely to be many additional serendipitous globular cluster stars in future data releases of APOGEE itself. 

Second, it will be interesting to see what can be learned applying the results of this investigation to the recently identified population of field stars with abundances similar to those of second generation globular cluster stars. \citet{2010A&A...519A..14M} discovered that approximately 3\% of Milky Way halo stars at distances of 4 to 40 kpc have high CN abundances, characteristic of anomalous globular cluster populations, a finding subsequently validated with APOGEE data \citep{2016ApJ...825..146M,2016ApJ...833..132F}. \cite{2017MNRAS.465..501S} found that the fraction of stars with such abundances is at least 7\% among metal-poor ([Fe/H] $\leq -1.0$) stars toward the inner $\sim$3 kpc of the Milky Way, and \citet{2017ApJ...846L...2F} have shown that the range of aluminum variations among these stars is variable. Even if we are to assume that this hypothesis of these stars' origin is correct, we cannot currently conclude  what fraction of these stars were evaporated from surviving globular clusters or from now dissolved globular clusters, let alone what the properties of this hypothetical population of dissolved globular clusters might have been. The relations between aluminum, nitrogen, and sodium abundance variations derived in this work may provide a way. 

We suspect, however, that theoretical work may contribute some of the most interesting follow-up investigations. We have shown that that the ratio of aluminum enrichment to sodium or nitrogen enrichment in globular clusters is correlated with the present-day stellar mass and metallicity of these clusters. These same data are consistent with no such correlation between sodium and nitrogen enrichment. This suggests one nucleosynthetic source for both of the CNO and Ne-Na nuclear processing, and a different source for the Al-Mg processing. We propose that this may be due to there having been two separate classes of non-supernovae chemical polluters that were common in the era of globular cluster formation, roughly $\sim$ 12 Gyr ago \citep{2000ApJ...533..215C,2009ApJ...694.1498M,2011ApJ...738...74D,2017ApJ...849..159D,2018ApJ...862...72V}, and that their relative contributions within globular clusters somehow correlate with globular cluster metallicity and present-day stellar mass. The nature of these two polluters, their relative contributions to other abundance trends, and why and how their effects were correlated with the metallicity and (then) future stellar mass of globular clusters, is a question begging exploration. 

\section*{Acknowledgments}

We thank Szabolcs M{\'e}sz{\'a}ros, Gail Zasowski, Jennifer Sobeck, Jo Bovy,  Katia Cunha, D. A. Garc\'{\i}a-Hern\'andez, Holger Baumgardt, Charli Sakari, Ryan Leaman, Long Wang, Sarah Martell, Henrik J{\"o}nsson, Thomas Masseron, Olga Zamora, Davide Massari, Livia Origlia and Francesco Ferraro for helpful comments.

We thank the anonymous referee, and the AAS statistics consultant, for constructive feedback which helped improve the manuscript. 

This research made use of Astropy \footnote{http://www.astropy.org} a community-developed core Python package for Astronomy, \citep{astropy:2013,astropy:2018}.

D.M.N was supported by the Allan C. and Dorothy H. Davis Fellowship. DM is supported by the BASAL Center for Astrophysics and Associated Technologies (CATA) through grant AFB 170002, by the Programa Iniciativa Científica Milenio grant IC120009, awarded to the Millennium Institute of Astrophysics (MAS), and by Proyecto FONDECYT No. 1170121. J.G.F-T is supported by FONDECYT No. 3180210 and the European COST Action CA16117 (ChETEC) project No 41736. SzM has been supported by the Premium Postdoctoral Research Program of the Hungarian Academy of Sciences, and by the Hungarian NKFI Grants K-119517 and GINOP-2.3.2-15-2016-00003 of the Hungarian National Research, Development and Innovation Office. D.G. gratefully acknowledges support from the Chilean Centro de Excelencia en Astrof{\'i}sica y Tecnolog{\'ias Afines (CATA) BASAL grant AFB-170002. D.G. also acknowledges financial support from the Direcci\'on de Investigaci\'on y Desarrollo de la Universidad de La Serena through the Programa de Incentivo a la Investigaci\'on de Acad\'emicos (PIA-DIDULS). DAGH acknowledges support provided by the Spanish Ministry of Economy and Competitiveness (MINECO) under grant AYA-2017-88254-P. DAGH acknowledge support provided by the Spanish Ministry
of Economy and Competitiveness (MINECO) under grant AYA-2017-88254-P. P.M.F., acknowledgesupport for this research from the National Science Foundation(AST-1311835 \& AST-1715662)

Funding for the Sloan Digital Sky Survey IV has been provided by the Alfred P. Sloan Foundation, the U.S. Department of Energy Office of Science, and the Participating Institutions. SDSS-IV acknowledges
support and resources from the Center for High-Performance Computing at
the University of Utah. The SDSS web site is www.sdss.org.

SDSS-IV is managed by the Astrophysical Research Consortium for the 
Participating Institutions of the SDSS Collaboration including the 
Brazilian Participation Group, the Carnegie Institution for Science, 
Carnegie Mellon University, the Chilean Participation Group, the French Participation Group, Harvard-Smithsonian Center for Astrophysics, 
Instituto de Astrof\'isica de Canarias, The Johns Hopkins University, Kavli Institute for the Physics and Mathematics of the Universe (IPMU) / 
University of Tokyo, the Korean Participation Group, Lawrence Berkeley National Laboratory, 
Leibniz Institut f\"ur Astrophysik Potsdam (AIP),  
Max-Planck-Institut f\"ur Astronomie (MPIA Heidelberg), 
Max-Planck-Institut f\"ur Astrophysik (MPA Garching), 
Max-Planck-Institut f\"ur Extraterrestrische Physik (MPE), 
National Astronomical Observatories of China, New Mexico State University, 
New York University, University of Notre Dame, 
Observat\'ario Nacional / MCTI, The Ohio State University, 
Pennsylvania State University, Shanghai Astronomical Observatory, 
United Kingdom Participation Group,
Universidad Nacional Aut\'onoma de M\'exico, University of Arizona, 
University of Colorado Boulder, University of Oxford, University of Portsmouth, 
University of Utah, University of Virginia, University of Washington, University of Wisconsin, 
Vanderbilt University, and Yale University.

\software{Astropy \citep{astropy:2013,astropy:2018}, SciPy \citep*{SciPy}, NumPy \citep{NumPy},
Matplotlib \citep{Hunter:2007}}

\bibliography{NatafApogeeBulgeNitrogen_V3}

\begin{thebibliography}{}
\expandafter\ifx\csname natexlab\endcsname\relax\def\natexlab#1{#1}\fi
\providecommand{\url}[1]{\href{#1}{#1}}
\providecommand{\dodoi}[1]{doi:~\href{http://doi.org/#1}{\nolinkurl{#1}}}
\providecommand{\doeprint}[1]{\href{http://ascl.net/#1}{\nolinkurl{http://ascl.net/#1}}}
\providecommand{\doarXiv}[1]{\href{https://arxiv.org/abs/#1}{\nolinkurl{https://arxiv.org/abs/#1}}}

\bibitem[{{Abolfathi} {et~al.}(2018){Abolfathi}, {Aguado}, {Aguilar}, {Allende
  Prieto}, {Almeida}, {Ananna}, {Anders}, {Anderson}, {Andrews}, {Anguiano}, \&
  et~al.}]{2018ApJS..235...42A}
{Abolfathi}, B., {Aguado}, D.~S., {Aguilar}, G., {et~al.} 2018, \apjs, 235, 42,
  \dodoi{10.3847/1538-4365/aa9e8a}

\bibitem[{{Alam} {et~al.}(2015){Alam}, {Albareti}, {Allende Prieto}, {Anders},
  {Anderson}, {Anderton}, {Andrews}, {Armengaud}, {Aubourg}, {Bailey}, \&
  et~al.}]{2015ApJS..219...12A}
{Alam}, S., {Albareti}, F.~D., {Allende Prieto}, C., {et~al.} 2015, \apjs, 219,
  12, \dodoi{10.1088/0067-0049/219/1/12}

\bibitem[{{Alves-Brito} {et~al.}(2012){Alves-Brito}, {Yong}, {Mel{\'e}ndez},
  {V{\'a}squez}, \& {Karakas}}]{2012A&A...540A...3A}
{Alves-Brito}, A., {Yong}, D., {Mel{\'e}ndez}, J., {V{\'a}squez}, S., \&
  {Karakas}, A.~I. 2012, \aap, 540, A3, \dodoi{10.1051/0004-6361/201118623}

\bibitem[{{Astropy Collaboration} {et~al.}(2013){Astropy Collaboration},
  {Robitaille}, {Tollerud}, {Greenfield}, {Droettboom}, {Bray}, {Aldcroft},
  {Davis}, {Ginsburg}, {Price-Whelan}, {Kerzendorf}, {Conley}, {Crighton},
  {Barbary}, {Muna}, {Ferguson}, {Grollier}, {Parikh}, {Nair}, {Unther},
  {Deil}, {Woillez}, {Conseil}, {Kramer}, {Turner}, {Singer}, {Fox}, {Weaver},
  {Zabalza}, {Edwards}, {Azalee Bostroem}, {Burke}, {Casey}, {Crawford},
  {Dencheva}, {Ely}, {Jenness}, {Labrie}, {Lim}, {Pierfederici}, {Pontzen},
  {Ptak}, {Refsdal}, {Servillat}, \& {Streicher}}]{astropy:2013}
{Astropy Collaboration}, {Robitaille}, T.~P., {Tollerud}, E.~J., {et~al.} 2013,
  aap, 558, A33, \dodoi{10.1051/0004-6361/201322068}

\bibitem[{{Barbuy} {et~al.}(2009){Barbuy}, {Zoccali}, {Ortolani}, {Hill},
  {Minniti}, {Bica}, {Renzini}, \& {G{\'o}mez}}]{2009A&A...507..405B}
{Barbuy}, B., {Zoccali}, M., {Ortolani}, S., {et~al.} 2009, \aap, 507, 405,
  \dodoi{10.1051/0004-6361/200912748}

\bibitem[{{Barbuy} {et~al.}(2014){Barbuy}, {Chiappini}, {Cantelli}, {Depagne},
  {Pignatari}, {Hirschi}, {Cescutti}, {Ortolani}, {Hill}, {Zoccali}, {Minniti},
  {Trevisan}, {Bica}, \& {G{\'o}mez}}]{2014A&A...570A..76B}
{Barbuy}, B., {Chiappini}, C., {Cantelli}, E., {et~al.} 2014, \aap, 570, A76,
  \dodoi{10.1051/0004-6361/201424311}

\bibitem[{{Bastian} \& {Lardo}(2018)}]{2018ARA&A..56...83B}
{Bastian}, N., \& {Lardo}, C. 2018, \araa, 56, 83,
  \dodoi{10.1146/annurev-astro-081817-051839}

\bibitem[{{Baumgardt}(2017)}]{2017MNRAS.464.2174B}
{Baumgardt}, H. 2017, \mnras, 464, 2174, \dodoi{10.1093/mnras/stw2488}

\bibitem[{{Baumgardt} \& {Hilker}(2018)}]{2018MNRAS.478.1520B}
{Baumgardt}, H., \& {Hilker}, M. 2018, \mnras, 478, 1520,
  \dodoi{10.1093/mnras/sty1057}

\bibitem[{{Baumgardt} {et~al.}(2019){Baumgardt}, {Hilker}, {Sollima}, \&
  {Bellini}}]{2019MNRAS.482.5138B}
{Baumgardt}, H., {Hilker}, M., {Sollima}, A., \& {Bellini}, A. 2019, \mnras,
  482, 5138, \dodoi{10.1093/mnras/sty2997}

\bibitem[{{Bekki}(2018)}]{2018arXiv180702309B}
{Bekki}, K. 2018, ArXiv e-prints.
\newblock \doarXiv{1807.02309}

\bibitem[{{Blanton} {et~al.}(2017){Blanton}, {Bershady}, {Abolfathi},
  {Albareti}, {Allende Prieto}, {Almeida}, {Alonso-Garc{\'{\i}}a}, {Anders},
  {Anderson}, {Andrews}, \& et~al.}]{2017AJ....154...28B}
{Blanton}, M.~R., {Bershady}, M.~A., {Abolfathi}, B., {et~al.} 2017, \aj, 154,
  28, \dodoi{10.3847/1538-3881/aa7567}

\bibitem[{{Bragaglia} {et~al.}(2015){Bragaglia}, {Carretta}, {Sollima},
  {Donati}, {D'Orazi}, {Gratton}, {Lucatello}, \&
  {Sneden}}]{2015A&A...583A..69B}
{Bragaglia}, A., {Carretta}, E., {Sollima}, A., {et~al.} 2015, \aap, 583, A69,
  \dodoi{10.1051/0004-6361/201526592}

\bibitem[{{Bressan} {et~al.}(2012){Bressan}, {Marigo}, {Girardi}, {Salasnich},
  {Dal Cero}, {Rubele}, \& {Nanni}}]{2012MNRAS.427..127B}
{Bressan}, A., {Marigo}, P., {Girardi}, L., {et~al.} 2012, \mnras, 427, 127,
  \dodoi{10.1111/j.1365-2966.2012.21948.x}

\bibitem[{{Buder} {et~al.}(2018){Buder}, {Asplund}, {Duong}, {Kos}, {Lind},
  {Ness}, {Sharma}, {Bland-Hawthorn}, {Casey}, {De Silva}, {D'Orazi},
  {Freeman}, {Lewis}, {Lin}, {Martell}, {Schlesinger}, {Simpson}, {Zucker},
  {Zwitter}, {Amarsi}, {Anguiano}, {Carollo}, {Casagrande}, {{\v C}otar},
  {Cottrell}, {Da Costa}, {Gao}, {Hayden}, {Horner}, {Ireland}, {Kafle},
  {Munari}, {Nataf}, {Nordlander}, {Stello}, {Ting}, {Traven}, {Watson},
  {Wittenmyer}, {Wyse}, {Yong}, {Zinn}, \& {{\v Z}erjal}}]{2018MNRAS.478.4513B}
{Buder}, S., {Asplund}, M., {Duong}, L., {et~al.} 2018, \mnras, 478, 4513,
  \dodoi{10.1093/mnras/sty1281}

\bibitem[{{Cabrera-Ziri} {et~al.}(2016){Cabrera-Ziri}, {Lardo}, {Davies},
  {Bastian}, {Beccari}, {Larsen}, \& {Hernandez}}]{2016MNRAS.460.1869C}
{Cabrera-Ziri}, I., {Lardo}, C., {Davies}, B., {et~al.} 2016, \mnras, 460,
  1869, \dodoi{10.1093/mnras/stw1090}

\bibitem[{{Carretta}(2015)}]{2015ApJ...810..148C}
{Carretta}, E. 2015, \apj, 810, 148, \dodoi{10.1088/0004-637X/810/2/148}

\bibitem[{{Carretta} \& {Bragaglia}(2018)}]{2018A&A...614A.109C}
{Carretta}, E., \& {Bragaglia}, A. 2018, \aap, 614, A109,
  \dodoi{10.1051/0004-6361/201832660}

\bibitem[{{Carretta} {et~al.}(2009{\natexlab{a}}){Carretta}, {Bragaglia},
  {Gratton}, \& {Lucatello}}]{2009A&A...505..139C}
{Carretta}, E., {Bragaglia}, A., {Gratton}, R., \& {Lucatello}, S.
  2009{\natexlab{a}}, \aap, 505, 139, \dodoi{10.1051/0004-6361/200912097}

\bibitem[{{Carretta} {et~al.}(2010){Carretta}, {Bragaglia}, {Gratton},
  {Recio-Blanco}, {Lucatello}, {D'Orazi}, \& {Cassisi}}]{2010A&A...516A..55C}
{Carretta}, E., {Bragaglia}, A., {Gratton}, R.~G., {et~al.} 2010, \aap, 516,
  A55, \dodoi{10.1051/0004-6361/200913451}

\bibitem[{{Carretta} {et~al.}(2018){Carretta}, {Bragaglia}, {Lucatello},
  {Gratton}, {D'Orazi}, \& {Sollima}}]{2018A&A...615A..17C}
{Carretta}, E., {Bragaglia}, A., {Lucatello}, S., {et~al.} 2018, \aap, 615,
  A17, \dodoi{10.1051/0004-6361/201732324}

\bibitem[{{Carretta} {et~al.}(2012){Carretta}, {D'Orazi}, {Gratton}, \&
  {Lucatello}}]{2012A&A...543A.117C}
{Carretta}, E., {D'Orazi}, V., {Gratton}, R.~G., \& {Lucatello}, S. 2012, \aap,
  543, A117, \dodoi{10.1051/0004-6361/201219277}

\bibitem[{{Carretta} {et~al.}(2013{\natexlab{a}}){Carretta}, {Gratton},
  {Bragaglia}, {D'Orazi}, \& {Lucatello}}]{2013A&A...550A..34C}
{Carretta}, E., {Gratton}, R.~G., {Bragaglia}, A., {D'Orazi}, V., \&
  {Lucatello}, S. 2013{\natexlab{a}}, \aap, 550, A34,
  \dodoi{10.1051/0004-6361/201220470}

\bibitem[{{Carretta} {et~al.}(2000){Carretta}, {Gratton}, {Clementini}, \&
  {Fusi Pecci}}]{2000ApJ...533..215C}
{Carretta}, E., {Gratton}, R.~G., {Clementini}, G., \& {Fusi Pecci}, F. 2000,
  \apj, 533, 215, \dodoi{10.1086/308629}

\bibitem[{{Carretta} {et~al.}(2007{\natexlab{a}}){Carretta}, {Bragaglia},
  {Gratton}, {Momany}, {Recio-Blanco}, {Cassisi}, {Fran{\c c}ois}, {James},
  {Lucatello}, \& {Moehler}}]{2007A&A...464..967C}
{Carretta}, E., {Bragaglia}, A., {Gratton}, R.~G., {et~al.} 2007{\natexlab{a}},
  \aap, 464, 967, \dodoi{10.1051/0004-6361:20066065}

\bibitem[{{Carretta} {et~al.}(2007{\natexlab{b}}){Carretta}, {Bragaglia},
  {Gratton}, {Catanzaro}, {Leone}, {Sabbi}, {Cassisi}, {Claudi}, {D'Antona},
  {Fran{\c c}ois}, {James}, \& {Piotto}}]{2007A&A...464..939C}
---. 2007{\natexlab{b}}, \aap, 464, 939, \dodoi{10.1051/0004-6361:20065730}

\bibitem[{{Carretta} {et~al.}(2009{\natexlab{b}}){Carretta}, {Bragaglia},
  {Gratton}, {Lucatello}, {Catanzaro}, {Leone}, {Bellazzini}, {Claudi},
  {D'Orazi}, {Momany}, {Ortolani}, {Pancino}, {Piotto}, {Recio-Blanco}, \&
  {Sabbi}}]{2009A&A...505..117C}
---. 2009{\natexlab{b}}, \aap, 505, 117, \dodoi{10.1051/0004-6361/200912096}

\bibitem[{{Carretta} {et~al.}(2013{\natexlab{b}}){Carretta}, {Bragaglia},
  {Gratton}, {Lucatello}, {D'Orazi}, {Bellazzini}, {Catanzaro}, {Leone},
  {Momany}, \& {Sollima}}]{2013A&A...557A.138C}
---. 2013{\natexlab{b}}, \aap, 557, A138, \dodoi{10.1051/0004-6361/201321905}

\bibitem[{{Carretta} {et~al.}(2015){Carretta}, {Bragaglia}, {Gratton},
  {D'Orazi}, {Lucatello}, {Sollima}, {Momany}, {Catanzaro}, \&
  {Leone}}]{2015A&A...578A.116C}
---. 2015, \aap, 578, A116, \dodoi{10.1051/0004-6361/201525951}

\bibitem[{{Casey} {et~al.}(2016){Casey}, {Hogg}, {Ness}, {Rix}, {Ho}, \&
  {Gilmore}}]{2016arXiv160303040C}
{Casey}, A.~R., {Hogg}, D.~W., {Ness}, M., {et~al.} 2016, ArXiv e-prints.
\newblock \doarXiv{1603.03040}

\bibitem[{{Cassisi} \& {Salaris}(1997)}]{1997MNRAS.285..593C}
{Cassisi}, S., \& {Salaris}, M. 1997, \mnras, 285, 593,
  \dodoi{10.1093/mnras/285.3.593}

\bibitem[{{Cavanna} {et~al.}(2015){Cavanna}, {Depalo}, {Aliotta}, {Anders},
  {Bemmerer}, {Best}, {Boeltzig}, {Broggini}, {Bruno}, {Caciolli},
  {Corvisiero}, {Davinson}, {di Leva}, {Elekes}, {Ferraro}, {Formicola},
  {F{\"u}l{\"o}p}, {Gervino}, {Guglielmetti}, {Gustavino}, {Gy{\"u}rky},
  {Imbriani}, {Junker}, {Menegazzo}, {Mossa}, {Pantaleo}, {Prati}, {Scott},
  {Somorjai}, {Straniero}, {Strieder}, {Sz{\"u}cs}, {Tak{\'a}cs}, {Trezzi}, \&
  {LUNA Collaboration}}]{2015PhRvL.115y2501C}
{Cavanna}, F., {Depalo}, R., {Aliotta}, M., {et~al.} 2015, Physical Review
  Letters, 115, 252501, \dodoi{10.1103/PhysRevLett.115.252501}

  \bibitem[{{Chen} {et~al.}(2018){Chen}, {D`Onghia}, {Pardy}, {Pasquali},
    {Bertelli Motta}, {Hanlon}, \& {Grebel}}]{2018ApJ...860...70C}
  {Chen}, B., {D'Onghia}, E., {Pardy}, S.~A., {et~al.} 2018, \apj, 860,
    70, \dodoi{10.3847/1538-4357/aac325}

\bibitem[{{Choplin} {et~al.}(2016){Choplin}, {Maeder}, {Meynet}, \&
  {Chiappini}}]{2016A&A...593A..36C}
{Choplin}, A., {Maeder}, A., {Meynet}, G., \& {Chiappini}, C. 2016, \aap, 593,
  A36, \dodoi{10.1051/0004-6361/201628083}

\bibitem[{{Cohen} {et~al.}(2002){Cohen}, {Briley}, \&
  {Stetson}}]{2002AJ....123.2525C}
{Cohen}, J.~G., {Briley}, M.~M., \& {Stetson}, P.~B. 2002, \aj, 123, 2525,
  \dodoi{10.1086/340179}

\bibitem[{{Conroy}(2012)}]{2012ApJ...758...21C}
{Conroy}, C. 2012, \apj, 758, 21, \dodoi{10.1088/0004-637X/758/1/21}

\bibitem[{{Cordero} {et~al.}(2014){Cordero}, {Pilachowski}, {Johnson},
  {McDonald}, {Zijlstra}, \& {Simmerer}}]{2014ApJ...780...94C}
{Cordero}, M.~J., {Pilachowski}, C.~A., {Johnson}, C.~I., {et~al.} 2014, \apj,
  780, 94, \dodoi{10.1088/0004-637X/780/1/94}

\bibitem[{{Cottrell} \& {Da Costa}(1981)}]{1981ApJ...245L..79C}
{Cottrell}, P.~L., \& {Da Costa}, G.~S. 1981, \apjl, 245, L79,
  \dodoi{10.1086/183527}

\bibitem[{{D'Antona} \& {Caloi}(2004)}]{2004ApJ...611..871D}
{D'Antona}, F., \& {Caloi}, V. 2004, \apj, 611, 871, \dodoi{10.1086/422334}

\bibitem[{{D'Antona} \& {Caloi}(2008)}]{2008MNRAS.390..693D}
---. 2008, \mnras, 390, 693, \dodoi{10.1111/j.1365-2966.2008.13760.x}

\bibitem[{{de Mink} {et~al.}(2009){de Mink}, {Pols}, {Langer}, \&
  {Izzard}}]{2009A&A...507L...1D}
{de Mink}, S.~E., {Pols}, O.~R., {Langer}, N., \& {Izzard}, R.~G. 2009, \aap,
  507, L1, \dodoi{10.1051/0004-6361/200913205}

\bibitem[{{De Silva} {et~al.}(2015){De Silva}, {Freeman}, {Bland-Hawthorn},
  {Martell}, {de Boer}, {Asplund}, {Keller}, {Sharma}, {Zucker}, {Zwitter},
  {Anguiano}, {Bacigalupo}, {Bayliss}, {Beavis}, {Bergemann}, {Campbell},
  {Cannon}, {Carollo}, {Casagrande}, {Casey}, {Da Costa}, {D'Orazi}, {Dotter},
  {Duong}, {Heger}, {Ireland}, {Kafle}, {Kos}, {Lattanzio}, {Lewis}, {Lin},
  {Lind}, {Munari}, {Nataf}, {O'Toole}, {Parker}, {Reid}, {Schlesinger},
  {Sheinis}, {Simpson}, {Stello}, {Ting}, {Traven}, {Watson}, {Wittenmyer},
  {Yong}, \& {{\v Z}erjal}}]{2015MNRAS.449.2604D}
{De Silva}, G.~M., {Freeman}, K.~C., {Bland-Hawthorn}, J., {et~al.} 2015,
  \mnras, 449, 2604, \dodoi{10.1093/mnras/stv327}

\bibitem[{{Decressin} {et~al.}(2007){Decressin}, {Meynet}, {Charbonnel},
  {Prantzos}, \& {Ekstr{\"o}m}}]{2007A&A...464.1029D}
{Decressin}, T., {Meynet}, G., {Charbonnel}, C., {Prantzos}, N., \&
  {Ekstr{\"o}m}, S. 2007, \aap, 464, 1029, \dodoi{10.1051/0004-6361:20066013}

\bibitem[{{Dell'Agli} {et~al.}(2018){Dell'Agli}, {Garc{\'{\i}}a-Hern{\'a}ndez},
  {Ventura}, {M{\'e}sz{\'a}ros}, {Masseron}, {Fern{\'a}ndez-Trincado}, {Tang},
  {Shetrone}, {Zamora}, \& {Lucatello}}]{2018MNRAS.475.3098D}
{Dell'Agli}, F., {Garc{\'{\i}}a-Hern{\'a}ndez}, D.~A., {Ventura}, P., {et~al.}
  2018, \mnras, 475, 3098, \dodoi{10.1093/mnras/stx3249}

\bibitem[{{Denissenkov} {et~al.}(2017){Denissenkov}, {VandenBerg}, {Kopacki},
  \& {Ferguson}}]{2017ApJ...849..159D}
{Denissenkov}, P.~A., {VandenBerg}, D.~A., {Kopacki}, G., \& {Ferguson}, J.~W.
  2017, \apj, 849, 159, \dodoi{10.3847/1538-4357/aa92c9}

\bibitem[{{D'Ercole} {et~al.}(2008){D'Ercole}, {Vesperini}, {D'Antona},
  {McMillan}, \& {Recchi}}]{2008MNRAS.391..825D}
{D'Ercole}, A., {Vesperini}, E., {D'Antona}, F., {McMillan}, S.~L.~W., \&
  {Recchi}, S. 2008, \mnras, 391, 825, \dodoi{10.1111/j.1365-2966.2008.13915.x}

\bibitem[{{di Criscienzo} {et~al.}(2010){di Criscienzo}, {Ventura}, {D'Antona},
  {Milone}, \& {Piotto}}]{2010MNRAS.408..999D}
{di Criscienzo}, M., {Ventura}, P., {D'Antona}, F., {Milone}, A., \& {Piotto},
  G. 2010, \mnras, 408, 999, \dodoi{10.1111/j.1365-2966.2010.17168.x}

\bibitem[{{Dias} {et~al.}(2016){Dias}, {Barbuy}, {Saviane}, {Held}, {Da Costa},
  {Ortolani}, {Gullieuszik}, \& {V{\'a}squez}}]{2016A&A...590A...9D}
{Dias}, B., {Barbuy}, B., {Saviane}, I., {et~al.} 2016, \aap, 590, A9,
  \dodoi{10.1051/0004-6361/201526765}

\bibitem[{{D'Orazi} {et~al.}(2015){D'Orazi}, {Gratton}, {Angelou}, {Bragaglia},
  {Carretta}, {Lattanzio}, {Lucatello}, {Momany}, {Sollima}, \&
  {Beccari}}]{2015MNRAS.449.4038D}
{D'Orazi}, V., {Gratton}, R.~G., {Angelou}, G.~C., {et~al.} 2015, \mnras, 449,
  4038, \dodoi{10.1093/mnras/stv612}

\bibitem[{{Dotter} {et~al.}(2008){Dotter}, {Chaboyer}, {Jevremovi{\'c}},
  {Kostov}, {Baron}, \& {Ferguson}}]{2008ApJS..178...89D}
{Dotter}, A., {Chaboyer}, B., {Jevremovi{\'c}}, D., {et~al.} 2008, \apjs, 178,
  89, \dodoi{10.1086/589654}

\bibitem[{{Dotter} {et~al.}(2018){Dotter}, {Milone}, {Conroy}, {Marino}, \&
  {Sarajedini}}]{2018ApJ...865L..10D}
{Dotter}, A., {Milone}, A.~P., {Conroy}, C., {Marino}, A.~F., \& {Sarajedini},
  A. 2018, \apjl, 865, L10, \dodoi{10.3847/2041-8213/aae08f}

\bibitem[{{Dotter} {et~al.}(2011){Dotter}, {Sarajedini}, \&
  {Anderson}}]{2011ApJ...738...74D}
{Dotter}, A., {Sarajedini}, A., \& {Anderson}, J. 2011, \apj, 738, 74,
  \dodoi{10.1088/0004-637X/738/1/74}

\bibitem[{{Eisenstein} {et~al.}(2011){Eisenstein}, {Weinberg}, {Agol},
  {Aihara}, {Allende Prieto}, {Anderson}, {Arns}, {Aubourg}, {Bailey},
  {Balbinot}, \& et~al.}]{2011AJ....142...72E}
{Eisenstein}, D.~J., {Weinberg}, D.~H., {Agol}, E., {et~al.} 2011, \aj, 142,
  72, \dodoi{10.1088/0004-6256/142/3/72}

\bibitem[{{Fern{\'a}ndez-Trincado} {et~al.}(2016){Fern{\'a}ndez-Trincado},
  {Robin}, {Moreno}, {Schiavon}, {Garc{\'{\i}}a P{\'e}rez}, {Vieira}, {Cunha},
  {Zamora}, {Sneden}, {Souto}, {Carrera}, {Johnson}, {Shetrone}, {Zasowski},
  {Garc{\'{\i}}a-Hern{\'a}ndez}, {Majewski}, {Reyl{\'e}}, {Blanco-Cuaresma},
  {Martinez-Medina}, {P{\'e}rez-Villegas}, {Valenzuela}, {Pichardo}, {Meza},
  {M{\'e}sz{\'a}ros}, {Sobeck}, {Geisler}, {Anders}, {Schultheis}, {Tang},
  {Roman-Lopes}, {Mennickent}, {Pan}, {Nitschelm}, \&
  {Allard}}]{2016ApJ...833..132F}
{Fern{\'a}ndez-Trincado}, J.~G., {Robin}, A.~C., {Moreno}, E., {et~al.} 2016,
  \apj, 833, 132, \dodoi{10.3847/1538-4357/833/2/132}

\bibitem[{{Fern{\'a}ndez-Trincado} {et~al.}(2017){Fern{\'a}ndez-Trincado},
  {Zamora}, {Garc{\'{\i}}a-Hern{\'a}ndez}, {Souto}, {Dell'Agli}, {Schiavon},
  {Geisler}, {Tang}, {Villanova}, {Hasselquist}, {Mennickent}, {Cunha},
  {Shetrone}, {Allende Prieto}, {Vieira}, {Zasowski}, {Sobeck}, {Hayes},
  {Majewski}, {Placco}, {Beers}, {Schleicher}, {Robin}, {M{\'e}sz{\'a}ros},
  {Masseron}, {Garc{\'{\i}}a P{\'e}rez}, {Anders}, {Meza}, {Alves-Brito},
  {Carrera}, {Minniti}, {Lane}, {Fern{\'a}ndez-Alvar}, {Moreno}, {Pichardo},
  {P{\'e}rez-Villegas}, {Schultheis}, {Roman-Lopes}, {Fuentes}, {Nitschelm},
  {Harding}, {Bizyaev}, {Pan}, {Oravetz}, {Simmons}, {Ivans},
  {Blanco-Cuaresma}, {Hern{\'a}ndez}, {Alonso-Garc{\'{\i}}a}, {Valenzuela}, \&
  {Chanam{\'e}}}]{2017ApJ...846L...2F}
{Fern{\'a}ndez-Trincado}, J.~G., {Zamora}, O., {Garc{\'{\i}}a-Hern{\'a}ndez},
  D.~A., {et~al.} 2017, \apjl, 846, L2, \dodoi{10.3847/2041-8213/aa8032}

\bibitem[{{Fern{\'a}ndez-Trincado} {et~al.}(2018){Fern{\'a}ndez-Trincado},
  {Zamora}, {Souto}, {Cohen}, {Dell'Agli}, {Garc{\'{\i}}a-Hern{\'a}ndez},
  {Masseron}, {Schiavon}, {M{\'e}sz{\'a}ros}, {Cunha}, {Hasselquist},
  {Shetrone}, {Schiappacasse Ulloa}, {Tang}, {Geisler}, {Schleicher},
  {Villanova}, {Mennickent}, {Minniti}, {Alonso-Garcia}, {Manchado}, {Beers},
  {Sobeck}, {Zasowski}, {Schultheis}, {Majewski}, {Rojas-Arriagada}, {Almeida},
  {Santana}, {Oelkers}, {Longa-Pe{\~n}a}, {Carrera}, {Burgasser}, {Lane},
  {Roman-Lopes}, {Ivans}, \& {Hearty}}]{2018arXiv180107136F}
{Fern{\'a}ndez-Trincado}, J.~G., {Zamora}, O., {Souto}, D., {et~al.} 2018,
  arXiv e-prints.
\newblock \doarXiv{1801.07136}

\bibitem[{{Ferraro} {et~al.}(2018){Ferraro}, {Tak{\'a}cs}, {Piatti}, {Cavanna},
  {Depalo}, {Aliotta}, {Bemmerer}, {Best}, {Boeltzig}, {Broggini}, {Bruno},
  {Caciolli}, {Chillery}, {Ciani}, {Corvisiero}, {Davinson}, {D'Erasmo}, {Di
  Leva}, {Elekes}, {Fiore}, {Formicola}, {F{\"u}l{\"o}p}, {Gervino},
  {Guglielmetti}, {Gustavino}, {Gy{\"u}rky}, {Imbriani}, {Junker}, {Karakas},
  {Kochanek}, {Lugaro}, {Marigo}, {Menegazzo}, {Mossa}, {Pantaleo},
  {Paticchio}, {Perrino}, {Prati}, {Schiavulli}, {St{\"o}ckel}, {Straniero},
  {Sz{\"u}cs}, {Trezzi}, {Zavatarelli}, \& {LUNA
  Collaboration}}]{2018PhRvL.121q2701F}
{Ferraro}, F., {Tak{\'a}cs}, M.~P., {Piatti}, D., {et~al.} 2018, Physical
  Review Letters, 121, 172701, \dodoi{10.1103/PhysRevLett.121.172701}

\bibitem[{{Ferraro} {et~al.}(2016){Ferraro}, {Massari}, {Dalessandro},
  {Lanzoni}, {Origlia}, {Rich}, \& {Mucciarelli}}]{2016ApJ...828...75F}
{Ferraro}, F.~R., {Massari}, D., {Dalessandro}, E., {et~al.} 2016, \apj, 828,
  75, \dodoi{10.3847/0004-637X/828/2/75}

\bibitem[{{Ferraro} {et~al.}(2009){Ferraro}, {Dalessandro}, {Mucciarelli},
  {Beccari}, {Rich}, {Origlia}, {Lanzoni}, {Rood}, {Valenti}, {Bellazzini},
  {Ransom}, \& {Cocozza}}]{2009Natur.462..483F}
{Ferraro}, F.~R., {Dalessandro}, E., {Mucciarelli}, A., {et~al.} 2009, \nat,
  462, 483, \dodoi{10.1038/nature08581}

\bibitem[{{Gaia Collaboration} {et~al.}(2018){Gaia Collaboration}, {Brown},
  {Vallenari}, {Prusti}, {de Bruijne}, {Babusiaux}, {Bailer-Jones}, {Biermann},
  {Evans}, {Eyer}, \& et~al.}]{2018A&A...616A...1G}
{Gaia Collaboration}, {Brown}, A.~G.~A., {Vallenari}, A., {et~al.} 2018, \aap,
  616, A1, \dodoi{10.1051/0004-6361/201833051}

\bibitem[{{Garc{\'{\i}}a P{\'e}rez} {et~al.}(2016){Garc{\'{\i}}a P{\'e}rez},
  {Allende Prieto}, {Holtzman}, {Shetrone}, {M{\'e}sz{\'a}ros}, {Bizyaev},
  {Carrera}, {Cunha}, {Garc{\'{\i}}a-Hern{\'a}ndez}, {Johnson}, {Majewski},
  {Nidever}, {Schiavon}, {Shane}, {Smith}, {Sobeck}, {Troup}, {Zamora},
  {Weinberg}, {Bovy}, {Eisenstein}, {Feuillet}, {Frinchaboy}, {Hayden},
  {Hearty}, {Nguyen}, {O'Connell}, {Pinsonneault}, {Wilson}, \&
  {Zasowski}}]{2016AJ....151..144G}
{Garc{\'{\i}}a P{\'e}rez}, A.~E., {Allende Prieto}, C., {Holtzman}, J.~A.,
  {et~al.} 2016, \aj, 151, 144, \dodoi{10.3847/0004-6256/151/6/144}

\bibitem[{{Gerber} {et~al.}(2018){Gerber}, {Friel}, \&
  {Vesperini}}]{2018AJ....156....6G}
{Gerber}, J.~M., {Friel}, E.~D., \& {Vesperini}, E. 2018, \aj, 156, 6,
  \dodoi{10.3847/1538-3881/aac2d4}

\bibitem[{{Gieles} {et~al.}(2018){Gieles}, {Charbonnel}, {Krause},
  {H{\'e}nault-Brunet}, {Agertz}, {Lamers}, {Bastian}, {Gualandris}, {Zocchi},
  \& {Petts}}]{2018MNRAS.478.2461G}
{Gieles}, M., {Charbonnel}, C., {Krause}, M.~G.~H., {et~al.} 2018, \mnras, 478,
  2461, \dodoi{10.1093/mnras/sty1059}

\bibitem[{{Gilmore} {et~al.}(2012){Gilmore}, {Randich}, {Asplund}, {Binney},
  {Bonifacio}, {Drew}, {Feltzing}, {Ferguson}, {Jeffries}, {Micela}, \&
  et~al.}]{2012Msngr.147...25G}
{Gilmore}, G., {Randich}, S., {Asplund}, M., {et~al.} 2012, The Messenger, 147,
  25

\bibitem[{{Gonz{\'a}lez Hern{\'a}ndez} \&
  {Bonifacio}(2009)}]{2009A&A...497..497G}
{Gonz{\'a}lez Hern{\'a}ndez}, J.~I., \& {Bonifacio}, P. 2009, \aap, 497, 497,
  \dodoi{10.1051/0004-6361/200810904}

\bibitem[{{Gratton} {et~al.}(2006){Gratton}, {Lucatello}, {Bragaglia},
  {Carretta}, {Momany}, {Pancino}, \& {Valenti}}]{2006A&A...455..271G}
{Gratton}, R.~G., {Lucatello}, S., {Bragaglia}, A., {et~al.} 2006, \aap, 455,
  271, \dodoi{10.1051/0004-6361:20064957}

\bibitem[{{Gratton} {et~al.}(2000){Gratton}, {Sneden}, {Carretta}, \&
  {Bragaglia}}]{2000A&A...354..169G}
{Gratton}, R.~G., {Sneden}, C., {Carretta}, E., \& {Bragaglia}, A. 2000, \aap,
  354, 169

\bibitem[{{Grevesse} \& {Noels}(1993)}]{1993oee..conf...15G}
{Grevesse}, N., \& {Noels}, A. 1993, in Origin and Evolution of the Elements,
  ed. N.~{Prantzos}, E.~{Vangioni-Flam}, \& M.~{Casse}, 15--25

\bibitem[{{Gunn} {et~al.}(2006){Gunn}, {Siegmund}, {Mannery}, {Owen}, {Hull},
  {Leger}, {Carey}, {Knapp}, {York}, {Boroski}, {Kent}, {Lupton}, {Rockosi},
  {Evans}, {Waddell}, {Anderson}, {Annis}, {Barentine}, {Bartoszek}, {Bastian},
  {Bracker}, {Brewington}, {Briegel}, {Brinkmann}, {Brown}, {Carr},
  {Czarapata}, {Drennan}, {Dombeck}, {Federwitz}, {Gillespie}, {Gonzales},
  {Hansen}, {Harvanek}, {Hayes}, {Jordan}, {Kinney}, {Klaene}, {Kleinman},
  {Kron}, {Kresinski}, {Lee}, {Limmongkol}, {Lindenmeyer}, {Long}, {Loomis},
  {McGehee}, {Mantsch}, {Neilsen}, {Neswold}, {Newman}, {Nitta}, {Peoples},
  {Pier}, {Prieto}, {Prosapio}, {Rivetta}, {Schneider}, {Snedden}, \&
  {Wang}}]{2006AJ....131.2332G}
{Gunn}, J.~E., {Siegmund}, W.~A., {Mannery}, E.~J., {et~al.} 2006, \aj, 131,
  2332, \dodoi{10.1086/500975}

\bibitem[{{Hale} {et~al.}(2004){Hale}, {Champagne}, {Iliadis}, {Hansper},
  {Powell}, \& {Blackmon}}]{2004PhRvC..70d5802H}
{Hale}, S.~E., {Champagne}, A.~E., {Iliadis}, C., {et~al.} 2004, \prc, 70,
  045802, \dodoi{10.1103/PhysRevC.70.045802}

\bibitem[{{Harris}(1996, 2010 edition)}]{1996AJ....112.1487H}
{Harris}, W.~E. 1996, 2010 edition, \aj, 112, 1487, \dodoi{10.1086/118116}

\bibitem[{{Ho} {et~al.}(2017){Ho}, {Rix}, {Ness}, {Hogg}, {Liu}, \&
  {Ting}}]{2017ApJ...841...40H}
{Ho}, A.~Y.~Q., {Rix}, H.-W., {Ness}, M.~K., {et~al.} 2017, \apj, 841, 40,
  \dodoi{10.3847/1538-4357/aa6db3}

\bibitem[{{Hogg} \& {Foreman-Mackey}(2018)}]{2018ApJS..236...11H}
{Hogg}, D.~W., \& {Foreman-Mackey}, D. 2018, \apjs, 236, 11,
  \dodoi{10.3847/1538-4365/aab76e}

\bibitem[{{Holtzman} {et~al.}(2015){Holtzman}, {Shetrone}, {Johnson}, {Allende
  Prieto}, {Anders}, {Andrews}, {Beers}, {Bizyaev}, {Blanton}, {Bovy},
  {Carrera}, {Chojnowski}, {Cunha}, {Eisenstein}, {Feuillet}, {Frinchaboy},
  {Galbraith-Frew}, {Garc{\'{\i}}a P{\'e}rez}, {Garc{\'{\i}}a-Hern{\'a}ndez},
  {Hasselquist}, {Hayden}, {Hearty}, {Ivans}, {Majewski}, {Martell},
  {Meszaros}, {Muna}, {Nidever}, {Nguyen}, {O'Connell}, {Pan}, {Pinsonneault},
  {Robin}, {Schiavon}, {Shane}, {Sobeck}, {Smith}, {Troup}, {Weinberg},
  {Wilson}, {Wood-Vasey}, {Zamora}, \& {Zasowski}}]{2015AJ....150..148H}
{Holtzman}, J.~A., {Shetrone}, M., {Johnson}, J.~A., {et~al.} 2015, \aj, 150,
  148, \dodoi{10.1088/0004-6256/150/5/148}

\bibitem[{Hunter(2007)}]{Hunter:2007}
Hunter, J.~D. 2007, Computing In Science \& Engineering, 9, 90,
  \dodoi{10.1109/MCSE.2007.55}

\bibitem[{{Iliadis} {et~al.}(2001){Iliadis}, {D'Auria}, {Starrfield},
  {Thompson}, \& {Wiescher}}]{2001ApJS..134..151I}
{Iliadis}, C., {D'Auria}, J.~M., {Starrfield}, S., {Thompson}, W.~J., \&
  {Wiescher}, M. 2001, \apjs, 134, 151, \dodoi{10.1086/320364}

\bibitem[{{Iliadis} {et~al.}(2010){Iliadis}, {Longland}, {Champagne}, {Coc}, \&
  {Fitzgerald}}]{2010NuPhA.841...31I}
{Iliadis}, C., {Longland}, R., {Champagne}, A.~E., {Coc}, A., \& {Fitzgerald},
  R. 2010, Nuclear Physics A, 841, 31, \dodoi{10.1016/j.nuclphysa.2010.04.009}

\bibitem[{{Ivans} {et~al.}(1999){Ivans}, {Sneden}, {Kraft}, {Suntzeff},
  {Smith}, {Langer}, \& {Fulbright}}]{1999AJ....118.1273I}
{Ivans}, I.~I., {Sneden}, C., {Kraft}, R.~P., {et~al.} 1999, \aj, 118, 1273,
  \dodoi{10.1086/301017}

\bibitem[{{Jofr{\'e}} {et~al.}(2018){Jofr{\'e}}, {Heiter}, \&
  {Soubiran}}]{2018arXiv181108041J}
{Jofr{\'e}}, P., {Heiter}, U., \& {Soubiran}, C. 2018, arXiv e-prints.
\newblock \doarXiv{1811.08041}

\bibitem[{{Jofr{\'e}} {et~al.}(2014){Jofr{\'e}}, {Heiter}, {Soubiran},
  {Blanco-Cuaresma}, {Worley}, {Pancino}, {Cantat-Gaudin}, {Magrini},
  {Bergemann}, {Gonz{\'a}lez Hern{\'a}ndez}, {Hill}, {Lardo}, {de Laverny},
  {Lind}, {Masseron}, {Montes}, {Mucciarelli}, {Nordlander}, {Recio Blanco},
  {Sobeck}, {Sordo}, {Sousa}, {Tabernero}, {Vallenari}, \& {Van
  Eck}}]{2014A&A...564A.133J}
{Jofr{\'e}}, P., {Heiter}, U., {Soubiran}, C., {et~al.} 2014, \aap, 564, A133,
  \dodoi{10.1051/0004-6361/201322440}

\bibitem[{{Johnson} {et~al.}(2017{\natexlab{a}}){Johnson}, {Caldwell}, {Rich},
  {Mateo}, {Bailey}, {Clarkson}, {Olszewski}, \&
  {Walker}}]{2017ApJ...836..168J}
{Johnson}, C.~I., {Caldwell}, N., {Rich}, R.~M., {et~al.} 2017{\natexlab{a}},
  \apj, 836, 168, \dodoi{10.3847/1538-4357/836/2/168}

\bibitem[{{Johnson} {et~al.}(2017{\natexlab{b}}){Johnson}, {Caldwell}, {Rich},
  {Mateo}, {Bailey}, {Olszewski}, \& {Walker}}]{2017ApJ...842...24J}
---. 2017{\natexlab{b}}, \apj, 842, 24, \dodoi{10.3847/1538-4357/aa7414}

\bibitem[{{Johnson} {et~al.}(2017{\natexlab{c}}){Johnson}, {Caldwell}, {Rich},
  \& {Walker}}]{2017AJ....154..155J}
{Johnson}, C.~I., {Caldwell}, N., {Rich}, R.~M., \& {Walker}, M.~G.
  2017{\natexlab{c}}, \aj, 154, 155, \dodoi{10.3847/1538-3881/aa86ac}

\bibitem[{{Johnson} {et~al.}(2005){Johnson}, {Kraft}, {Pilachowski}, {Sneden},
  {Ivans}, \& {Benman}}]{2005PASP..117.1308J}
{Johnson}, C.~I., {Kraft}, R.~P., {Pilachowski}, C.~A., {et~al.} 2005, \pasp,
  117, 1308, \dodoi{10.1086/497435}

\bibitem[{{Johnson} \& {Pilachowski}(2010)}]{2010ApJ...722.1373J}
{Johnson}, C.~I., \& {Pilachowski}, C.~A. 2010, \apj, 722, 1373,
  \dodoi{10.1088/0004-637X/722/2/1373}

\bibitem[{{Johnson} {et~al.}(2018){Johnson}, {Rich}, {Caldwell}, {Mateo},
  {Bailey}, {Olszewski}, \& {Walker}}]{2018AJ....155...71J}
{Johnson}, C.~I., {Rich}, R.~M., {Caldwell}, N., {et~al.} 2018, \aj, 155, 71,
  \dodoi{10.3847/1538-3881/aaa294}

\bibitem[{{Jones} {et~al.}(2001--){Jones}, {Olyphant}, \& {Peterson}}]{SciPy}
{Jones}, E., {Olyphant}, T., \& {Peterson}, P. 2001--, {SciPy}: Open source
  scientific tools for {Python}.
\newblock \url{http://www.scipy.org/}

\bibitem[{{J{\"o}nsson} {et~al.}(2018){J{\"o}nsson}, {Allende Prieto},
  {Holtzman}, {Feuillet}, {Hawkins}, {Cunha}, {M{\'e}sz{\'a}ros},
  {Hasselquist}, {Fern{\'a}ndez-Trincado}, {Garc{\'{\i}}a-Hern{\'a}ndez},
  {Bizyaev}, {Carrera}, {Majewski}, {Pinsonneault}, {Shetrone}, {Smith},
  {Sobeck}, {Souto}, {Stringfellow}, {Teske}, \&
  {Zamora}}]{2018AJ....156..126J}
{J{\"o}nsson}, H., {Allende Prieto}, C., {Holtzman}, J.~A., {et~al.} 2018, \aj,
  156, 126, \dodoi{10.3847/1538-3881/aad4f5}

\bibitem[{{Kamann} {et~al.}(2018){Kamann}, {Husser}, {Dreizler}, {Emsellem},
  {Weilbacher}, {Martens}, {Bacon}, {den Brok}, {Giesers}, {Krajnovi{\'c}},
  {Roth}, {Wendt}, \& {Wisotzki}}]{2018MNRAS.473.5591K}
{Kamann}, S., {Husser}, T.-O., {Dreizler}, S., {et~al.} 2018, \mnras, 473,
  5591, \dodoi{10.1093/mnras/stx2719}

\bibitem[{{Karakas} {et~al.}(2018){Karakas}, {Lugaro}, {Carlos}, {Cseh},
  {Kamath}, \& {Garc{\'{\i}}a-Hern{\'a}ndez}}]{2018MNRAS.477..421K}
{Karakas}, A.~I., {Lugaro}, M., {Carlos}, M., {et~al.} 2018, \mnras, 477, 421,
  \dodoi{10.1093/mnras/sty625}

\bibitem[{{Karakas} {et~al.}(2014){Karakas}, {Marino}, \&
  {Nataf}}]{2014ApJ...784...32K}
{Karakas}, A.~I., {Marino}, A.~F., \& {Nataf}, D.~M. 2014, \apj, 784, 32,
  \dodoi{10.1088/0004-637X/784/1/32}

\bibitem[{{Kemp} {et~al.}(2018){Kemp}, {Casey}, {Miles}, {Norfolk},
  {Lattanzio}, {Karakas}, {Schlaufman}, {Ho}, {Tout}, {Ness}, \&
  {Ji}}]{2018MNRAS.tmp.1822K}
{Kemp}, A.~J., {Casey}, A.~R., {Miles}, M.~T., {et~al.} 2018, \mnras,
  \dodoi{10.1093/mnras/sty1915}

\bibitem[{{Kim} \& {Lee}(2018)}]{2018ApJ...869...35K}
{Kim}, J.~J., \& {Lee}, Y.-W. 2018, \apj, 869, 35,
  \dodoi{10.3847/1538-4357/aaec67}

\bibitem[{{Koch} \& {McWilliam}(2008)}]{2008AJ....135.1551K}
{Koch}, A., \& {McWilliam}, A. 2008, \aj, 135, 1551,
  \dodoi{10.1088/0004-6256/135/4/1551}

\bibitem[{{Koch} \& {McWilliam}(2014)}]{2014A&A...565A..23K}
---. 2014, \aap, 565, A23, \dodoi{10.1051/0004-6361/201323119}

\bibitem[{{Lagioia} {et~al.}(2018){Lagioia}, {Milone}, {Marino}, {Cassisi},
  {Aparicio}, {Piotto}, {Anderson}, {Barbuy}, {Bedin}, {Bellini}, {Brown},
  {D'Antona}, {Nardiello}, {Ortolani}, {Pietrinferni}, {Renzini}, {Salaris},
  {Sarajedini}, {van der Marel}, \& {Vesperini}}]{2018MNRAS.475.4088L}
{Lagioia}, E.~P., {Milone}, A.~P., {Marino}, A.~F., {et~al.} 2018, \mnras, 475,
  4088, \dodoi{10.1093/mnras/sty083}

\bibitem[{{Lanzoni} {et~al.}(2010){Lanzoni}, {Ferraro}, {Dalessandro},
  {Mucciarelli}, {Beccari}, {Miocchi}, {Bellazzini}, {Rich}, {Origlia},
  {Valenti}, {Rood}, \& {Ransom}}]{2010ApJ...717..653L}
{Lanzoni}, B., {Ferraro}, F.~R., {Dalessandro}, E., {et~al.} 2010, \apj, 717,
  653, \dodoi{10.1088/0004-637X/717/2/653}

\bibitem[{{Lapenna} {et~al.}(2015){Lapenna}, {Mucciarelli}, {Ferraro},
  {Origlia}, {Lanzoni}, {Massari}, \& {Dalessandro}}]{2015ApJ...813...97L}
{Lapenna}, E., {Mucciarelli}, A., {Ferraro}, F.~R., {et~al.} 2015, \apj, 813,
  97, \dodoi{10.1088/0004-637X/813/2/97}

\bibitem[{{Lattanzio} {et~al.}(2000){Lattanzio}, {Forestini}, \&
  {Charbonnel}}]{2000MmSAI..71..737L}
{Lattanzio}, J., {Forestini}, M., \& {Charbonnel}, C. 2000, \memsai, 71, 737

\bibitem[{{Leung} \& {Bovy}(2019)}]{2019MNRAS.483.3255L}
{Leung}, H.~W., \& {Bovy}, J. 2019, \mnras, 483, 3255,
  \dodoi{10.1093/mnras/sty3217}

\bibitem[{{Luo} {et~al.}(2015){Luo}, {Zhao}, {Zhao}, {Deng}, {Liu}, {Jing},
  {Wang}, {Zhang}, {Shi}, {Cui}, {Chu}, {Li}, {Bai}, {Wu}, {Cai}, {Cao}, {Cao},
  {Carlin}, {Chen}, {Chen}, {Chen}, {Chen}, {Chen}, {Chen}, {Chen},
  {Christlieb}, {Chu}, {Cui}, {Dong}, {Du}, {Fan}, {Feng}, {Fu}, {Gao}, {Gong},
  {Gu}, {Guo}, {Han}, {He}, {Hou}, {Hou}, {Hou}, {Hu}, {Hu}, {Hu}, {Huo},
  {Jia}, {Jiang}, {Jiang}, {Jiang}, {Jin}, {Kong}, {Kong}, {Lei}, {Li}, {Li},
  {Li}, {Li}, {Li}, {Li}, {Li}, {Li}, {Li}, {Li}, {Li}, {Li}, {Liang}, {Lin},
  {Liu}, {Liu}, {Liu}, {Liu}, {Lu}, {Luo}, {Mao}, {Newberg}, {Ni}, {Qi}, {Qi},
  {Shen}, {Shi}, {Song}, {Song}, {Su}, {Su}, {Tang}, {Tao}, {Tian}, {Wang},
  {Wang}, {Wang}, {Wang}, {Wang}, {Wang}, {Wang}, {Wang}, {Wang}, {Wang},
  {Wang}, {Wang}, {Wang}, {Wang}, {Wang}, {Wang}, {Wang}, {Wang}, {Wang},
  {Wang}, {Wei}, {Wei}, {Wu}, {Wu}, {Wu}, {Wu}, {Xing}, {Xu}, {Xu}, {Xu},
  {Yan}, {Yang}, {Yang}, {Yang}, {Yang}, {Yao}, {Yu}, {Yuan}, {Yuan}, {Yuan},
  {Yuan}, {Zhai}, {Zhang}, {Zhang}, {Zhang}, {Zhang}, {Zhang}, {Zhang},
  {Zhang}, {Zhang}, {Zhao}, {Zhou}, {Zhou}, {Zhu}, {Zhu}, {Zou}, \&
  {Zuo}}]{2015RAA....15.1095L}
{Luo}, A.-L., {Zhao}, Y.-H., {Zhao}, G., {et~al.} 2015, Research in Astronomy
  and Astrophysics, 15, 1095, \dodoi{10.1088/1674-4527/15/8/002}

\bibitem[{{MacLean} {et~al.}(2018){MacLean}, {Campbell}, {De Silva},
  {Lattanzio}, {D'Orazi}, {Cottrell}, {Momany}, \&
  {Casagrande}}]{2018MNRAS.475..257M}
{MacLean}, B.~T., {Campbell}, S.~W., {De Silva}, G.~M., {et~al.} 2018, \mnras,
  475, 257, \dodoi{10.1093/mnras/stx3217}

\bibitem[{{Maeder} \& {Meynet}(2006)}]{2006A&A...448L..37M}
{Maeder}, A., \& {Meynet}, G. 2006, \aap, 448, L37,
  \dodoi{10.1051/0004-6361:200600012}

\bibitem[{{Majewski} {et~al.}(2017){Majewski}, {Schiavon}, {Frinchaboy},
  {Allende Prieto}, {Barkhouser}, {Bizyaev}, {Blank}, {Brunner}, {Burton},
  {Carrera}, {Chojnowski}, {Cunha}, {Epstein}, {Fitzgerald}, {Garc{\'{\i}}a
  P{\'e}rez}, {Hearty}, {Henderson}, {Holtzman}, {Johnson}, {Lam}, {Lawler},
  {Maseman}, {M{\'e}sz{\'a}ros}, {Nelson}, {Nguyen}, {Nidever}, {Pinsonneault},
  {Shetrone}, {Smee}, {Smith}, {Stolberg}, {Skrutskie}, {Walker}, {Wilson},
  {Zasowski}, {Anders}, {Basu}, {Beland}, {Blanton}, {Bovy}, {Brownstein},
  {Carlberg}, {Chaplin}, {Chiappini}, {Eisenstein}, {Elsworth}, {Feuillet},
  {Fleming}, {Galbraith-Frew}, {Garc{\'{\i}}a}, {Garc{\'{\i}}a-Hern{\'a}ndez},
  {Gillespie}, {Girardi}, {Gunn}, {Hasselquist}, {Hayden}, {Hekker}, {Ivans},
  {Kinemuchi}, {Klaene}, {Mahadevan}, {Mathur}, {Mosser}, {Muna}, {Munn},
  {Nichol}, {O'Connell}, {Parejko}, {Robin}, {Rocha-Pinto}, {Schultheis},
  {Serenelli}, {Shane}, {Silva Aguirre}, {Sobeck}, {Thompson}, {Troup},
  {Weinberg}, \& {Zamora}}]{2017AJ....154...94M}
{Majewski}, S.~R., {Schiavon}, R.~P., {Frinchaboy}, P.~M., {et~al.} 2017, \aj,
  154, 94, \dodoi{10.3847/1538-3881/aa784d}

\bibitem[{{Mar{\'{\i}}n-Franch} {et~al.}(2009){Mar{\'{\i}}n-Franch},
  {Aparicio}, {Piotto}, {Rosenberg}, {Chaboyer}, {Sarajedini}, {Siegel},
  {Anderson}, {Bedin}, {Dotter}, {Hempel}, {King}, {Majewski}, {Milone},
  {Paust}, \& {Reid}}]{2009ApJ...694.1498M}
{Mar{\'{\i}}n-Franch}, A., {Aparicio}, A., {Piotto}, G., {et~al.} 2009, \apj,
  694, 1498, \dodoi{10.1088/0004-637X/694/2/1498}

\bibitem[{{Marino} {et~al.}(2008){Marino}, {Villanova}, {Piotto}, {Milone},
  {Momany}, {Bedin}, \& {Medling}}]{2008A&A...490..625M}
{Marino}, A.~F., {Villanova}, S., {Piotto}, G., {et~al.} 2008, \aap, 490, 625,
  \dodoi{10.1051/0004-6361:200810389}

\bibitem[{{Marino} {et~al.}(2011){Marino}, {Sneden}, {Kraft}, {Wallerstein},
  {Norris}, {Da Costa}, {Milone}, {Ivans}, {Gonzalez}, {Fulbright}, {Hilker},
  {Piotto}, {Zoccali}, \& {Stetson}}]{2011A&A...532A...8M}
{Marino}, A.~F., {Sneden}, C., {Kraft}, R.~P., {et~al.} 2011, \aap, 532, A8,
  \dodoi{10.1051/0004-6361/201116546}

\bibitem[{{Marino} {et~al.}(2012){Marino}, {Milone}, {Piotto}, {Cassisi},
  {D'Antona}, {Anderson}, {Aparicio}, {Bedin}, {Renzini}, \&
  {Villanova}}]{2012ApJ...746...14M}
{Marino}, A.~F., {Milone}, A.~P., {Piotto}, G., {et~al.} 2012, \apj, 746, 14,
  \dodoi{10.1088/0004-637X/746/1/14}

\bibitem[{{Marino} {et~al.}(2014){Marino}, {Milone}, {Przybilla}, {Bergemann},
  {Lind}, {Asplund}, {Cassisi}, {Catelan}, {Casagrande}, {Valcarce}, {Bedin},
  {Cort{\'e}s}, {D'Antona}, {Jerjen}, {Piotto}, {Schlesinger}, {Zoccali}, \&
  {Angeloni}}]{2014MNRAS.437.1609M}
{Marino}, A.~F., {Milone}, A.~P., {Przybilla}, N., {et~al.} 2014, \mnras, 437,
  1609, \dodoi{10.1093/mnras/stt1993}

\bibitem[{{Marino} {et~al.}(2016){Marino}, {Milone}, {Casagrande}, {Collet},
  {Dotter}, {Johnson}, {Lind}, {Bedin}, {Jerjen}, {Aparicio}, \&
  {Sbordone}}]{2016MNRAS.459..610M}
{Marino}, A.~F., {Milone}, A.~P., {Casagrande}, L., {et~al.} 2016, \mnras, 459,
  610, \dodoi{10.1093/mnras/stw611}

\bibitem[{{Martell} \& {Grebel}(2010)}]{2010A&A...519A..14M}
{Martell}, S.~L., \& {Grebel}, E.~K. 2010, \aap, 519, A14,
  \dodoi{10.1051/0004-6361/201014135}

\bibitem[{{Martell} {et~al.}(2016){Martell}, {Shetrone}, {Lucatello},
  {Schiavon}, {M{\'e}sz{\'a}ros}, {Allende Prieto},
  {Garc{\'{\i}}a-Hern{\'a}ndez}, {Beers}, \& {Nidever}}]{2016ApJ...825..146M}
{Martell}, S.~L., {Shetrone}, M.~D., {Lucatello}, S., {et~al.} 2016, \apj, 825,
  146, \dodoi{10.3847/0004-637X/825/2/146}

\bibitem[{{Massari} {et~al.}(2014){Massari}, {Mucciarelli}, {Ferraro},
  {Origlia}, {Rich}, {Lanzoni}, {Dalessandro}, {Valenti}, {Ibata}, {Lovisi},
  {Bellazzini}, \& {Reitzel}}]{2014ApJ...795...22M}
{Massari}, D., {Mucciarelli}, A., {Ferraro}, F.~R., {et~al.} 2014, \apj, 795,
  22, \dodoi{10.1088/0004-637X/795/1/22}

\bibitem[{{Massari} {et~al.}(2017){Massari}, {Mucciarelli}, {Dalessandro},
  {Bellazzini}, {Cassisi}, {Fiorentino}, {Ibata}, {Lardo}, \&
  {Salaris}}]{2017MNRAS.468.1249M}
{Massari}, D., {Mucciarelli}, A., {Dalessandro}, E., {et~al.} 2017, \mnras,
  468, 1249, \dodoi{10.1093/mnras/stx549}

\bibitem[{{Masseron} {et~al.}(2016){Masseron}, {Merle}, \&
  {Hawkins}}]{2016ascl.soft05004M}
{Masseron}, T., {Merle}, T., \& {Hawkins}, K. 2016, {BACCHUS: Brussels
  Automatic Code for Characterizing High accUracy Spectra}, Astrophysics Source
  Code Library, \dodoi{10.20356/C4TG6R}

\bibitem[{{Masseron} {et~al.}(2018){Masseron}, {Garc{\'{\i}}a-Hern{\'a}ndez},
  {M{\'e}sz{\'a}ros}, {Zamora}, {Dell'Agli}, {Allende Prieto}, {Edvardsson},
  {Shetrone}, {Plez}, {Fern{\'a}ndez-Trincado}, {Cunha}, {J{\"o}nsson},
  {Geisler}, {Beers}, \& {Cohen}}]{2018arXiv181208817M}
{Masseron}, T., {Garc{\'{\i}}a-Hern{\'a}ndez}, D.~A., {M{\'e}sz{\'a}ros}, S.,
  {et~al.} 2018, arXiv e-prints.
\newblock \doarXiv{1812.08817}

\bibitem[{{Masseron} {et~al.}(2019){Masseron}, {Garc{\'{\i}}a-Hern{\'a}ndez},
  {M{\'e}sz{\'a}ros}, {Zamora}, {Dell'Agli}, {Allende Prieto}, {Edvardsson},
  {Shetrone}, {Plez}, {Fern{\'a}ndez-Trincado}, {Cunha}, {J{\"o}nsson},
  {Geisler}, {Beers}, \& {Cohen}}]{2019A&A...622A.191M}
---. 2019, \aap, 622, A191, \dodoi{10.1051/0004-6361/201834550}

\bibitem[{{M{\'e}sz{\'a}ros} {et~al.}(2015){M{\'e}sz{\'a}ros}, {Martell},
  {Shetrone}, {Lucatello}, {Troup}, {Bovy}, {Cunha},
  {Garc{\'{\i}}a-Hern{\'a}ndez}, {Overbeek}, {Allende Prieto}, {Beers},
  {Frinchaboy}, {Garc{\'{\i}}a P{\'e}rez}, {Hearty}, {Holtzman}, {Majewski},
  {Nidever}, {Schiavon}, {Schneider}, {Sobeck}, {Smith}, {Zamora}, \&
  {Zasowski}}]{2015AJ....149..153M}
{M{\'e}sz{\'a}ros}, S., {Martell}, S.~L., {Shetrone}, M., {et~al.} 2015, \aj,
  149, 153, \dodoi{10.1088/0004-6256/149/5/153}

\bibitem[{{Milone} {et~al.}(2015){Milone}, {Marino}, {Piotto}, {Renzini},
  {Bedin}, {Anderson}, {Cassisi}, {D'Antona}, {Bellini}, {Jerjen},
  {Pietrinferni}, \& {Ventura}}]{2015ApJ...808...51M}
{Milone}, A.~P., {Marino}, A.~F., {Piotto}, G., {et~al.} 2015, \apj, 808, 51,
  \dodoi{10.1088/0004-637X/808/1/51}

\bibitem[{{Milone} {et~al.}(2017){Milone}, {Piotto}, {Renzini}, {Marino},
  {Bedin}, {Vesperini}, {D'Antona}, {Nardiello}, {Anderson}, {King}, {Yong},
  {Bellini}, {Aparicio}, {Barbuy}, {Brown}, {Cassisi}, {Ortolani}, {Salaris},
  {Sarajedini}, \& {van der Marel}}]{2017MNRAS.464.3636M}
{Milone}, A.~P., {Piotto}, G., {Renzini}, A., {et~al.} 2017, \mnras, 464, 3636,
  \dodoi{10.1093/mnras/stw2531}

\bibitem[{{Mu{\~n}oz} {et~al.}(2017){Mu{\~n}oz}, {Villanova}, {Geisler},
  {Saviane}, {Dias}, {Cohen}, \& {Mauro}}]{2017A&A...605A..12M}
{Mu{\~n}oz}, C., {Villanova}, S., {Geisler}, D., {et~al.} 2017, \aap, 605, A12,
  \dodoi{10.1051/0004-6361/201730468}

\bibitem[{{Mu{\~n}oz} {et~al.}(2018){Mu{\~n}oz}, {Geisler}, {Villanova},
  {Saviane}, {Cort{\'e}s}, {Dias}, {Cohen}, {Mauro}, \& {Moni
  Bidin}}]{2018A&A...620A..96M}
{Mu{\~n}oz}, C., {Geisler}, D., {Villanova}, S., {et~al.} 2018, \aap, 620, A96,
  \dodoi{10.1051/0004-6361/201833373}

\bibitem[{{Mucciarelli} {et~al.}(2015{\natexlab{a}}){Mucciarelli},
  {Bellazzini}, {Merle}, {Plez}, {Dalessandro}, \&
  {Ibata}}]{2015ApJ...801...68M}
{Mucciarelli}, A., {Bellazzini}, M., {Merle}, T., {et~al.} 2015{\natexlab{a}},
  \apj, 801, 68, \dodoi{10.1088/0004-637X/801/1/68}

\bibitem[{{Mucciarelli} {et~al.}(2018){Mucciarelli}, {Lapenna}, {Ferraro}, \&
  {Lanzoni}}]{2018ApJ...859...75M}
{Mucciarelli}, A., {Lapenna}, E., {Ferraro}, F.~R., \& {Lanzoni}, B. 2018,
  \apj, 859, 75, \dodoi{10.3847/1538-4357/aaba80}

\bibitem[{{Mucciarelli} {et~al.}(2015{\natexlab{b}}){Mucciarelli}, {Lapenna},
  {Massari}, {Ferraro}, \& {Lanzoni}}]{2015ApJ...801...69M}
{Mucciarelli}, A., {Lapenna}, E., {Massari}, D., {Ferraro}, F.~R., \&
  {Lanzoni}, B. 2015{\natexlab{b}}, \apj, 801, 69,
  \dodoi{10.1088/0004-637X/801/1/69}

\bibitem[{{Mucciarelli} {et~al.}(2017){Mucciarelli}, {Merle}, \&
  {Bellazzini}}]{2017A&A...600A.104M}
{Mucciarelli}, A., {Merle}, T., \& {Bellazzini}, M. 2017, \aap, 600, A104,
  \dodoi{10.1051/0004-6361/201730410}

\bibitem[{{Mucciarelli} {et~al.}(2016){Mucciarelli}, {Dalessandro}, {Massari},
  {Bellazzini}, {Ferraro}, {Lanzoni}, {Lardo}, {Salaris}, \&
  {Cassisi}}]{2016ApJ...824...73M}
{Mucciarelli}, A., {Dalessandro}, E., {Massari}, D., {et~al.} 2016, \apj, 824,
  73, \dodoi{10.3847/0004-637X/824/2/73}

\bibitem[{{Nardiello} {et~al.}(2015){Nardiello}, {Piotto}, {Milone}, {Marino},
  {Bedin}, {Anderson}, {Aparicio}, {Bellini}, {Cassisi}, {D'Antona}, {Hidalgo},
  {Ortolani}, {Pietrinferni}, {Renzini}, {Salaris}, {Marel}, \&
  {Vesperini}}]{2015MNRAS.451..312N}
{Nardiello}, D., {Piotto}, G., {Milone}, A.~P., {et~al.} 2015, \mnras, 451,
  312, \dodoi{10.1093/mnras/stv971}

\bibitem[{{Nataf} {et~al.}(2013){Nataf}, {Gould}, {Pinsonneault}, \&
  {Udalski}}]{2013ApJ...766...77N}
{Nataf}, D.~M., {Gould}, A.~P., {Pinsonneault}, M.~H., \& {Udalski}, A. 2013,
  \apj, 766, 77, \dodoi{10.1088/0004-637X/766/2/77}

\bibitem[{{Ness} {et~al.}(2015){Ness}, {Hogg}, {Rix}, {Ho}, \&
  {Zasowski}}]{2015ApJ...808...16N}
{Ness}, M., {Hogg}, D.~W., {Rix}, H.-W., {Ho}, A.~Y.~Q., \& {Zasowski}, G.
  2015, \apj, 808, 16, \dodoi{10.1088/0004-637X/808/1/16}

\bibitem[{{Nidever} {et~al.}(2015){Nidever}, {Holtzman}, {Allende Prieto},
  {Beland}, {Bender}, {Bizyaev}, {Burton}, {Desphande}, {Fleming},
  {Garc{\'{\i}}a P{\'e}rez}, {Hearty}, {Majewski}, {M{\'e}sz{\'a}ros}, {Muna},
  {Nguyen}, {Schiavon}, {Shetrone}, {Skrutskie}, {Sobeck}, \&
  {Wilson}}]{2015AJ....150..173N}
{Nidever}, D.~L., {Holtzman}, J.~A., {Allende Prieto}, C., {et~al.} 2015, \aj,
  150, 173, \dodoi{10.1088/0004-6256/150/6/173}

\bibitem[{{Norris}(2004)}]{2004ApJ...612L..25N}
{Norris}, J.~E. 2004, \apjl, 612, L25, \dodoi{10.1086/423986}

\bibitem[{Oliphant(2006--)}]{NumPy}
Oliphant, T. 2006--, {NumPy}: A guide to {NumPy}, USA: Trelgol Publishing.
\newblock \url{http://www.numpy.org/}

\bibitem[{{O'Malley} \& {Chaboyer}(2018)}]{2018ApJ...856..130O}
{O'Malley}, E.~M., \& {Chaboyer}, B. 2018, \apj, 856, 130,
  \dodoi{10.3847/1538-4357/aab554}

\bibitem[{{O'Malley} {et~al.}(2017){O'Malley}, {Knaizev}, {McWilliam}, \&
  {Chaboyer}}]{2017ApJ...846...23O}
{O'Malley}, E.~M., {Knaizev}, A., {McWilliam}, A., \& {Chaboyer}, B. 2017,
  \apj, 846, 23, \dodoi{10.3847/1538-4357/aa7b72}

\bibitem[{{Origlia} {et~al.}(2008){Origlia}, {Valenti}, \&
  {Rich}}]{2008MNRAS.388.1419O}
{Origlia}, L., {Valenti}, E., \& {Rich}, R.~M. 2008, \mnras, 388, 1419,
  \dodoi{10.1111/j.1365-2966.2008.13492.x}

\bibitem[{{Origlia} {et~al.}(2011){Origlia}, {Rich}, {Ferraro}, {Lanzoni},
  {Bellazzini}, {Dalessandro}, {Mucciarelli}, {Valenti}, \&
  {Beccari}}]{2011ApJ...726L..20O}
{Origlia}, L., {Rich}, R.~M., {Ferraro}, F.~R., {et~al.} 2011, \apjl, 726, L20,
  \dodoi{10.1088/2041-8205/726/2/L20}

\bibitem[{{Origlia} {et~al.}(2018){Origlia}, {Mucciarelli}, {Fiorentino},
  {Ferraro}, {Dalessandro}, {Lanzoni}, {Rich}, {Massari}, {Contreras},
  {Matsunaga}, \& {-}}]{2018arXiv181204325O}
{Origlia}, L., {Mucciarelli}, A., {Fiorentino}, G., {et~al.} 2018, arXiv
  e-prints.
\newblock \doarXiv{1812.04325}

\bibitem[{{Osborn}(1971)}]{1971Obs....91..223O}
{Osborn}, W. 1971, The Observatory, 91, 223

\bibitem[{{Pancino} {et~al.}(2017){Pancino}, {Romano}, {Tang}, {Tautvai{\v
  s}ien{\.e}}, {Casey}, {Gruyters}, {Geisler}, {San Roman}, {Randich},
  {Alfaro}, {Bragaglia}, {Flaccomio}, {Korn}, {Recio-Blanco}, {Smiljanic},
  {Carraro}, {Bayo}, {Costado}, {Damiani}, {Jofr{\'e}}, {Lardo}, {de Laverny},
  {Monaco}, {Morbidelli}, {Sbordone}, {Sousa}, \&
  {Villanova}}]{2017A&A...601A.112P}
{Pancino}, E., {Romano}, D., {Tang}, B., {et~al.} 2017, \aap, 601, A112,
  \dodoi{10.1051/0004-6361/201730474}

\bibitem[{{Piotto} {et~al.}(2005){Piotto}, {Villanova}, {Bedin}, {Gratton},
  {Cassisi}, {Momany}, {Recio-Blanco}, {Lucatello}, {Anderson}, {King},
  {Pietrinferni}, \& {Carraro}}]{2005ApJ...621..777P}
{Piotto}, G., {Villanova}, S., {Bedin}, L.~R., {et~al.} 2005, \apj, 621, 777,
  \dodoi{10.1086/427796}

\bibitem[{{Piotto} {et~al.}(2007){Piotto}, {Bedin}, {Anderson}, {King},
  {Cassisi}, {Milone}, {Villanova}, {Pietrinferni}, \&
  {Renzini}}]{2007ApJ...661L..53P}
{Piotto}, G., {Bedin}, L.~R., {Anderson}, J., {et~al.} 2007, \apjl, 661, L53,
  \dodoi{10.1086/518503}

\bibitem[{{Portegies Zwart} {et~al.}(2004){Portegies Zwart}, {Baumgardt},
  {Hut}, {Makino}, \& {McMillan}}]{2004Natur.428..724P}
{Portegies Zwart}, S.~F., {Baumgardt}, H., {Hut}, P., {Makino}, J., \&
  {McMillan}, S.~L.~W. 2004, \nat, 428, 724, \dodoi{10.1038/nature02448}

\bibitem[{{Prager} {et~al.}(2017){Prager}, {Ransom}, {Freire}, {Hessels},
  {Stairs}, {Arras}, \& {Cadelano}}]{2017ApJ...845..148P}
{Prager}, B.~J., {Ransom}, S.~M., {Freire}, P.~C.~C., {et~al.} 2017, \apj, 845,
  148, \dodoi{10.3847/1538-4357/aa7ed7}

\bibitem[{{Prantzos} {et~al.}(2007){Prantzos}, {Charbonnel}, \&
  {Iliadis}}]{2007A&A...470..179P}
{Prantzos}, N., {Charbonnel}, C., \& {Iliadis}, C. 2007, \aap, 470, 179,
  \dodoi{10.1051/0004-6361:20077205}

\bibitem[{{Price-Whelan} {et~al.}(2018){Price-Whelan}, {Sip{'{o}}cz},
  {G{"u}nther}, {Lim}, {Crawford}, {Conseil}, {Shupe}, {Craig}, {Dencheva},
  {Ginsburg}, {VanderPlas}, {Bradley}, {P{'e}rez-Su{'a}rez}, {de Val-Borro},
  {Paper Contributors}, {Aldcroft}, {Cruz}, {Robitaille}, {Tollerud},
  {Coordination Committee}, {Ardelean}, {Babej}, {Bach}, {Bachetti}, {Bakanov},
  {Bamford}, {Barentsen}, {Barmby}, {Baumbach}, {Berry}, {Biscani}, {Boquien},
  {Bostroem}, {Bouma}, {Brammer}, {Bray}, {Breytenbach}, {Buddelmeijer},
  {Burke}, {Calderone}, {Cano Rodr{'i}guez}, {Cara}, {Cardoso}, {Cheedella},
  {Copin}, {Corrales}, {Crichton}, {D{ extquoteright}Avella}, {Deil},
  {Depagne}, {Dietrich}, {Donath}, {Droettboom}, {Earl}, {Erben}, {Fabbro},
  {Ferreira}, {Finethy}, {Fox}, {Garrison}, {Gibbons}, {Goldstein}, {Gommers},
  {Greco}, {Greenfield}, {Groener}, {Grollier}, {Hagen}, {Hirst}, {Homeier},
  {Horton}, {Hosseinzadeh}, {Hu}, {Hunkeler}, {Ivezi{'c}}, {Jain}, {Jenness},
  {Kanarek}, {Kendrew}, {Kern}, {Kerzendorf}, {Khvalko}, {King}, {Kirkby},
  {Kulkarni}, {Kumar}, {Lee}, {Lenz}, {Littlefair}, {Ma}, {Macleod},
  {Mastropietro}, {McCully}, {Montagnac}, {Morris}, {Mueller}, {Mumford},
  {Muna}, {Murphy}, {Nelson}, {Nguyen}, {Ninan}, {N{"o}the}, {Ogaz}, {Oh},
  {Parejko}, {Parley}, {Pascual}, {Patil}, {Patil}, {Plunkett}, {Prochaska},
  {Rastogi}, {Reddy Janga}, {Sabater}, {Sakurikar}, {Seifert}, {Sherbert},
  {Sherwood-Taylor}, {Shih}, {Sick}, {Silbiger}, {Singanamalla}, {Singer},
  {Sladen}, {Sooley}, {Sornarajah}, {Streicher}, {Teuben}, {Thomas},
  {Tremblay}, {Turner}, {Terr{'o}n}, {van Kerkwijk}, {de la Vega}, {Watkins},
  {Weaver}, {Whitmore}, {Woillez}, {Zabalza}, \& {Contributors}}]{astropy:2018}
{Price-Whelan}, A.~M., {Sip{'{o}}cz}, B.~M., {G{"u}nther}, H.~M., {et~al.}
  2018, aj, 156, 123, \dodoi{10.3847/1538-3881/aabc4f}

\bibitem[{{Queiroz} {et~al.}(2018){Queiroz}, {Anders}, {Santiago}, {Chiappini},
  {Steinmetz}, {Dal Ponte}, {Stassun}, {da Costa}, {Maia}, {Crestani}, {Beers},
  {Fern{\'a}ndez-Trincado}, {Garc{\'{\i}}a-Hern{\'a}ndez}, {Roman-Lopes}, \&
  {Zamora}}]{2018MNRAS.476.2556Q}
{Queiroz}, A.~B.~A., {Anders}, F., {Santiago}, B.~X., {et~al.} 2018, \mnras,
  476, 2556, \dodoi{10.1093/mnras/sty330}

\bibitem[{{Ram{\'{\i}}rez} {et~al.}(2012){Ram{\'{\i}}rez}, {Mel{\'e}ndez}, \&
  {Chanam{\'e}}}]{2012ApJ...757..164R}
{Ram{\'{\i}}rez}, I., {Mel{\'e}ndez}, J., \& {Chanam{\'e}}, J. 2012, \apj, 757,
  164, \dodoi{10.1088/0004-637X/757/2/164}

\bibitem[{{Renzini}(2008)}]{2008MNRAS.391..354R}
{Renzini}, A. 2008, \mnras, 391, 354, \dodoi{10.1111/j.1365-2966.2008.13892.x}

\bibitem[{{Renzini} {et~al.}(2015){Renzini}, {D'Antona}, {Cassisi}, {King},
  {Milone}, {Ventura}, {Anderson}, {Bedin}, {Bellini}, {Brown}, {Piotto}, {van
  der Marel}, {Barbuy}, {Dalessandro}, {Hidalgo}, {Marino}, {Ortolani},
  {Salaris}, \& {Sarajedini}}]{2015MNRAS.454.4197R}
{Renzini}, A., {D'Antona}, F., {Cassisi}, S., {et~al.} 2015, \mnras, 454, 4197,
  \dodoi{10.1093/mnras/stv2268}

\bibitem[{{Roederer} \& {Thompson}(2015)}]{2015MNRAS.449.3889R}
{Roederer}, I.~U., \& {Thompson}, I.~B. 2015, \mnras, 449, 3889,
  \dodoi{10.1093/mnras/stv546}

\bibitem[{{Schiavon} {et~al.}(2017{\natexlab{a}}){Schiavon}, {Johnson},
  {Frinchaboy}, {Zasowski}, {M{\'e}sz{\'a}ros}, {Garc{\'{\i}}a-Hern{\'a}ndez},
  {Cohen}, {Tang}, {Villanova}, {Geisler}, {Beers}, {Fern{\'a}ndez-Trincado},
  {Garc{\'{\i}}a P{\'e}rez}, {Lucatello}, {Majewski}, {Martell}, {O'Connell},
  {Allende Prieto}, {Bizyaev}, {Carrera}, {Lane}, {Malanushenko},
  {Malanushenko}, {Mu{\~n}oz}, {Nitschelm}, {Oravetz}, {Pan}, {Roman-Lopes},
  {Schultheis}, \& {Simmons}}]{2017MNRAS.466.1010S}
{Schiavon}, R.~P., {Johnson}, J.~A., {Frinchaboy}, P.~M., {et~al.}
  2017{\natexlab{a}}, \mnras, 466, 1010, \dodoi{10.1093/mnras/stw3093}

\bibitem[{{Schiavon} {et~al.}(2017{\natexlab{b}}){Schiavon}, {Zamora},
  {Carrera}, {Lucatello}, {Robin}, {Ness}, {Martell}, {Smith},
  {Garc{\'{\i}}a-Hern{\'a}ndez}, {Manchado}, {Sch{\"o}nrich}, {Bastian},
  {Chiappini}, {Shetrone}, {Mackereth}, {Williams}, {M{\'e}sz{\'a}ros},
  {Allende Prieto}, {Anders}, {Bizyaev}, {Beers}, {Chojnowski}, {Cunha},
  {Epstein}, {Frinchaboy}, {Garc{\'{\i}}a P{\'e}rez}, {Hearty}, {Holtzman},
  {Johnson}, {Kinemuchi}, {Majewski}, {Muna}, {Nidever}, {Nguyen}, {O'Connell},
  {Oravetz}, {Pan}, {Pinsonneault}, {Schneider}, {Schultheis}, {Simmons},
  {Skrutskie}, {Sobeck}, {Wilson}, \& {Zasowski}}]{2017MNRAS.465..501S}
{Schiavon}, R.~P., {Zamora}, O., {Carrera}, R., {et~al.} 2017{\natexlab{b}},
  \mnras, 465, 501, \dodoi{10.1093/mnras/stw2162}

\bibitem[{{Sharma}(2017)}]{2017ARA&A..55..213S}
{Sharma}, S. 2017, \araa, 55, 213, \dodoi{10.1146/annurev-astro-082214-122339}

\bibitem[{{Shetrone}(1996)}]{1996AJ....112.2639S}
{Shetrone}, M.~D. 1996, \aj, 112, 2639, \dodoi{10.1086/118208}

\bibitem[{{Shingles} {et~al.}(2015){Shingles}, {Doherty}, {Karakas},
  {Stancliffe}, {Lattanzio}, \& {Lugaro}}]{2015MNRAS.452.2804S}
{Shingles}, L.~J., {Doherty}, C.~L., {Karakas}, A.~I., {et~al.} 2015, \mnras,
  452, 2804, \dodoi{10.1093/mnras/stv1489}

\bibitem[{{Skrutskie} {et~al.}(2006){Skrutskie}, {Cutri}, {Stiening},
  {Weinberg}, {Schneider}, {Carpenter}, {Beichman}, {Capps}, {Chester},
  {Elias}, {Huchra}, {Liebert}, {Lonsdale}, {Monet}, {Price}, {Seitzer},
  {Jarrett}, {Kirkpatrick}, {Gizis}, {Howard}, {Evans}, {Fowler}, {Fullmer},
  {Hurt}, {Light}, {Kopan}, {Marsh}, {McCallon}, {Tam}, {Van Dyk}, \&
  {Wheelock}}]{2006AJ....131.1163S}
{Skrutskie}, M.~F., {Cutri}, R.~M., {Stiening}, R., {et~al.} 2006, \aj, 131,
  1163, \dodoi{10.1086/498708}

\bibitem[{{Slemer} {et~al.}(2017){Slemer}, {Marigo}, {Piatti}, {Aliotta},
  {Bemmerer}, {Best}, {Boeltzig}, {Bressan}, {Broggini}, {Bruno}, {Caciolli},
  {Cavanna}, {Ciani}, {Corvisiero}, {Davinson}, {Depalo}, {Di Leva}, {Elekes},
  {Ferraro}, {Formicola}, {F{\"u}l{\"o}p}, {Gervino}, {Guglielmetti},
  {Gustavino}, {Gy{\"u}rky}, {Imbriani}, {Junker}, {Menegazzo}, {Mossa},
  {Pantaleo}, {Prati}, {Straniero}, {Sz{\"u}cs}, {Tak{\'a}cs}, \&
  {Trezzi}}]{2017MNRAS.465.4817S}
{Slemer}, A., {Marigo}, P., {Piatti}, D., {et~al.} 2017, \mnras, 465, 4817,
  \dodoi{10.1093/mnras/stw3029}

\bibitem[{{Tang} {et~al.}(2017){Tang}, {Cohen}, {Geisler}, {Schiavon},
  {Majewski}, {Villanova}, {Carrera}, {Zamora}, {Garcia-Hernandez}, {Shetrone},
  {Frinchaboy}, {Meza}, {Fern{\'a}ndez-Trincado}, {Mu{\~n}oz}, {Lin}, {Lane},
  {Nitschelm}, {Pan}, {Bizyaev}, {Oravetz}, \& {Simmons}}]{2017MNRAS.465...19T}
{Tang}, B., {Cohen}, R.~E., {Geisler}, D., {et~al.} 2017, \mnras, 465, 19,
  \dodoi{10.1093/mnras/stw2739}

\bibitem[{{Tang} {et~al.}(2018){Tang}, {Fern{\'a}ndez-Trincado}, {Geisler},
  {Zamora}, {M{\'e}sz{\'a}ros}, {Masseron}, {Cohen},
  {Garc{\'{\i}}a-Hern{\'a}ndez}, {Dell'Agli}, {Beers}, {Schiavon}, {Sohn},
  {Hasselquist}, {Robin}, {Shetrone}, {Majewski}, {Villanova}, {Schiappacasse
  Ulloa}, {Lane}, {Minnti}, {Roman-Lopes}, {Almeida}, \&
  {Moreno}}]{2018ApJ...855...38T}
{Tang}, B., {Fern{\'a}ndez-Trincado}, J.~G., {Geisler}, D., {et~al.} 2018,
  \apj, 855, 38, \dodoi{10.3847/1538-4357/aaaaea}

\bibitem[{{Ting} {et~al.}(2018){Ting}, {Conroy}, {Rix}, \&
  {Cargile}}]{2018arXiv180401530T}
{Ting}, Y.-S., {Conroy}, C., {Rix}, H.-W., \& {Cargile}, P. 2018, ArXiv
  e-prints.
\newblock \doarXiv{1804.01530}

\bibitem[{{Uttenthaler} {et~al.}(2007){Uttenthaler}, {Hron}, {Lebzelter},
  {Busso}, {Schultheis}, \& {K{\"a}ufl}}]{2007A&A...463..251U}
{Uttenthaler}, S., {Hron}, J., {Lebzelter}, T., {et~al.} 2007, \aap, 463, 251,
  \dodoi{10.1051/0004-6361:20065463}

\bibitem[{{Uttenthaler} {et~al.}(2018){Uttenthaler}, {McDonald}, {Bernhard},
  {Cristallo}, \& {Gobrecht}}]{2018arXiv181207434U}
{Uttenthaler}, S., {McDonald}, I., {Bernhard}, K., {Cristallo}, S., \&
  {Gobrecht}, D. 2018, arXiv e-prints.
\newblock \doarXiv{1812.07434}

\bibitem[{{Valcarce} \& {Catelan}(2011)}]{2011A&A...533A.120V}
{Valcarce}, A.~A.~R., \& {Catelan}, M. 2011, \aap, 533, A120,
  \dodoi{10.1051/0004-6361/201116955}

\bibitem[{{VandenBerg} {et~al.}(2013){VandenBerg}, {Brogaard}, {Leaman}, \&
  {Casagrande}}]{2013ApJ...775..134V}
{VandenBerg}, D.~A., {Brogaard}, K., {Leaman}, R., \& {Casagrande}, L. 2013,
  \apj, 775, 134, \dodoi{10.1088/0004-637X/775/2/134}

\bibitem[{{VandenBerg} \& {Denissenkov}(2018)}]{2018ApJ...862...72V}
{VandenBerg}, D.~A., \& {Denissenkov}, P.~A. 2018, \apj, 862, 72,
  \dodoi{10.3847/1538-4357/aaca9b}

\bibitem[{{Ventura} \& {D'Antona}(2009)}]{2009A&A...499..835V}
{Ventura}, P., \& {D'Antona}, F. 2009, \aap, 499, 835,
  \dodoi{10.1051/0004-6361/200811139}

\bibitem[{{Ventura} {et~al.}(2001){Ventura}, {D'Antona}, {Mazzitelli}, \&
  {Gratton}}]{2001ApJ...550L..65V}
{Ventura}, P., {D'Antona}, F., {Mazzitelli}, I., \& {Gratton}, R. 2001, \apjl,
  550, L65, \dodoi{10.1086/319496}

\bibitem[{{Ventura} {et~al.}(2016){Ventura}, {Garc{\'{\i}}a-Hern{\'a}ndez},
  {Dell'Agli}, {D'Antona}, {M{\'e}sz{\'a}ros}, {Lucatello}, {Di Criscienzo},
  {Shetrone}, {Tailo}, {Tang}, \& {Zamora}}]{2016ApJ...831L..17V}
{Ventura}, P., {Garc{\'{\i}}a-Hern{\'a}ndez}, D.~A., {Dell'Agli}, F., {et~al.}
  2016, \apjl, 831, L17, \dodoi{10.3847/2041-8205/831/2/L17}

\bibitem[{{Villanova} {et~al.}(2013){Villanova}, {Geisler}, {Carraro}, {Moni
  Bidin}, \& {Mu{\~n}oz}}]{2013ApJ...778..186V}
{Villanova}, S., {Geisler}, D., {Carraro}, G., {Moni Bidin}, C., \&
  {Mu{\~n}oz}, C. 2013, \apj, 778, 186, \dodoi{10.1088/0004-637X/778/2/186}

\bibitem[{{Villanova} {et~al.}(2016){Villanova}, {Monaco}, {Moni Bidin}, \&
  {Assmann}}]{2016MNRAS.460.2351V}
{Villanova}, S., {Monaco}, L., {Moni Bidin}, C., \& {Assmann}, P. 2016, \mnras,
  460, 2351, \dodoi{10.1093/mnras/stw1146}

\bibitem[{{Villanova} {et~al.}(2017){Villanova}, {Moni Bidin}, {Mauro},
  {Munoz}, \& {Monaco}}]{2017MNRAS.464.2730V}
{Villanova}, S., {Moni Bidin}, C., {Mauro}, F., {Munoz}, C., \& {Monaco}, L.
  2017, \mnras, 464, 2730, \dodoi{10.1093/mnras/stw2509}

\bibitem[{{Wang} {et~al.}(2016){Wang}, {Spurzem}, {Aarseth}, {Giersz}, {Askar},
  {Berczik}, {Naab}, {Schadow}, \& {Kouwenhoven}}]{2016MNRAS.458.1450W}
{Wang}, L., {Spurzem}, R., {Aarseth}, S., {et~al.} 2016, \mnras, 458, 1450,
  \dodoi{10.1093/mnras/stw274}

\bibitem[{{Watkins} {et~al.}(2015){Watkins}, {van der Marel}, {Bellini}, \&
  {Anderson}}]{2015ApJ...803...29W}
{Watkins}, L.~L., {van der Marel}, R.~P., {Bellini}, A., \& {Anderson}, J.
  2015, \apj, 803, 29, \dodoi{10.1088/0004-637X/803/1/29}

\bibitem[{{Webb} {et~al.}(2013){Webb}, {Harris}, {Sills}, \&
  {Hurley}}]{2013ApJ...764..124W}
{Webb}, J.~J., {Harris}, W.~E., {Sills}, A., \& {Hurley}, J.~R. 2013, \apj,
  764, 124, \dodoi{10.1088/0004-637X/764/2/124}

\bibitem[{{Wiescher} {et~al.}(2010){Wiescher}, {G{\"o}rres}, {Uberseder},
  {Imbriani}, \& {Pignatari}}]{2010ARNPS..60..381W}
{Wiescher}, M., {G{\"o}rres}, J., {Uberseder}, E., {Imbriani}, G., \&
  {Pignatari}, M. 2010, Annual Review of Nuclear and Particle Science, 60, 381,
  \dodoi{10.1146/annurev.nucl.012809.104505}

\bibitem[{{Wilson} {et~al.}(2012){Wilson}, {Hearty}, {Skrutskie}, {Majewski},
  {Schiavon}, {Eisenstein}, {Gunn}, {Holtzman}, {Nidever}, {Gillespie},
  {Weinberg}, {Blank}, {Henderson}, {Smee}, {Barkhouser}, {Harding}, {Hope},
  {Fitzgerald}, {Stolberg}, {Arns}, {Nelson}, {Brunner}, {Burton}, {Walker},
  {Lam}, {Maseman}, {Barr}, {Leger}, {Carey}, {MacDonald}, {Ebelke}, {Beland},
  {Horne}, {Young}, {Rieke}, {Rieke}, {O'Brien}, {Crane}, {Carr}, {Harrison},
  {Stoll}, {Vernieri}, {Shetrone}, {Allende-Prieto}, {Johnson}, {Frinchaboy},
  {Zasowski}, {Garcia Perez}, {Bizyaev}, {Cunha}, {Smith}, {Meszaros}, {Zhao},
  {Hayden}, {Chojnowski}, {Andrews}, {Loomis}, {Owen}, {Klaene}, {Brinkmann},
  {Stauffer}, {Long}, {Jordan}, {Holder}, {Cope}, {Naugle}, {Pfaffenberger},
  {Schlegel}, {Blanton}, {Muna}, {Weaver}, {Snedden}, {Pan}, {Brewington},
  {Malanushenko}, {Malanushenko}, {Simmons}, {Oravetz}, {Mahadevan}, \&
  {Halverson}}]{2012SPIE.8446E..0HW}
{Wilson}, J.~C., {Hearty}, F., {Skrutskie}, M.~F., {et~al.} 2012, in \procspie,
  Vol. 8446, Ground-based and Airborne Instrumentation for Astronomy IV, 84460H

\bibitem[{{Yong} {et~al.}(2009){Yong}, {Grundahl}, {D'Antona}, {Karakas},
  {Lattanzio}, \& {Norris}}]{2009ApJ...695L..62Y}
{Yong}, D., {Grundahl}, F., {D'Antona}, F., {et~al.} 2009, \apjl, 695, L62,
  \dodoi{10.1088/0004-637X/695/1/L62}

\bibitem[{{Yong} {et~al.}(2008{\natexlab{a}}){Yong}, {Grundahl}, {Johnson}, \&
  {Asplund}}]{2008ApJ...684.1159Y}
{Yong}, D., {Grundahl}, F., {Johnson}, J.~A., \& {Asplund}, M.
  2008{\natexlab{a}}, \apj, 684, 1159, \dodoi{10.1086/590658}

\bibitem[{{Yong} {et~al.}(2003){Yong}, {Grundahl}, {Lambert}, {Nissen}, \&
  {Shetrone}}]{2003A&A...402..985Y}
{Yong}, D., {Grundahl}, F., {Lambert}, D.~L., {Nissen}, P.~E., \& {Shetrone},
  M.~D. 2003, \aap, 402, 985, \dodoi{10.1051/0004-6361:20030296}

\bibitem[{{Yong} {et~al.}(2015){Yong}, {Grundahl}, \&
  {Norris}}]{2015MNRAS.446.3319Y}
{Yong}, D., {Grundahl}, F., \& {Norris}, J.~E. 2015, \mnras, 446, 3319,
  \dodoi{10.1093/mnras/stu2334}

\bibitem[{{Yong} {et~al.}(2008{\natexlab{b}}){Yong}, {Mel{\'e}ndez}, {Cunha},
  {Karakas}, {Norris}, \& {Smith}}]{2008ApJ...689.1020Y}
{Yong}, D., {Mel{\'e}ndez}, J., {Cunha}, K., {et~al.} 2008{\natexlab{b}}, \apj,
  689, 1020, \dodoi{10.1086/592229}

\bibitem[{{Zasowski} {et~al.}(2013){Zasowski}, {Johnson}, {Frinchaboy},
  {Majewski}, {Nidever}, {Rocha Pinto}, {Girardi}, {Andrews}, {Chojnowski},
  {Cudworth}, {Jackson}, {Munn}, {Skrutskie}, {Beaton}, {Blake}, {Covey},
  {Deshpande}, {Epstein}, {Fabbian}, {Fleming}, {Garcia Hernandez}, {Herrero},
  {Mahadevan}, {M{\'e}sz{\'a}ros}, {Schultheis}, {Sellgren}, {Terrien}, {van
  Saders}, {Allende Prieto}, {Bizyaev}, {Burton}, {Cunha}, {da Costa},
  {Hasselquist}, {Hearty}, {Holtzman}, {Garc{\'{\i}}a P{\'e}rez}, {Maia},
  {O'Connell}, {O'Donnell}, {Pinsonneault}, {Santiago}, {Schiavon}, {Shetrone},
  {Smith}, \& {Wilson}}]{2013AJ....146...81Z}
{Zasowski}, G., {Johnson}, J.~A., {Frinchaboy}, P.~M., {et~al.} 2013, \aj, 146,
  81, \dodoi{10.1088/0004-6256/146/4/81}

\bibitem[{{Zasowski} {et~al.}(2017){Zasowski}, {Cohen}, {Chojnowski},
  {Santana}, {Oelkers}, {Andrews}, {Beaton}, {Bender}, {Bird}, {Bovy},
  {Carlberg}, {Covey}, {Cunha}, {Dell'Agli}, {Fleming}, {Frinchaboy},
  {Garc{\'{\i}}a-Hern{\'a}ndez}, {Harding}, {Holtzman}, {Johnson}, {Kollmeier},
  {Majewski}, {M{\'e}sz{\'a}ros}, {Munn}, {Mu{\~n}oz}, {Ness}, {Nidever},
  {Poleski}, {Rom{\'a}n-Z{\'u}{\~n}iga}, {Shetrone}, {Simon}, {Smith},
  {Sobeck}, {Stringfellow}, {Szigeti{\'a}ros}, {Tayar}, \&
  {Troup}}]{2017AJ....154..198Z}
{Zasowski}, G., {Cohen}, R.~E., {Chojnowski}, S.~D., {et~al.} 2017, \aj, 154,
  198, \dodoi{10.3847/1538-3881/aa8df9}

\bibitem[{{Zasowski} {et~al.}(2019){Zasowski}, {Schultheis}, {Hasselquist},
  {Cunha}, {Sobeck}, {Johnson}, {Rojas-Arriagada}, {Majewski}, {Andrews},
  {J{\"o}nsson}, {Beers}, {Chojnowski}, {Frinchaboy}, {Holtzman}, {Minniti},
  {Nidever}, \& {Nitschelm}}]{2019ApJ...870..138Z}
{Zasowski}, G., {Schultheis}, M., {Hasselquist}, S., {et~al.} 2019, \apj, 870,
  138, \dodoi{10.3847/1538-4357/aaeff4}

\end{thebibliography}
	
	
\end{document}